%% file: main.tex
\documentclass[aps,amssymb,amsmath,pra,twocolumn,showpacs,superscriptaddress,longbibliography]{revtex4-2}

\usepackage{graphicx}
\graphicspath{{figs_main/}{figs_SI/}}

\usepackage{dcolumn}
\usepackage{bm}
\usepackage{subfigure}
\usepackage{color}
\usepackage{soul}

\usepackage{amsfonts,verbatim}

\usepackage[margin=0.6in]{geometry}

\usepackage{tikz}

\usepackage[english]{babel}
\usepackage{times}
\usepackage{dsfont}
\usepackage{latexsym}
\usepackage{float}
\usepackage{afterpage}
\usepackage{enumitem}
\usepackage{mleftright}

\definecolor{lR}{rgb}{1, 0.8, 0.79}

\usepackage{eso-pic, graphicx}

\usepackage{listings}
\usepackage{multirow}
\usepackage{xcolor,colortbl}

\usepackage{bbm}
\usepackage{upgreek}
\usepackage{newtxtext,newtxmath}

\newcommand{\nocontentsline}[3]{}
\newcommand{\tocless}[2]{\bgroup\let\addcontentsline=\nocontentsline#1{#2}\egroup}

\definecolor{Ablue}{rgb}{0.96,0.24,0.00}

\definecolor{Abluetitle}{rgb}{0.,0.24,0.51}

\definecolor{orange}{rgb}{0.96,0.24,0.00}

\definecolor{darkred}{rgb}{0.55, 0.0, 0.0}

\definecolor{darksalmon}{rgb}{0.91, 0.59, 0.48}
\definecolor{maroon}{cmyk}{0,0.87,0.68,0.32}

\setcitestyle{numbers,square,citesep={,\kern-.24em}}

\definecolor{mustard}{rgb}{1.0, 0.86, 0.35}

\addto\captionsenglish{\renewcommand{\figurename}{Fig.}}

\definecolor{Gray}{gray}{0.85}
\definecolor{LightCyan}{rgb}{0.88,1,1}
\newcolumntype{a}{$>${\columncolor{Gray}}c}
\newcolumntype{b}{$>${\columncolor{white}}c}

\usepackage{array}
\newcolumntype{L}[1]{$>${\raggedright\let\newline\\\arraybackslash\hspace{0pt}}m{#1}}
\newcolumntype{C}[1]{$>${\centering\let\newline\\\arraybackslash\hspace{0pt}}m{#1}}
\newcolumntype{R}[1]{$>${\raggedleft\let\newline\\\arraybackslash\hspace{0pt}}m{#1}}

\newcolumntype{P}[1]{>{\centering\arraybackslash}p{#1}}
\newcolumntype{M}[1]{>{\centering\arraybackslash}m{#1}}

\usepackage[colorlinks=true , citecolor=black,urlcolor=blue]{hyperref}

\input{./Commands3.tex}

\mleftright
\makeatletter
\@addtoreset{section}{part}
\makeatother

\newcommand{\beginsupplement}{%
        \setcounter{table}{0}
        \renewcommand{\thetable}{S\arabic{table}}%
        \setcounter{figure}{0}
        \renewcommand{\thefigure}{S\arabic{figure}}%
				
     }

\newcommand{\affA}{Department of Chemistry, University of California, Berkeley, Berkeley, CA 94720, USA.}
\newcommand{\affB}{Department of Physics, KTH Royal Institute of Technology, SE-106 91 Stockholm, Sweden.}
\newcommand{\affC}{Department of Physics, St. Kliment Ohridski University of Sofia, 5 James Bourchier Blvd, 1164 Sofia, Bulgaria.}
\newcommand{\affD}{Max Planck Institute for the Physics of Complex Systems, N\"othnitzer Str.~38, 01187 Dresden, Germany.}
\newcommand{\affE}{Chemical Sciences Division,  Lawrence Berkeley National Laboratory,  Berkeley, CA 94720, USA.}

\begin{document}
\title{Observation of a critical prethermal discrete time crystal created by two-frequency driving }
\author{William Beatrez}\thanks{Equal contribution}\affiliation{\affA}
\author{Christoph Fleckenstein}\thanks{Equal contribution}\affiliation{\affB}
\author{Arjun Pillai}\affiliation{\affA}
\author{Erica Sanchez}\affiliation{\affA}
\author{Amala Akkiraju}\affiliation{\affA}
\author{Jesus Alcala}\affiliation{\affA}
\author{Sophie Conti}\affiliation{\affA}
\author{Paul Reshetikhin}\affiliation{\affA}
\author{Emanuel Druga}\affiliation{\affA}
\author{Marin Bukov}\email{mgbukov@phys.uni-sofia.bg}\affiliation{\affC}\affiliation{\affD}
\author{Ashok Ajoy}\email{ashokaj@berkeley.edu}\affiliation{\affA}\affiliation{\affE}

\begin{abstract}
We report the observation of long-lived Floquet prethermal discrete time crystalline (PDTC) order in a three-dimensional position-disordered lattice of interacting dipolar-coupled $\Cs$ nuclei in diamond at room temperature. We demonstrate a novel strategy of \I{"two-frequency"} driving, involving an interleaved application of slow and fast drives that simultaneously prethermalize the spins with an emergent quasi-conserved magnetization along the $\xhat$-axis, while enabling continuous and highly resolved observation of their dynamic evolution when periodically kicked away from $\xhat$. The PDTC order manifests itself in a robust period doubling response of this drive-induced quasi-conserved spin magnetization interchanging between $\xhat$ and -$\xhat$; our experiments allow a unique means to study the formation and melting of PDTC order. We obtain movies of the time-crystalline response with a clarity and throughput orders of magnitude greater than previous experiments. Parametric control over the drive frequencies allows us to reach PDTC lifetimes up to 396 Floquet cycles which we measure in a single-shot experiment. Such rapid measurement enables detailed characterization of the entire PDTC phase diagram, rigidity and lifetime, informing on the role of prethermalization towards stabilizing the DTC response.  The two-frequency drive approach represents the simplest generalization of DTCs to multi-frequency drives; it expands the toolkit for realizing and investigating long-lived non-equilibrium phases of matter stabilized by emergent quasi-conservation laws.
\end{abstract}

\maketitle
%

 \vspace{-3mm}
\tocless\section{Introduction}
  \vspace{-1mm}
Periodically driven (Floquet) quantum systems enable the realization of novel phases of matter far from thermal equilibrium. An interesting example are discrete time crystals (DTC)~\cite{sacha2017time,khemani2019brief,Else2020} -- infinitely long-lived non-equilibrium phases of matter characterized by the dynamical breaking of time-translation symmetry~\cite{Sacha2015,khemani2016phase,else2016floquet,keyserlingk2016absolute,yao2017discrete,ho2017critical,Liao2019}, and stabilized by many-body localization, signatures of which were observed in recent experiments~\cite{Choi17,Zhang17,pal2018temporal,Smits2018,Rovny2018,Randall21,Mi21}. In the absence of localization and at high drive frequencies, energy absorption in Floquet systems is suppressed to exceedingly long times -- a phenomenon known as \I{``Floquet prethermalization”}~\cite{singh2019quantifying,abadal2020floquet,peng2021floquet}, wherein the periodically driven many-body state has a lifetime ($T_2’$) that is significantly enhanced with respect to the natural system free induction decay time $T_2^{\ast}$. The heating rate $\propto\!(T_2^{\prime})^{-1}$ is exponentially suppressed for sufficiently local interactions and large driving frequencies $\xo_f{\gg} J$, where $J$ is the intrinsic spin coupling strength~\cite{abanin_15,mori_15,Machado2020}. Recently, it has been suggested that Floquet drives can yield prethermal discrete time crystals (PDTCs)~\cite{Else17}, non-equilibrium metastable states characterized by a robust subharmonic response in the drive frequency, and with a parametrically long lifetime that harnesses prethermalization. This has also inspired extensions of time crystal phenomena to classical systems~\cite{Pizzi21,ye2021floquet,yao2020classical}. Nevertheless, characterizing the full phase diagram of the emergent prethermal DTC order, rigidity, and elucidating its thermalization dynamics towards infinite temperature, remains a challenging task~\cite{Rovny18,Kyprianidis21}. At the same time, this is critical for advancing our fundamental understanding of non-equilibrium order, and for leveraging such collective phenomena in applications, such as in quantum simulation and sensing. 

\begin{figure}[t!]
  \centering
  {\includegraphics[width=0.5\textwidth]{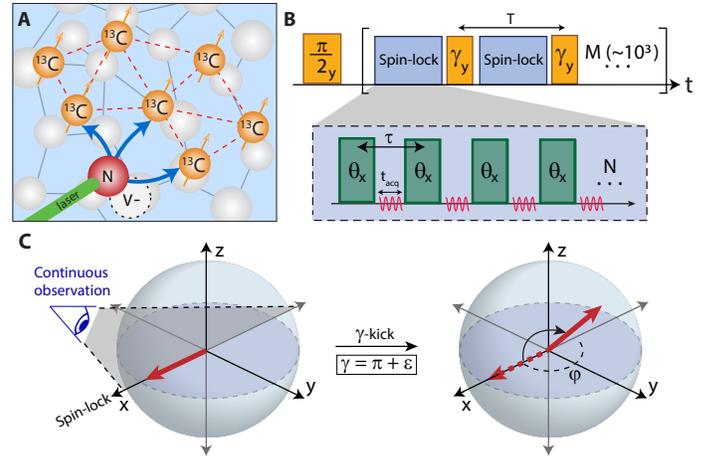}}
  \caption{\T{Experimental Implementation.} (A) \I{System.} Dipolar lattice of $\Cs$ nuclei in diamond. Dashed lines display representative interspin couplings. Optically pumped NV centers hyperpolarize the $\Cs$ nuclei (blue arrows). (B) \I{Concatenated Floquet drive} at 7T consists of interleaved application of a fast-drive (blue) along $\xhat$ and slow kicks (yellow) along $\yhat$ separated by period $T$. Fast drive is composed of a train of $\xt ({\neq}\pi)$ pulses (green) separated by period $\qt$ that spin-lock the nuclei along $\xhat$. Kick interval $T{=}N\qt$, and the ratio $N$ is tunable within the range 1-$10^3$. System undergoes evolution under interspin couplings in intervals between the pulses. Spins are interrogated by RF induction in $\tacq{=}64\mu$s windows between the pulses (red zig-zag lines); Larmor precession is sampled every 1ns. This effectively yields the ability to track projections along $\xhat$ and $\yhat$ quasi-continuously with period $\qt$. (C) \I{Schematic Bloch sphere depiction} of the reduced density matrix of a single spin in the rotating frame. Following a kick, spins prethermalize towards the quasi-stationary state along $\xhat$.  Continuous tracking of the spins in $\xy$ plane during this process is highlighted. Kicks shown are for the example $\xg{=}\pi+\vxe$, where a PDTC state manifests as a periodic switching of the spins between $+\xhat$ and $-\xhat$ between successive kicks.}
	\zfl{fig1}
\end{figure}

In this paper, we report on a novel experimental approach for the high-throughput characterization of the formation and melting of PDTCs, which permits measuring their phase diagram with unprecedented high resolution. Our experiments are performed on an interacting system of driven, hyperpolarized, $\Cs$ nuclear spins in diamond (\zfr{fig1}A), endowed with long prethermal lifetimes under Floquet driving. We propose a novel use of a \I{"two-frequency"} Floquet drive \footnote{The term frequency refers to the inverse periods of the two superimposed drives, rather than their Fourier decompositions.} applied to the $\Cs$ nuclei (see \zfr{fig1}B): the spins are made to prethermalize to an effective Hamiltonian $\ov{\mH}$ (featuring a quasi-conserved $\xhat$-magnetization~\cite{SM}), under a fast drive (with period $\qt$), while also being periodically kicked away from the prethermal state by a slower drive with period $T$ (${=}N\qt$). The deviation away from the $\xhat$ axis can be continuously monitored in periods between the fast drive, allowing a means to track the full system dynamics for long periods ({>}14s), corresponding to 450 Floquet cycles or ${>}1.35\times10^5$ fast pulses, \textit{without collapsing the quantum state}. Compared to point-by-point measurements in previous experiments, this allows a dimension reduction for mapping the PDTC phase diagram. In concert with the multiple-minute-long $T_2^{\prime}$ lifetimes of the $\Cs$ nuclei, we are able to unravel the emergent prethermal DTC order with significantly higher clarity than previous measurements. 

The two-frequency drive developed here represents a generalization of conventional Floquet driving; at once it allows the dynamical engineering of an emergent quasi-conserved quantity to stabilize non-equilibrium order at arbitrary temperatures, while simultaneously exciting the DTC order. Here we analyze the effect of driving on the long-time melting of the time-crystalline metastable state and show that the associated heating rates match well to a theoretical model. Overall therefore, this work enhances the Floquet engineering toolbox for the experimental realization and observation of non-equilibrium order beyond the paradigmatic DTC.

\begin{figure*}[t!]
  \centering
  {\includegraphics[width=1.0\textwidth]{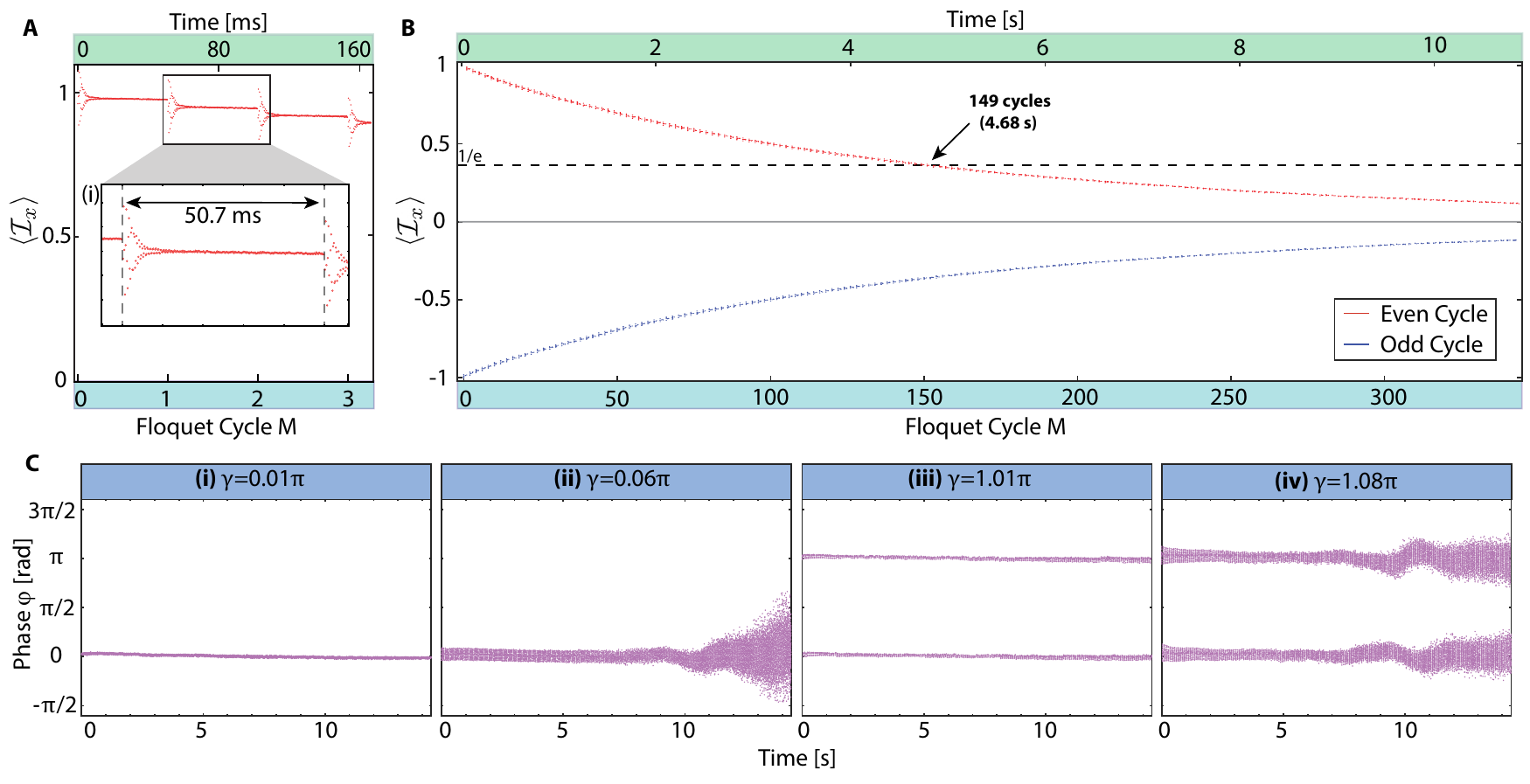}}
  \caption{\T{Continuously observed prethermalization and PDTC}. (A) \I{Floquet prethermalization} of the kicked $\Cs$ spins, shown for $\xt{=}\pi/2, \qt{=}169\mu$s, $N{=}301$, and $\xg{=}0.08\pi$. Data (points) depict single-shot measurement of $\expec{\mathcal{I}_x}$, for four kicks (165ms). Oscillations depict transient approach to prethermal plateaus (flat regions). Here, we normalize the signal upon reaching the initial prethermal state. Due to a slight frequency offset in the pulses applied, the spin-lock axis is slightly tilted from $\xhat$ and deviation from it results in the signal being apparently greater than unity during the transient. Upper axis denotes time $t$, while lower axis denotes Floquet cycle number $M$. \I{Inset:} Zoom into region between two kicks (dashed lines) separated by 50.7ms. First kick is preceded by a 1s-long prethermalization along $\xhat$ (not shown). For movie of full dataset, see ref.~\cite{PDTC_zoomed} for link. (B) \I{Single-shot PDTC measurement} for $\xt{=}\pi/2, \qt{=}105\mu$s, $N{=}300$, and $\xg{=}0.97\pi$. Panel shows the signal $\expec{\mathcal{I}_x}$; signals from alternate kicks are depicted by red and blue points.  Total time here corresponds to $1.35\zt 10^5$ periods of the fast drive, and ${\sim}$400 slow kicks (see \zfr{fig3}A for full data). PTDC decay envelope is approximately monoexponential with $1/\mathrm e$ time constant ${\app}$4.68s (dashed line). For movie of full dataset, see ref.~\cite{PDTC_video55} for link. (C) \I{PDTC phase response.} For characteristic vertical slices in \zfr{fig3}(A), we plot the phase of the spins $\xph$ on the Bloch sphere (see \zfr{fig1}C) for the entire 14s period. \I{(i)} Phase response at $\xg{=}0.01\pi$ showing the quasi-equilibrium of the spins along $\xhat$.  \I{(ii)}.  At slight deviations, $\xg{=}0.06\pi$, the spins start leaving the prethermal axis $\xhat$, and undergo heating, manifesting in the randomization of their phase at long times (here $t\gtrsim10$s). \I{(iii)} PDTC response at $\xg{=}1.01\pi$ showing stable period doubled oscillations, represented by the $\pi$-phase shift between successive $\gamma$-kicks. \I{(iv)} Melting of the PDTC order at $\xg{=}1.08\pi$, observable by the randomization of the phase at long times.
}
	\zfl{fig2}
\end{figure*}

 \vspace{4mm}
\tocless\section{System}
 We consider a lattice of $\Cs$ nuclei (\zfr{fig1}A) in diamond, optically hyperpolarized for $\tpol{=}60$s by surrounding Nitrogen Vacancy (NV) centers~\cite{Ajoy17, Ajoy18}. Hyperpolarization yields a ${\app}680$-fold enhancement in $\Cs$ magnetization over thermal equilibrium, yielding a starting density matrix $\xr_0{\sim}\xe \mathcal{I}_z$, where $\xe{=}0.68\%$ (see Methods). Here, $\mathcal{I}_{\nu}{=} \sum_j I_{j\nu}$ with $\nu {\in} \{x,y,z\}$ and $I_{j\nu}$ refer to spin-1/2 Pauli operators associated with nuclear spin $j$~\cite{Duer04}. The natural abundance (1\%) nuclei are not spatially ordered, and are coupled via dipolar interactions, $\mH_{\R{dd}}{=}\sum_{k<\ell} b_{k\ell} (3I_{kz}I_{\ell z}-\vec{I_k}\cdot \vec{I_\ell})$. One can define a characteristic energy scale via the median interspin coupling, $J{=}$0.66kHz~\cite{Beatrez21}. They are also subject to electron-mediated random on-site fields from surrounding NV and P1 paramagnetic defects, $\mH_{z}{=}\sum_j c_j I_{zj}$~\cite{Reynhardt03a,Ajoy19relax}. Despite the concomitant position disorder and random on-site fields, the three-dimensional long-range nature of the interactions precludes many-body localization. The long-range interactions also make simulating the exact Floquet dynamics for a large number of spins inaccessible using classical computers.

\zfr{fig1}B describes the experimental protocol — hyperpolarized $\Cs$ nuclei are tipped along $\xhat$, and subject to concatenated ``slow” and ``fast” Floquet drives, characterized by periods $T$ and $\qt$ respectively, with $T{=}N\qt$. First, the fast drive, consisting of a train of  $\xt ({\neq} l\pi)$ pulses (\zfr{fig1}B), engineers the internuclear Hamiltonian so that spins, initially aligned along $\xhat$ are rendered quasi-stationary~\cite{Beatrez21}. This is accomplished by arranging $\mHdd{\rt}\ov{\mH} + d_N \mathcal{I}_x$ to leading order in the Magnus expansion, such that [$\ov{\mH} + d_N \mathcal{I}_x, \mathcal{I}_x]{=}0$~\cite{SM}. This quasi-conservation causes the spins to prethermalize along $\xhat$ due to the nonintegrable character of $\ov{\mH}$. By contrast, in the absence of the fast drive, evolution under $\mHdd$ causes system observables to rapidly (in $J^{-1}$) become indistinguishable from a featureless infinite temperature state. 

Interspersed at period $T$, the slower drive \I{kicks} the spins along the $\yhat$ (or $\zhat$) axis with angle $\gamma$ (see \zfr{fig1}B). The spins are allowed to prethermalize back along $\xhat$ between successive kicks (see \zfr{fig2}A).  \zfr{fig1}C shows this visually on the Bloch sphere (in the rotating frame) for a kick of angle $\gamma{=}\pi {+} \vxe$. Therefore, in the prethermal plateau, the system is governed by an effective Hamiltonian obtained through an inverse frequency expansion~\cite{SM}. For $\gamma {=} \{0, \pi\}$, although the slow $\yhat$-kicks do not cause any extra heating, they give rise to non-equilibrium ordered states. At $\gamma{=}\pi$ the two-cycle time-evolution operator, $U_F^2{=}\exp[-i2T\overline{\mathcal{H}}]$, is governed by the $\mathbb{Z}_2$ symmetric many-body Hamiltonian $\overline{\mathcal{H}}$, where the $\mathbb{Z}_2$ symmetry is implemented by flipping the $\xhat$-direction of all spins. This drive-induced symmetry of $U_F^2$, together with the discrete time-translation invariance, creates a \textit{spatio-temporal} eigenstate order in $U_F$. Any initial state that breaks this symmetry is forced to oscillate with period $2T$, forming a PDTC state. In the experiment, we additionally observe that interactions stabilize a finite region near $\gamma{=}\pi$, where a stable PDTC period doubling response arises, with the spins flipping from $+\xhat$ to $-\xhat$ between successive kicks (\zfr{fig1}C).

A distinguishing feature of our experiments is the ability to quasi-continuously track the prethermalization dynamics after each kick. The spins are \textit{non-destructively} interrogated by Nuclear Magnetic Resonance (NMR) in $\tacq ({>}0.6\qt)$ windows between the fast pulses (\zfr{fig1}B). The magnitude and phase of the $\Cs$ Larmor precession is sampled every 1ns, allowing one to reconstruct the instantaneous projections $\expec{\mathcal{I}_x(t)}$, as well as the phase $\phi(t)$ of the spin vector in the $\xy$ plane (see Methods). In reality, the quasi-continuous time variable $t$ is discretized in units of $\qt$. Hyperpolarization enables high signal-to-noise (SNR${\gtrsim}10^2$) signal acquisition per measurement point.

It is worth emphasizing that, although rotating-frame DTCs are also observable in our system under a suitable single-frequency drive, either no continuous measurement can be performed or the spins rapidly decay to infinite temperature near the DTC point $\xt = \pi$~\cite{SM}. Two-frequency driving circumvents this problem since $\xt$ can be arbitrarily chosen (except for $\xt{=}\{0,\pi\}$) and DTCs instead appear conditioned on $\gamma$. Concatenated driving, therefore, engineers a separation between the interaction driven spin-dynamics and breaking of a discrete spatio-temporal symmetry~\cite{SM}.  Additionally, the ratio $N{=}T/\qt$ is tunable, affording flexibility in exploring dynamical regimes at small and large $N$. In typical experiments, $NM{>}1.35\times 10^5$ fast pulses are applied, corresponding to $M{>}$450 Floquet cycles. Our approach portends observing the long-time and intra-period time-crystalline dynamics continuously and without state reinitialization. This constitutes a vast improvement in throughput with respect to contemporary experiments, and yields access to \I{movies} that elucidate the formation and melting of the DTC state (see SI~\cite{SM}). The resulting measurement speedup is at least $NM{>}10^5$; compared to experiments probing lab-frame ($\zhat$) DTCs~\cite{Rovny18}, these gains could be as much as $NM(T_1/\tpol){\sim}10^7$.

\begin{figure*}[t!]
  \centering
  {\includegraphics[width=1.0\textwidth]{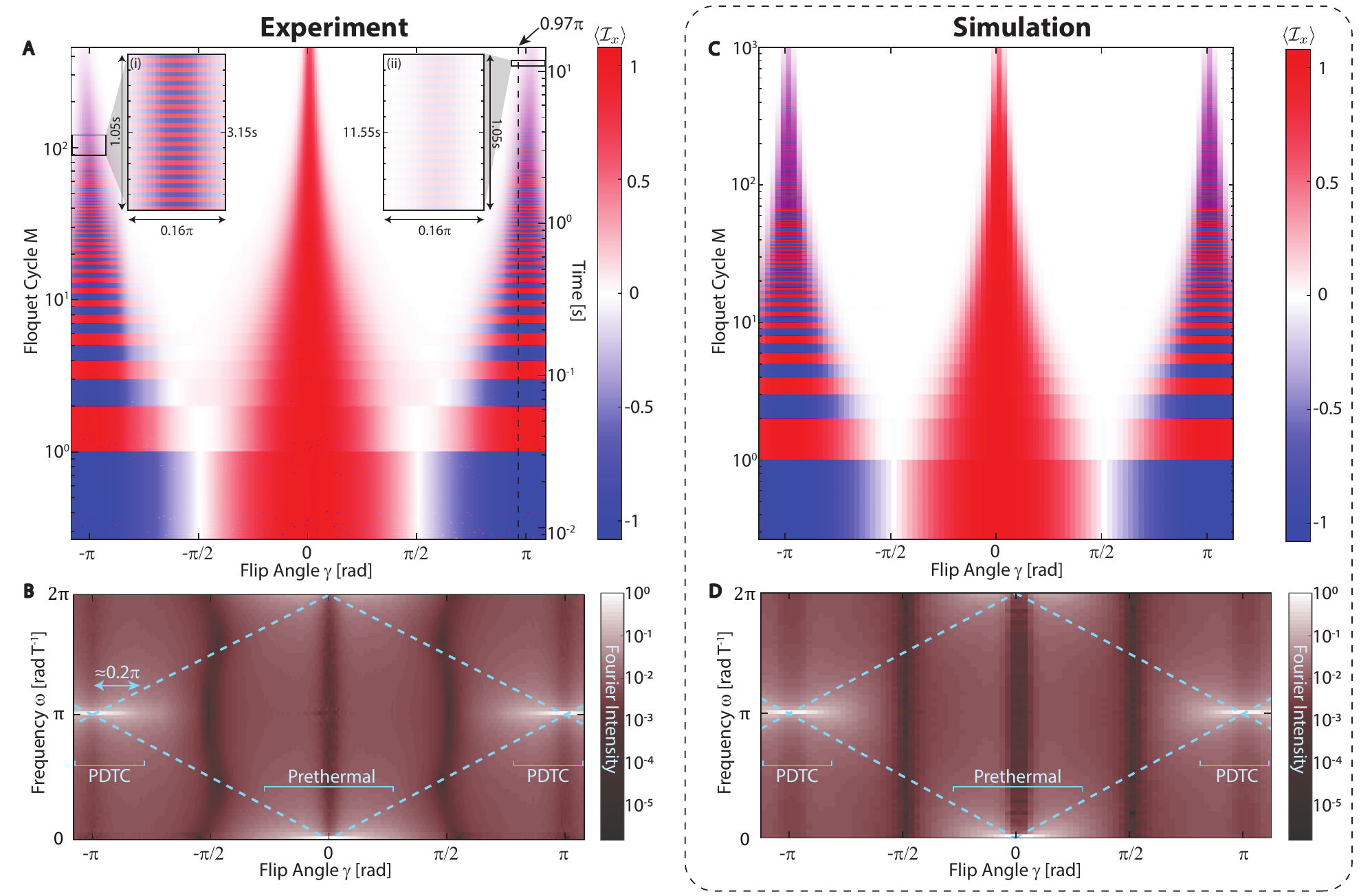}}
  \caption{\T{Prethermal DTC phase diagram}. (A) \I{Movie} showing emergence of prethermal DTCs. 285 traces similar to \zfr{fig2}B are plotted stacked for different values of kick angle $\xg$ in $[-1.1\pi,1.1\pi]$. Colors represent signal $\expec{\mathcal{I}_x}$ (see colorbar). Time $N\qt$ and Floquet cycle number $M$ run vertically and are plotted on a logarithmic scale. Data are taken to 14s ($M{=}450$ cycles). Central feature (near $\xg{=}0$) shows stabilization of long-time spin survival as a result of drive-induced quasiconservation of $\xhat$-magnetization due to Floquet prethermalization. PDTC response (striped regions) are visible near ${\xg}={\pm}\pi$. Striped signal denotes spins inverting between $\xhat$ and $-\xhat$ in a period-doubled fashion. Spins flip every $N{=}300$ fast pulses, and rigid PDTC response (peaks) persist to three decades of the fast drive. Remaining regions are characterized by rapid spin decay due to dipolar interactions, and are indistinguishable from the infinite-temperature state. \I{Inset:} Zoom into PDTC response in two 1s-windows centered at $t{=}3.15$s, and $t{=}11.55$s respectively plotted on a linear scale. Rigid DTC behavior is observable for over $14$s, even at $\xg{=}0.97\pi$. See ref.~\cite{PDTC_video55} for movie version of same dataset. (B) \I{Fourier transform} of the data in (A), plotted with respect to the inverse period of the slow drive $\omega=2\pi~T^{-1}$ radians. Colors represent strength of Fourier peak intensity spanning six orders of magnitude (colorbar). PDTC (period-doubled) order is evident by the extended white peaks at $(\gamma,\omega)=(\pm\pi,\pi)$, while the trivial (prethermal) phase arises for $(\gamma,\omega)=(0,0)$. Dashed lines indicate expected Fourier peak pattern in the absence of interactions. (C-D) \I{Numerical simulations} analogous to the experiments in A-B for $L{=}14$ interacting spins on a pseudo random-graph with $J\tau =0.07$ and $N=300$ (see ~\cite{SM}). There is excellent qualitative agreement with experimental observations.
  }
	\zfl{fig3}
\end{figure*}

 \vspace{4mm}
\tocless\section{Prethermalization and discrete time crystals} 
 \zfr{fig2}A first clarifies the prethermalization process during the spin kicks, essential to ultimately generate the PDTCs. Shown is a single-shot trace plotting $\expec{\mathcal{I}_x}$ after every fast pulse. We display the Floquet cycle number $M$ on the lower $\xhat$-axis and absolute time on the upper $\xhat$-axis for an exemplary 165ms window. Here $\xt{=}\pi/2$, and the $\gamma$-kicks are applied every 50.7ms (denoted by the dashed lines). Each $\gamma$-kick is associated with transient dynamics of the coupled $\Cs$ nuclei as they prethermalize along $\xhat$, producing the flat plateau-like regions shown. The inset \zfr{fig2}A\I{(i)} zooms into one representative transient; the oscillation period here is set approximately by the number of fast pulses required to complete a $2\pi$ rotation. The high temporal resolution revealed by the high SNR, and the ability of the $\Cs$ nuclear system to sustain a large number of $\gamma$-kicks (see also \zfr{fig3}A) allows us to track the kicked prethermalization dynamics for long periods.

Employing an exemplary choice of flip-angle $\xg{=}0.97\pi$ slightly away from the perfect DTC point, \zfr{fig2}B demonstrates generation of stable DTC order, exhibiting period doubling in $\expec{\mathcal{I}_x(t)}$ during the application of $M{>}$450 $\xg$-kicks (see \zfr{fig3}A for full data). The data were collected in a single run of the experiment. Red (blue) colors here represent odd (even) $\xg$-kicks respectively, and prethermal plateaus separate successive kicks. These data comprise ${>}1.35\zt 10^5$ fast pulses, and \I{rigid} DTC behavior is observable for ${>}$14s (see also \zfr{fig3}A). The decay of the signal is approximately mono-exponential with a $1/\mathrm e$ lifetime $t{\app}4.68$s (corresponding to $M_e{\app} 149$ Floquet cycles at $\gamma{=}0.97\pi$), making it amongst the longest DTCs observed in the literature. Moreover, the $Jt{\app}10^4$ value here is considerably beyond state-of-art for systems exhibiting DTC order~\cite{Zhang17, Choi17,Randall21,Mi21,Kyprianidis21}, demonstrating an ability to probe long-time dynamics in our system. This long-time stability can be attributed to the emergent quasi-conservation of $\xhat$-magnetization under the evolution engineered by the two-frequency drive. In particular, our driving protocol allows the formation and observation of DTC order, even if temperature in the prethermal plateau is (close to) infinite~\cite{luitz2020prethermalization}, but also at room temperature, which may be well above the critical temperature associated with symmetry breaking~\cite{SM}. The long-range nature of the spin-spin interaction suggests critical DTC order \cite{ho2017critical}. On the other hand, the lifetime of the DTC order is parametrically controlled by the frequency (of switching) of the employed drives (see sec.~\ref{sec:melting} and SI \cite{SM}). Thus, the observed DTC order corresponds to a \textit{critical prethermal} DTC.

Continuous observation yields an insightful view into the thermalization dynamics away from the stable points, a challenging task in other experimental systems. This is demonstrated in data focusing on the instantaneous phase $\xph(t) {=}\tan^{-1}(\expec{\mathcal{I}_y}/\expec{\mathcal{I}_x})$ of the spins in the $\xhat\tm\yhat$ plane after every fast pulse. This is captured by the points in \zfr{fig2}C for the full 14s experiment, and considering four representative constant-$\gamma$ values (see \zfr{fig3}A for full data). When $\xg{=}0.01\pi$, the spins are locked at $\xph{\app}0$, reflecting prethermalization along $\xhat$. A slight deviation $\xg{=}0.06\pi$, reveals transient oscillations in $\xph$ with every $\xg$-kick but no sign inversion. The transients result in the observed data spreading around $\xph{=}0$. \zfr{fig2}C(ii) therefore permits visualization of the "melting" of the prethermal $\xhat$-magnetization order to infinite temperature, where $\xph$ becomes random; this is observable at $t{\gtrsim}10$s. Analogously, the PDTC order (\zfr{fig2}C(iii)), here at $\xg{=}1.01\pi$, is visible as characteristic $\pi$-phase switching between successive kicks. The last panel (\zfr{fig2}C(iv)) denotes $\xg{=}1.08\pi$, when again PDTC melting can be observed via the phase randomization at long times. 

Collating 285 such data traces while varying angle $\xg{\in}([-1.1\pi,1.1\pi])$, it is possible to construct a movie of the kicked prethermalizing spins (see ancillary .gif files in SI~\cite{SM}). The result is plotted in \zfr{fig3}A, where we display $\expec{\mathcal{I}_x}$ (cf.~colorbar), with each vertical slice corresponding to measurement data as in \zfr{fig2}B. Left and right vertical axes here refer to the Floquet cycle number $M$ and absolute time, respectively, on a logarithmic scale. The data highlight the almost three decades in the slow drive kicks, for which we observe the PDTC order. The transition into and out of the finite PDTC regions is clearly evident, and not easily accessible in other experiments, allowing the ability to precisely characterize the heating dynamics (see \zfr{fig4}), and map the entire non-equilibrium phase diagram.
Near $\gamma{=}0$, we observe a suppression of heating due to the emergent drive-induced quasiconservation of the $\xhat$-magnetization; the low-frequency kicks approximately wrap to the identity, and the effective Hamiltonian $\mHeff{\app} \ov{\mH}$ is built solely from the high-frequency $\xhat$-drive. 
Instead, near $\xg{=}\pm\pi$, we observe a region corresponding to PDTC order, where there is a regular switching of the magnetization from $\xhat$ to -$\xhat$ with every slow kick (here 300 fast pulses), a signature of period doubling response. Interspin interactions play a crucial role to preserve a uniform switching frequency away from $\gamma\!=\!\pi$.
The insets (\zfr{fig3}A\I{(i)} and \I{(ii)}) show zooms into PDTC regions in a 1.05s window at $t{=}3.15$s and $t{=}11.55$s respectively. We plot them here on a linear scale with time for clarity. While faint, the stable periodic DTC response is markedly clear even in \zfr{fig3}A\I{(ii)}, highlighting the high SNR in the experiment.

A complementary view of \zfr{fig3}A is presented in \zfr{fig3}B, where we consider the Fourier transform of $\expec{\mathcal{I}_x(M)}$, the mean signal value between successive $\xg$-kicks for each value of $\xg$ (vertical slices in \zfr{fig3}A). Plotted is the corresponding Fourier intensity on logarithmic scale, where a span to six orders of magnitude is visible. The PDTC response appears as a sharp period-doubling peak in frequency at $(\gamma,\omega)=(\pm\pi,\pi)$. Using a 20\% magnitude threshold, we estimate the finite $\gamma$-extent of the rigid PDTC regions to $\Delta\vxe{=}\pm0.2\pi$ about $\xg{=}\pm\pi$. Similarly, the long-lived prethermal phase at $\xg{=}0$ appears as a sharp peak at $(\gamma,\omega)=(0,0)$. For comparison, the dark regions correspond to rapid state decay near $\xg{=}\pm \pi/2$, where the effective Hamiltonian no longer features a quasi-conserved $\xhat$-magnetization~\cite{SM}. Finally, we note in the dashed lines in \zfr{fig3}B the expected position of the Fourier peaks in the absence of interactions, $J{=}0$. In this diagram, the peak positions trace a rhombic pattern (dashed lines). The difference in the experimental data is evident, which indicates the role played by interactions in our system.

Comparing the experimental measurements to corresponding numerical simulations (see \cite{SM} for details), we find an excellent qualitative agreement: In \zfr{fig3}C-D we display the exact simulation results matching experimental conditions in the data in \zfr{fig3}A-B. Interestingly, we observe that the long-range interactions, together with the random spin positions, induce a self-averaging effect in the simulations so that reliable theoretical results can already be obtained for moderately small system sizes (here $L{=}14$ spins). By contrast, the experimental platform comprises a cluster of about $L{\sim}10^4$ interacting spins~\cite{Ajoy19relax} which outcompetes the numerically reachable system sizes by three orders of magnitude. The presence of a large number of interacting degrees of freedom is crucial for experimentally observing collective statistical mechanics phenomena such as symmetry breaking and thermalization. Thus, our results indicate that Floquet-engineered $\Cs$ nuclei can serve as a competitive quantum simulator of thermalizing spin dynamics. 

The observed PDTC behavior is also insensitive to the initial state. While we lack microscopic control over the initial state, different and highly non-trivial initial states can be obtained from letting the initial hyperpolarized density matrix $\rho_0$ evolve under the system Hamiltonian $\mathcal{H}_\mathrm{dd}+\mathcal{H}_z$ up to times $\sim\! J^{-1}$. For a set of such states we find comparable results between experiment and simulation~\cite{SM}. Our system thus satisfies all required landmarks of PDTCs: a parametrically long lived prethermal window featuring spatio-temporal symmetry breaking, which is rigid over a finite $\gamma$-region, and insensitive to fine-tuned initial states.

\begin{figure}[t!]
  \centering
  {\includegraphics[width=0.49\textwidth]{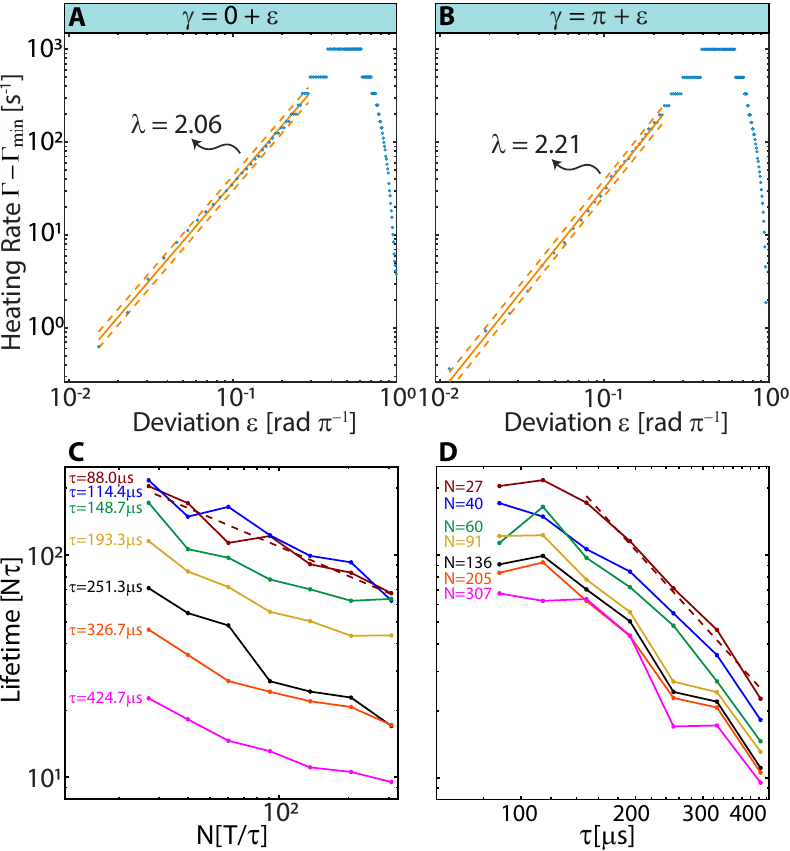}}
  \caption{\T{Experimental characterization of PDTC rigidity and prethermal lifetimes.} (A-B) \I{System heating rates} estimated by the inverse of a $1/\mathrm   e$-threshold lifetime. We observe the heating rates follow a power law $\xG{\propto} g\xe^\xl+\xG_{\mathrm{min}}$ [see text], with an exponent $\xl{=}2.06$ for prethermal phase (A) and $\xl{=}2.21$ for PDTC phase (B), agreeing well with numerical simulations  \cite{SM}. Solid line is a fit and dashed lines are error bars at two standard deviations. (C-D) \I{Prethermal lifetimes} for different values of $N$ and $\tau$. Dashed lines correspond to power-law fits with exponents $ -0.44 $ (C) and $ -1.90$ (D). Increasing the drive frequency leads to an increase in lifetime of the prethermal DTC state.
  }
	\zfl{fig4}
\end{figure}

 \vspace{4mm}
\tocless\section{ \label{sec:melting} Melting of prethermal order} 

To quantify the stability of the PDTC order away from the stable point at $\gamma{=}\pi$, we investigate the influence of finite $\varepsilon$ on its lifetime, which we define as the $1/\mathrm e$ decay time of the signal. In turn, the inverse lifetime defines the associated heating rate. The long-range character of the dipolar interactions leads to a logarithmically divergent total energy of the system in three dimensions that, when combined with the lack of Lieb-Robinson bounds~\cite{Machado2020}, makes the theoretical analysis of the heating rates for large systems difficult. For these reasons, to quantify the heating rates of the two-frequency drive, we perform a series of numerical simulations, and compare the results against experimental observations~\cite{SM}.

Note that, even at $\gamma {=}\pi$ (where the slow drive does not contribute to heating) the system slowly heats up due to energy absorption resulting from the fast (spin-locking) $\xhat$-drive. This provides a lower bound for relevant heating rates, $\Gamma_{\mathrm{min}}(\tau)$, which we find scales as a power-law of the $\xhat$-drive period, $\Gamma_{\mathrm{min}} {\propto} (J\tau)^{2}$~\cite{Beatrez21}. Finite $\varepsilon$ opens an additional channel for the system to absorb energy and, eventually, the heating rates of both drives conspire to yield a combined overall heating rate $\Gamma(\tau,\varepsilon)$. In \zfr{fig4} we extract the heating rates associated to finite $\vxe$, where we display $\Gamma - \Gamma_{\mathrm{min}}$ for $|\expec{\mathcal{I}_x(t)}|$ away from the stable points ($\xg{=}\{0,\pi\})$ obtained from \zfr{fig3}A-B, and plotted on a logarithmic scale with respect to both heating rate and deviation $\vxe$.
The curves indicate a parametrically controllable  power-law heating with an exponent ${\app}2.21$, consistent with Fermi's Golden rule. Intriguingly, we observe that the power law heating rates are close to identical for both stable points $\gamma{=}\{0,\pi\}$, where the relevant timescales are governed by a Lorentzian $\Gamma^{-1} {\propto} [g \vxe^2 +\Gamma_{\mathrm{min}}]^{-1}$ for some system-dependent constant $g$; this behavior is also borne out in numerical simulations \cite{SM}.

For the observed prethermal DTC lifetime to be \textit{controllable}, relevant life-times are expected to increase with increasing drive frequencies. Indeed, the heating time in units of Floquet cycles $\Gamma^{-1}/N$ depends sensitively on the frequency of the Floquet drive $\omega_F = 2\pi/(N\tau)$. In particular, at $\vxe =0$ our model predicts $\Gamma^{-1}/N \sim 1/(N J\tau^2)$. Increasing the frequency of switching by tuning $N$ ($\tau$), is expected to lead to a linear (quadratic) increase of $\Gamma^{-1}/N$, which we have confirmed numerically for our system \cite{SM}. In \zfr{fig4}C-D we display experimental measurements of the dependence of $\Gamma^{-1}/N$ as a function of $N$ and $\tau$, for finite $\vxe$. Although a crisp $1/N$ dependence is washed out in the presence of finite $\vxe$ (see also \cite{SM}), we observe a clear increase in the lifetime. For sufficiently large $\tau$ in the timescale-separated regime $T/\tau\gg 1$, $\Gamma^{-1}(\tau)/N$ as a function of $\tau$ agrees well with a power law with an exponent close to $-2$, while at very small $\tau$ the DTC lifetime increase comes to a halt. This is to be expected for fixed $\varepsilon$, since decreasing the value of $\tau$ reduces the many-body nature of the effective Hamiltonian (see \cite{SM}). Note that DTC order is a many-body effect that relies on spontaneous spatio-temporal symmetry breaking in interacting systems. Thus, when the many-body nature gets gradually reduced, the stability and rigidity of the DTC order are expected to decline accordingly.

\vspace{4mm}
\tocless\section{ \label{sec:small_N} Two-frequency Floquet engineering} 

When the condition $N {=} T/\tau {\gg} 1$ is not met, time-scale separation between the slow and the fast drive is violated, and a complex interplay between the two drives emerges. In this regime, the two drives cannot be treated independently of one another; instead, they mutually influence each other leading to novel effective Hamiltonians that sensitively depend on $N$. Consequently, the corresponding heating diagrams exhibit significant differences, as compared to the time-scale separated case, but also for different values of $N$. In \zfr{fig5} we show the time-evolution of the $\xhat$-magnetization as a function of $\gamma$ for $N=8$ and $N=9$. Three features are particularly noteworthy: 
(i) in contrast to the timescale separated case, the heating dynamics around $\gamma=\pi$ differs significantly from that around $\gamma=0$. 
In particular, 
(ii) far away from $\gamma = \pi$, heating depends sensitively on the specific value of $N$, where even a minimal change (in $N$) can induce completely different heating behaviours (compare \zfr{fig5}A-B). These features can be explained by the different effective Hamiltonians (see \cite{SM}) forming at different $N$, $\gamma$, respectively. 
(iii) Even though interference effects induce new effective Hamiltonians with case-specific heating properties, the formation of time-crystalline order is stable against these deformations (\zfr{fig5}C-D). 
However, the relevant lifetimes and regimes of rigidity depend sensitively on $N$ \cite{SM}: changing $N=8$ into $N=9$ amounts in a (late) lifetime increase from $270$ to $396$ Floquet cycles (cf. \zfr{fig5}C-D)).
These results provide a proof-of-principle example for the fine interplay between the two drives which offers a versatile tool to engineer new kinds of effective Hamiltonians with orchestrated physical properties, such as DTC order and beyond.

\begin{figure}[t!]
  \centering
  {\includegraphics[width=0.49\textwidth]{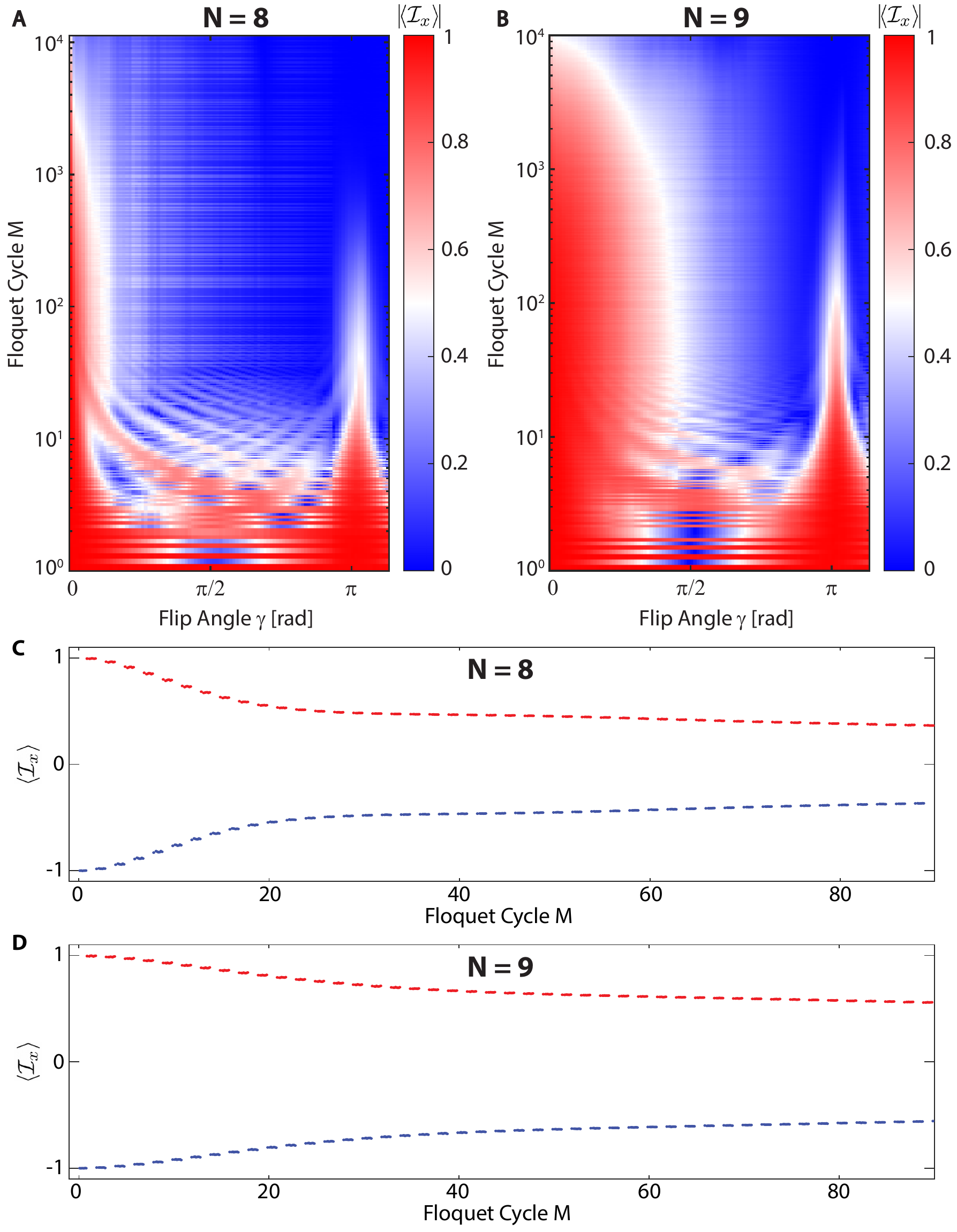}}
  \caption{\T{Experimental characterisation of the heating dynamics at small N.} 
  	Complex interplay between the fast and slow periodic drives is seen when the condition $N=T/\tau \gg 1$ is not met. (A) Movie showing time-series data from 103 experiments with different values of flip angle $\gamma$ in $[0,1.1\pi]$ and $N = 8$ $\xhat$-pulses between $\yhat$-pulses. 
	Colors represent absolute value of signal $\vert \langle \mathcal{I}_x\rangle \vert$ (see colorbar). Floquet cycle number $M$ runs vertically on a logarithmic scale. Data are taken past $10^4$ Floquet cycles. (B) Movie similar to (A) but with  $N=9$  $\xhat$-pulses between the $\yhat$-pulses. In both movies (A) and (B), $\xhat$-pulse flip angle was calibrated to $\vartheta = \pi/2$. (C-D) Line cut of (A-B), respectively, at $\gamma=1.011\pi$, shows a clear DTC signal with even (odd) cycles shown in red (blue). Late lifetimes (measured starting from Floquet cycle 100, after the initial transient regime) are 270 Floquet cycles for $N=8$ (C) and 396 Floquet cycles for $N=9$ (D). Due to the complexity of signal processing we plot absolute values in (A-B) and unwrap the phase information only for the line cuts shown in (C-D). 
}
	\zfl{fig5}
\end{figure}

 \vspace{4mm}
\tocless\section{Conclusions and Outlook} 

Summarizing, we have observed critical prethermal discrete time crystals in dipolar-coupled nuclear spins in a bulk three-dimensional solid. We developed a novel protocol to excite and observe the PDTC formation and melting using a concatenated two-frequency Floquet drive. Parametric control over both drive frequencies allows us to reach PDTC lifetimes up to 396 Floquet cycles, observable in a single run of the experiment. 
This experimental advance unveils properties of the PDTC with a high degree of clarity, including its rigidity and melting characteristics, and to map the entire prethermal phase diagram of the system. Our measurements are in excellent agreement with numerical simulations, and approximate theoretical predictions~\cite{SM}.

Our study greatly expands the Floquet engineering toolkit: it portends multi-frequency concatenated drives to excite and stabilize quantum matter far from equilibrium by engineering emergent quasi-conservation laws that offer protection against immediate high-temperature melting. The novel Floquet driving protocol we introduced paves the way to experimentally realizing intriguing non-equilibrium states, potentially including PDTCs with a stable fractional subharmonic response~\cite{pizzi2021higher}, or PDTCs and topological prethermal states stabilized by quasiperiodic driving~\cite{else2020longlived}. 
An extension of this work to multi-frequency driving is well within the scope of present-day experimental capability, and can be used to introduce more degrees of freedom to Floquet engineering. We envision application of these ideas in quantum simulation and sensing in AMO platforms as well as with hyperpolarized prethermal spins in solid-state systems.

\vspace{4mm}
\tocless\section{Acknowledgments}

We thank J.~Bardarson, M.~Heyl, C.~von Keyserlingk, D.~Luitz, R.~Moessner, J. Reimer, and D. Suter for valuable discussions. A.A. acknowledges funding from ONR under contract N00014-20-1-2806. C.F. acknowledges support from the European Research Council (ERC) under the European Union’s Horizon 2020 research and innovation program (grant agreements No. 679722 and No. 101001902). M.B. was supported by the Marie Sklodowska-Curie grant agreement No 890711, and the Bulgarian National Science Fund within National Science Program VIHREN, contract number KP-06-DV-5 (until 25.06.2021). Computational work reported on in this paper was performed on the W\"urzburg HPC cluster.

\noindent\T{\I{Data availability}} -- Raw data and scripts to process them is available upon request.

\noindent\T{\I{Author contributions}} -- 
W.B., C.F., M.B., and A.A.~conceived the research. 
W.B., A.P., E.S., A.Ak., J.A., S.C., P.R, E.D., and A.A. set up the experimental apparatus, performed measurements and analyzed the data.
C.F. performed the numerical simulations and the perturbative analysis. 
A.A.~and M.B. supervised the experiment and the theory work.

\noindent\T{\I{Competing interests}} -- Authors declare no competing interests.

\vspace{+5mm}
\tocless\section{\label{sec:methods}Methods}

\T{\I{1. Sample and hyperpolarization methodology:}} Here we employ a CVD grown single crystal of diamond with a defect density of ~1ppm NV centers and natural abundance $\Cs$. 
    The sample is placed with its [100] face parallel to the hyperpolarization and readout magnetic fields (~38mT and 7T respectively). $\Cs$ nuclei are hyperpolarized via the NV centers via a protocol described in Ref.~\cite{Beatrez21} involving continuous optical pumping and swept microwave (MW) irradiation. Polarization is transferred from the NV electrons to proximal $\Cs$ nuclei via a spin-ratchet mechanism that involves traversals of Landau-Zener anticrossings in the rotating frame of the swept MWs. The traversal probabilities are nuclear spin-state selective and are biased towards one nuclear spin orientation; this results in hyperpolarization. Spin diffusion serves to transfer this polarization to nuclear spins in the bulk lattice. Typical bulk polarization levels reached are about 0.6\%.\\
    
\T{\I{2. Data collection and processing}}: The data processing pipeline follows a similar approach as described in Ref.~\cite{Beatrez21}. The NMR signal is sampled continuously in $t_{\R{acq}}$ windows between the fast spin-lock pulses, at a sampling rate of 1GS/s via a Tabor Proteus arbitrary waveform transceiver. Continuous observation exploits the fact that the NMR coil produces no backaction on the spins. In typical experiments, the pulses are spaced apart by $\qt{=}105\;\mu$s and the acquisition windows are $t_{\R{acq}}{=}64\;\mu$s. The $\Cs$ Larmor precession (at ${\app}$75MHz) is heterodyned to 20MHz prior to digitization. For each $t_{\R{acq}}$ acquisition window, we take a Fourier transform and extract the magnitude and phase of the 20MHz peak. This corresponds to the application of a digital bandpass filter with a linewidth of $t_{\R{acq}}^{-1}{\app} 31.2$kHz. Fast digitization hence yields signal-to-noise (SNR) gains, and typical SNR per point (detection window) is ${\gtrsim}10^2$. For the 14s long acquisition in the paper, there are $\sim$135k such data collection windows.\\

\T{\I{3. Extraction of amplitude and phase in \zfr{fig2} and \zfr{fig3}}}:
The magnitude and phase of the Fourier transform of the heterodyned precession in each $t_{\R{acq}}$ readout window report respectively on the magnitude and phase of the spin vector in the $\xhat\tm\yhat$ plane in the lab frame. This corresponds to magnitude $S_L{=}[\expec{\mathcal{I}_x^2} +\expec{\mathcal{I}_y^2} ]^{1/2}$, and phase $\xph_L{=}\expec{\mathcal{I}_y}/\expec{\mathcal{I}_x}$, where subscript $L$ here refers to the lab frame. It is more convenient to to instead obtain the phase of the spins $\xph_R{=}\xph$ in the rotating frame. To do this, we note that the phase values $\xph_L$ obtained in successive $t_{\R{acq}}$ windows just differ by the (trivial) phase accrued during the $t_p$ spin-locking pulse. Subtracting this global phase allows us to extract $\xph$, which in combination with the magnitude signal then allows us to extract the survival probability along the $\xhat$ direction in the rotating frame, $\expec{\mathcal{I}_x(t)}$ that we display in \zfr{fig2} and \zfr{fig3} of the main text.\\

\T{\I{4. Numerical simulations}}: To numerically simulate the many-body dynamics of dipolar interacting $\Cs$ nuclei, we design random graphs of $L{=}14$ and $L{=}16$ interacting spins-$1/2$ and perform exact time evolution with up to $10^6$ $\xhat$-kicks (corresponding to $\approx 10^3$ $\yhat$-kicks) based on an OMP-parallelized Krylov method using the open-source python package QuSpin~\cite{weinberg2017quspin}. For further details we refer the reader to the supplementary material \cite{SM}.\\

\T{\I{5. Comparison with previous work}}: To date DTC order has been studied in various systems ranging from cold atoms \cite{Zhang17,Kyprianidis21} over superconducting qubits \cite{Mi21} to systems based on NV centers \cite{Choi17,Randall21}. The DTC order observed in these works can be separated into two groups: many-body localized DTCs and prethermal DTCs. While many-body localized DTCs are assumed to be infinitely long-lived in the absence of decoherence, prethermal DTCs, as in our experiments, are ultimately limited by the lifetime of the prethermal plateau. However, in reality also many-body localized DTCs are subject to decoherence due to technical limitations. Remarkably, even though our DTC order is of different physical origin, we find comparable lifetimes to state-of-the-art many-body-localized DTCs. Moreover, our lifetimes exceed those reported for prethermal DTCs, both, in units of Floquet cycles as well as absolute time.

Apart from these benchmark parameters, our system shares ingredients with the systems investigated in Refs.~\cite{Choi17} and \cite{Randall21}. Like in our work, these works investigate systems based on NV centers: Reference~\cite{Randall21} examines a quasi one-dimensional system of 9 dipolar coupled $^{13}$C nuclear spins in the many-body localized regime, while in Ref.~\cite{Choi17} effective two-level systems of electronic states in NV centers are used to implement DTC order. In contrast, our (three-dimensional) system consists of $10^3-10^4$ dipolar coupled $^{13}$C \textit{nuclear} spins. In particular, in comparison to Ref.~\cite{Choi17} the normalized interaction strength $\langle J\rangle/\gamma_n^2$ in our system is increased by a factor $4.5\times 10^4$ [here $\gamma_{e/n}$ denotes the gyromagnetic ratio of electronic/nuclear spins]; moreover, at comparable normalized driving strength, we obtain an improvement of normalized spin lifetimes by a factor of $5.7\times 10^2$ (see also \cite{SM} for a detailed table of relevant system parameters). Induced by the two-frequency drive, the DTC order we observe is prethermal, i.e., our DTC lifetime is parametrically controlled by the drive frequency, while no such feature is present in Ref.~\cite{Choi17} where a single-frequency drive is used.

\bibliography{main.bbl}
\pagebreak

\clearpage
\onecolumngrid
\begin{center}
\textbf{\large{\textit{Supplementary Information} \\ \smallskip
Observation of a critical prethermal discrete time crystal created by two-frequency driving}}\\
\hfill \break
\smallskip
William Beatrez$^{1,\ast}$, Christoph Fleckenstein$^{2,\ast}$, Arjun Pillai$^{1}$, Erica Sanchez$^{1}$, Amala Akkiraju$^{1}$, \\
Jesus Alcala,$^{1}$ Sophie Conti,$^{1}$ Paul Reshetikhin,$^{1}$ Emanuel Druga,$^{1}$, Marin Bukov$^{3,4}$, and Ashok Ajoy$^{1,5}$\\
${}^{1}$\I{{\small Department of Chemistry, University of California, Berkeley, Berkeley, CA 94720, USA.}}\\
${}^{2}$\I{{\small Department of Physics, KTH Royal Institute of Technology, SE-106 91 Stockholm, Sweden.}}\\
${}^{3}$\I{{\small Department of Physics, St. Kliment Ohridski University of Sofia, 5 James Bourchier Blvd, 1164 Sofia, Bulgaria.}}\\
${}^{4}$\I{{\small Max Planck Institute for the Physics of Complex Systems, N\"othnitzer Str.~38, 01187 Dresden, Germany.}}\\
${}^{5}$\I{{\small Chemical Sciences Division Lawrence Berkeley National Laboratory,  Berkeley, CA 94720, USA.}}\\
\end{center}

\twocolumngrid
\beginsupplement

\setcounter{section}{0}


In this supplementary material we provide additional experimental and theoretical details for the article \textit{``Observation of a long-lived prethermal discrete time crystal created by two-frequency driving"}. In Sec.~\ref{sec:two_freq}, we briefly introduce the Hamiltonian of the system. We continue with motivating the employed driving scheme in Sec.~\ref{sec:motivation}, details on the experimental implementation (Sec.~\ref{sec:experimental_details}) and numerical simulations (Sec.~\ref{sec:num_details}). 
Section~\ref{sec:theory_vs_exp} provides additional experimental data backed up by numerics that studies the thermalization behavior under $\zhat$ kicks, and also proves the equivalence of $\zhat$ and $\yhat$ slow kicks. Subsequently, in Sec.~\ref{sec:eff_H}, we derive the effective Hamiltonian associated with the fast $\xhat$-, and slow $\zhat$-concatenated drives. This is followed in Sec.~\ref{sec:FDTC} by a detailed analysis of the formation of time-crystalline order and the prethermalization dynamics in our system. We conclude in Sec.~\ref{sec:heating} with a detailed analysis of the heating timescales and the lifetime of the prethermal order. Throughout the text, we compare our theoretical predictions against observed experimental data; in particular, we compare our results in detail in Sec.~\ref{sec:theory_vs_exp}, Sec.~\ref{sec:FDTC} and Sec.~\ref{sec:heating}. 
In Sec.~\ref{sec:two-freqs} we investigate the implications of two-frequency Floquet engineering. Finally, Sec.~\ref{sec:random} we discuss the random graph dependence of our numerical simulations.

\tableofcontents

\section{ \label{sec:two_freq} Model}

\begin{figure}[t]
\includegraphics[scale=0.4]{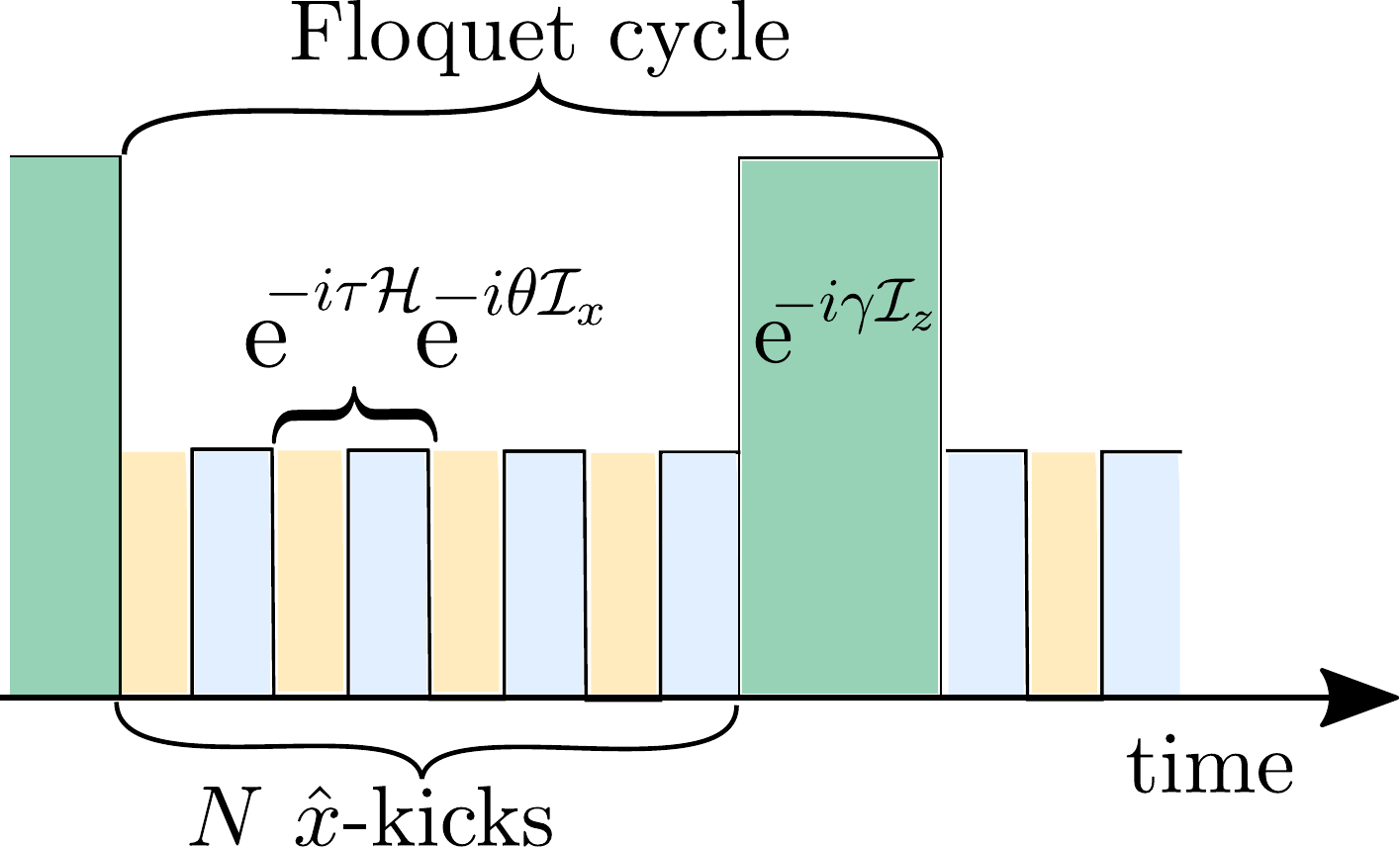}
\caption{\T{Schematic of the two-frequency drive}. An effective Hamiltonian is stabilized using a high-frequency drive with rapidly applied pulses along $\xhat$ direction.  Low-frequency $\zhat$-kicks are applied after $N$ fast kicks. This cycle is repeated so as to apply a total of $M$ slow kicks. Ultimately, the system is driven towards infinite temperature, and we are able to track this spin thermalization dynamics.}
\label{fig:two_freq_drive}
\end{figure}

In this section, we analyze the dynamics of the Floquet drive built out of two superimposed step drives, that we use in the main text. Since the periods of repetition, $N$ and $M$, of the two step drives may vary independently, we refer to this protocol as a ``two-frequency'' drive [cf.~Fig.~\ref{fig:two_freq_drive}].  Note that the term frequency here refers to the inverse periods of the two superimposed drives, rather than the support of their Fourier decompositions. Discrete~\cite{khodjasteh2005fault,witzel2007concatenated} and continuous~\cite{cai2012robust} versions of such two-frequency drives have been previously used for dynamical decoupling protocols and Hamiltonian engineering~\cite{hayes2014programmable}.

We investigate a hyperpolarized lattice of $\Cs$ nuclei in diamond \cite{Ajoy17,Ajoy18}. In the high-field (secular) approximation, the physics of the interactions in the system is captured by the dipolar spin Hamiltonian,
\begin{eqnarray}
\label{eq:Hdd}
\mathcal{H}_\mathrm{dd} = \sum_{j<k}^L b_{jk}\left(3 I_{jz}I_{kz}- \vec{I}_j\cdot\vec{I}_k\right),
\end{eqnarray}
where $b_{jk}= \frac{\mu_0}{4\pi} \hbar \gamma_n^2(3 \cos^2\alpha_{jk}-1)\frac{1}{|\vec{r}_{jk}|^3}$ with the gyromagnetic ratio $\gamma_n=10.7\mathrm{MHz/T}$; the angle of the interspin vector $\vec{r}_{jk}$ and the external magnetic field $\vec{B}_0$, is $\alpha_{jk}=\cos^{-1}\left(\frac{\vec{r}_{jk}\cdot \vec{B}_0}{|\vec r_{jk}||\vec B_0|}\right)$ . $I_{j\mu}$ describes a spin-1/2 operator in direction $\mu$ at some spatial position $j$. The number of spins in the system is $L$. In the main text, we had introduced the coupling strength $J=\expec{b_{jk}}$.

$\Cs$ spins located near paramagnetic impurities, such as NV-centers or other lattice paramagnetic defects (e.g., P1 centers), experience a hyperfine-mediated magnetic field localized from the defect site. Since this field falls off as $1/r^3$, different $\Cs$ atoms experience a different shift, based on their location and the distance to the nearest paramagnetic impurity. This effect can be approximately modeled by introducing an additional term in the Hamiltonian,
\begin{eqnarray}
\mathcal{H}_z= \sum_{j=1}^L c_j I_{jz},
\end{eqnarray}
where we assume that $c_j$ are Gaussian distributed random numbers with a standard-deviation on the order of the relevant energy scales in Eq.~\eqref{eq:Hdd} \cite{Ajoy19relax}.

The total system Hamiltonian in the laboratory frame is composed of $\mathcal{H}_\mathrm{dd}$, $\mathcal{H}_z$ and the trivial Zeeman Hamiltonian. Assuming the pulses applied are on-resonance, however, the system is conveniently described in the rotating frame by the Hamiltonian,
\begin{equation}
\mathcal{H}= \mathcal{H}_\mathrm{dd}+\mathcal{H}_z.
\end{equation}

Next, we additionally subject the system to a two-step periodic drive created by two independent external fields applied along (i) the $\xhat$ direction, $U_x$, and (ii) the $\zhat$-direction, $U_z$, cf.~Fig~\ref{fig:two_freq_drive} [in the theoretical analysis below, we consider $\zhat$ kicks, but in experiments $\yhat$-kicks are used as they are easier to implement; as we show in Sec.~\ref{sec:eff_H}, $\zhat$- and $\yhat$-kicks create the same effective Hamiltonian to leading order and thus yield the same effect on the dynamics, cf.~Fig.~\ref{fig:z_drive}]. The physical system, which generates the time evolution $U_\mathcal{H}$, is exposed to a rapid periodic application of the $\xhat$-drive; every $N$ repetitions of the $\xhat$-drive, we apply the $\zhat$-field $U_z$ once, which completes one Floquet period (or cycle). We note that the two constituent drives used here have commensurate periods. The entire evolution over one driving cycle is then described by the Floquet unitary $U_F$:
\begin{eqnarray}
\label{eq:two_freq}
 U_F &=& \left( U_x U_\mathcal{H}\right)^N U_z, \\
U_z &=& \mathrm e^{-i\gamma \mathcal{I}_z},\quad U_x=\mathrm e^{-i \vartheta \mathcal{I}_x}, \quad U_\mathcal{H} = \mathrm e^{-i\tau \mathcal{H}}, \nonumber
\end{eqnarray}
where we defined the net spin operator $\mathcal{I}_\nu = \sum_j I_{j\nu}$ with $\nu \in \{x,y,z\}$. Note here we have assumed that the pulse action can be described as instantaneous rotations ($\xd$-pulse approximation). The angles of the $\xhat$ and $\zhat$ rotations are denoted $\vartheta$ and $\gamma$, respectively [Fig~\ref{fig:two_freq_drive}]. 

The stroboscopic dynamics of any initial state $|\psi_0\rangle$ is thus 
determined by successive applications of $U_F$
\begin{equation}
|\psi(M)\rangle = U_F^M |\psi_0\rangle= \left[\left( U_x U_\mathcal{H}\right)^N U_z \right]^M|\psi_0\rangle,
\label{eq:psi_t}
\end{equation}
where $M$ denotes the stroboscopic Floquet cycle, and similarly for mixed initial
states.

\begin{figure}[t!]
	\centering
	\includegraphics[scale=1]{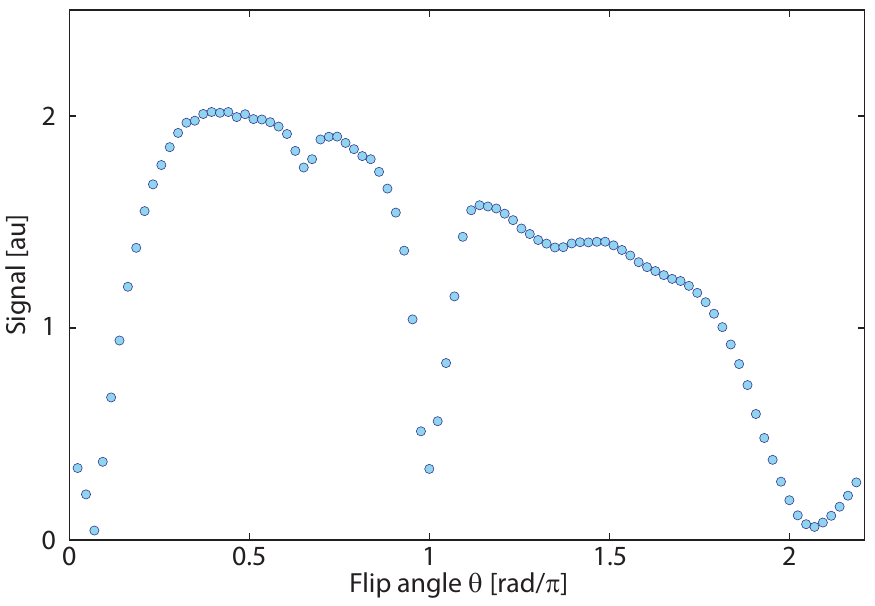}
	\caption{\T{Effect of a single high-frequency drive (experimental data).} The figure shows the net $^{13}$C NMR signal obtained using a single-frequency drive consisting of the application of a train of $\vartheta$-pulses following the initial $\pi$/2-pulse (adapted from Ref.~\cite{Beatrez21}). Here, $t_{\R{acq}}$=32$\mu$s and pulse spacing $\tau$=100$\mu$s. There are strong signal dips at $\vartheta {\approx} \{0, \pi,2\pi \}$ due to rapid decay from evolution under the dipolar Hamiltonian. This signal decay is avoided in this work through the use of the two-frequency drive (see ~Fig.~\ref{fig:two_freq_drive}). }
	\label{fig:flipangle_graph}
\end{figure}

\begin{figure*}[t]
    \centering
    \includegraphics[scale=0.875]{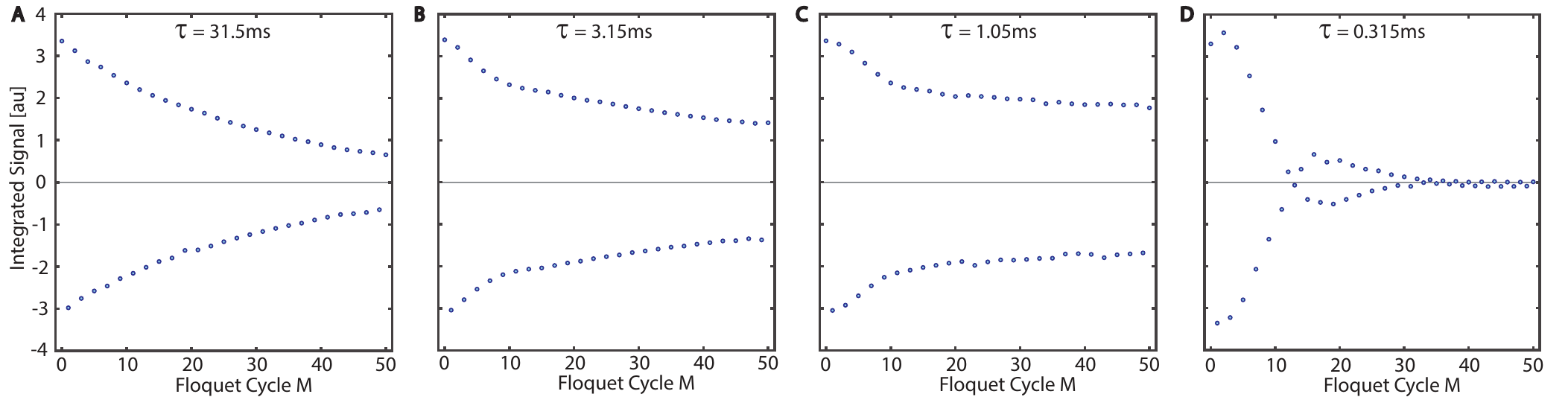}
    \caption{\T{Single-frequency driving PDTC. Experimental data} showing PDTC with traditional single-frequency driving after initializing the spins with a polarization along $\zhat$. Each data point represents the integrated signal $\expec{\mathcal{I}_z}$, readout by multiple-pulse spin-locking after the application of $M$ pulses with flip angle $\xg=0.97\pi$ and period $\tau={31.5\R{ms},3.15\R{ms},1.05\R{ms},0.315\R{ms}}$ (A-D). For $\tau=31.5ms$ (A), the DTC response persists for less than 30 Floquet cycles before reaching 1/e signal decay. The optimal PDTC lifetime is found for $\tau=1.05$ms (C) where the system sees a quick decay during the first 10 Floquet cycles followed by a long plateau. When driving is too fast, the Floquet Hamiltonian becomes effectively single-particle [cf. Eq.~\eqref{eq:H_F}] and a split beat pattern emerges, quickly destroying the DTC response (see (D), $\tau=0.315$ms). These data motivate the use of a two-frequency in our system which enables the observation of long-lived PDTC order.}
\label{fig:lab-frame-DTC}
\end{figure*}

 \vspace{-3mm}
\section{\label{sec:motivation} Motivation for using a two-frequency drive}
\vspace{-1mm}
 
Let us now elucidate upon the experimental and theoretical motivation for using the two-frequency drive in this work. We begin by noting that the $\Cs$ nuclei are dipolar coupled with the Hamiltonian $\mHdd$ above, and when prepared in an initial state $\rho_0\propto \mI_x$, undergo dipolar evolution under it. This results  in the rapid free induction decay, with a time constant $T_2^{\ast}\sim 1.5$ms. 

Consider now the application of a single frequency drive,  consisting of a train of $\xt$-pulses spin locked with the $\Cs$ nuclei. Figure~\ref{fig:flipangle_graph} shows the integrated signal obtained for different values of $\xt$. While the resulting spin decay lifetimes $T_2^{\prime}$ can be far in excess of $T_2^{\ast}$, the fastest signal decay occurs for $\xt=\{0,\pi, 2\pi\}$. This stems from the fact that the dipolar interaction is invariant under rotations by $\pi$; this yields an average Hamiltonian under the Floquet drive, $\bar{\mathcal{H}}^{(0)}=\mHdd$. For other angles of $\xt$, it is easy to show instead that the dipolar interaction is effectively engineered to a form that commutes with the initial state (see Sec.~\ref{sec:eff_H}). This fast decay at $\xt=\pi$ would normally prevent observation of a PDTC in the rotating frame with a single frequency drive alone. In our work, this problem is essentially circumvented by the use of the two frequency drive. 

We note that it is still possible to observe the DTC-like behavior in the laboratory frame with a single frequency drive. This is shown experimentally in Fig.~\ref{fig:lab-frame-DTC}. However, obtaining data similar to Fig.~2A of the main text then has to be constructed in a point-by-point fashion, and the experiment has to be re-initialized between successive measurement kicks.  This makes observation of the formation and melting of the PDTC order at high resolution extremely challenging (cf.~Fig.~3 of main text). In summary, our two-frequency drive approach allows the ability to excite the PDTC order while it simultaneously permits the tracking of the spin thermalization continuously, yielding a unique means to observe the driven interacting spin dynamics away from equilibrium.

From the theoretical perspective, multi-frequency drives are especially interesting as they allow to Floquet engineer novel kinds of effective Hamiltonians and non-equilibrium ordered states that might be inaccessible with single-frequency drives. To illustrate this idea, let us assume a two-step (i.e., single-frequency) drive composed of the Hamiltonians $H_1$ and $H_2$ repeatedly applied with amplitudes $J$ and $\vartheta$ such that time evolution over one Floquet period is given by
\begin{eqnarray}
\label{eq:basic_step_drive}
U_F = \exp\left(-iJH_1\right)\exp\left(-i\vartheta H_2\right).
\end{eqnarray}
Provided $\vert \vert J H_1 \vert \vert \sim \vert \vert \vartheta H_2 \vert \vert \ll 1$, the Baker-Campbell-Hausdorff expansion predicts a Floquet Hamiltonian of the form $H_F =( J H_1  + \vartheta H_2 )/(J+\vartheta)+\mathcal{O}(T')$ with $T'    \sim J, \vartheta $. Although, $H_F$ might provide a new effective Hamiltonian that mixes the properties of $H_1$ and $H_2$, there is only a limited amount of flexibility: the lowest order effective Hamiltonian will always be given by the average Hamiltonian of $JH_1$ and $\vartheta H_2$. In contrast, a much more interesting regime is found when $\vert \vert J H_1 \vert \vert \ll 1 $ but $ \vert \vert \vartheta H_2 \vert \vert \sim 1$. In this case, the standard Baker-Campbell-Hausdorff expansion leads to an infinite series of terms $\propto J$ (i.e., terms $\propto J\vartheta, J\vartheta^2, J\vartheta^3$, etc.), and more sophisticated techniques are required to find closed form expressions \cite{vajna2018replica}. Eventually, the series resummation leads to new types of (interaction) terms which can depend on $\vartheta$ in a non-linear way (see also Sec.~\ref{subsec:x_drive}). This ansatz allows us to engineer new effective Hamiltonians with case-specific properties beyond the average Hamiltonian: for instance, in this work, we engineer a many-body effective Hamiltonian with an emergent $U(1)$ and $\mathbb{Z}_2$ symmetry, which is subject to a second drive to implement non-equilibrium order -- in our case a prethermal discrete time-crystalline order -- in the system. 



\begin{figure}
    \centering
    \includegraphics[scale=0.55]{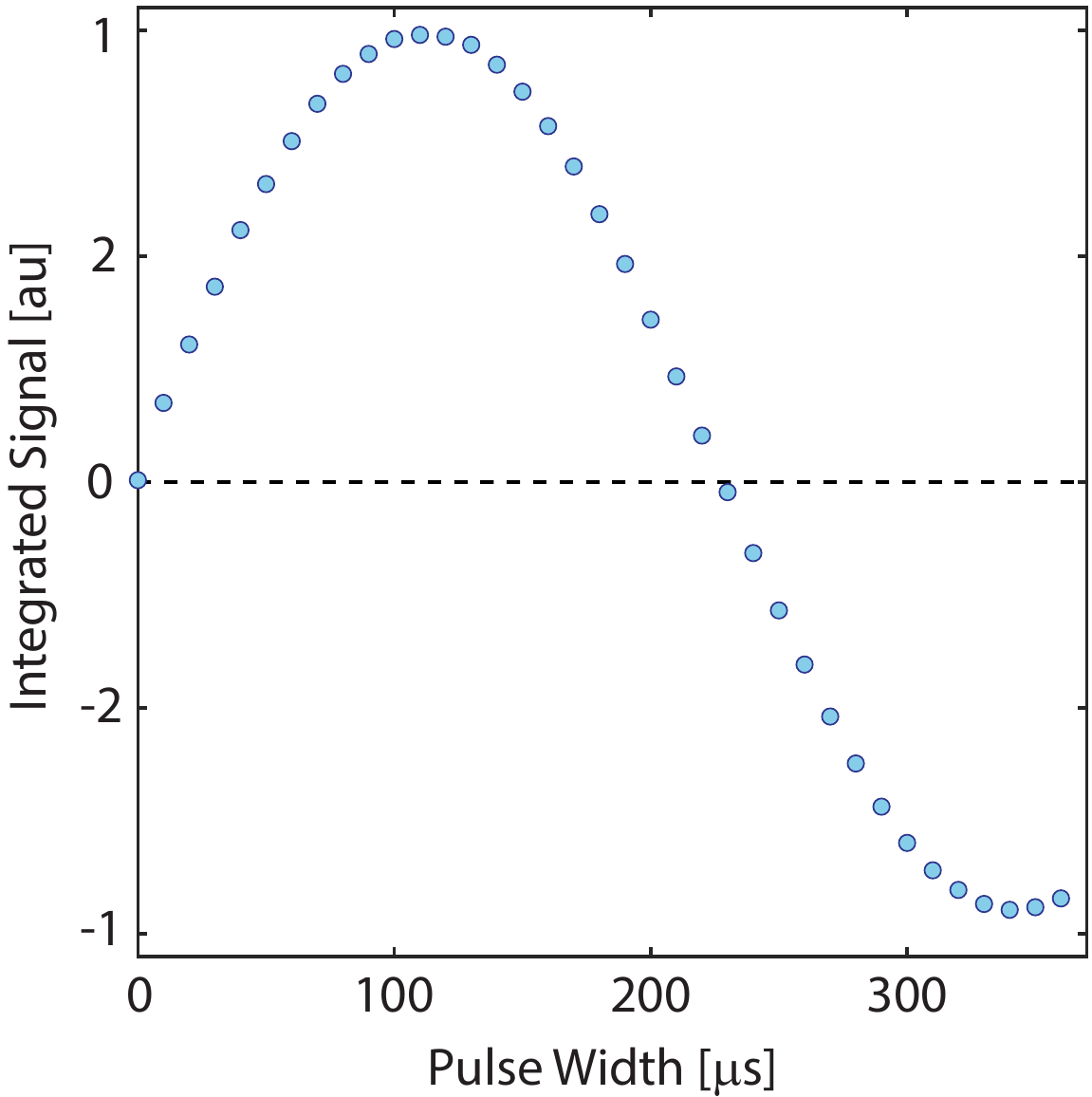}
   \caption{\T{Calibration of flip angle. Experimental data} points represent integrated signal readout via spin-locking following initial excitation pulse of finite width. Experiments were performed for different pulse widths and plot represents a Rabi oscillation, with flip angle $\pi$ being calibrated to the zero-crossing and the rest of the flip angles scaling linearly with pulse width. }
\label{fig:rabi-oscillation}
\end{figure}

\vspace{-3mm}
\section{\label{sec:experimental_details} Details of the experimental implementation}

 \vspace{-3mm}
\subsection{Pulse calibration and parameter regime}
\begin{figure}
    \centering
    \includegraphics[scale=1.05]{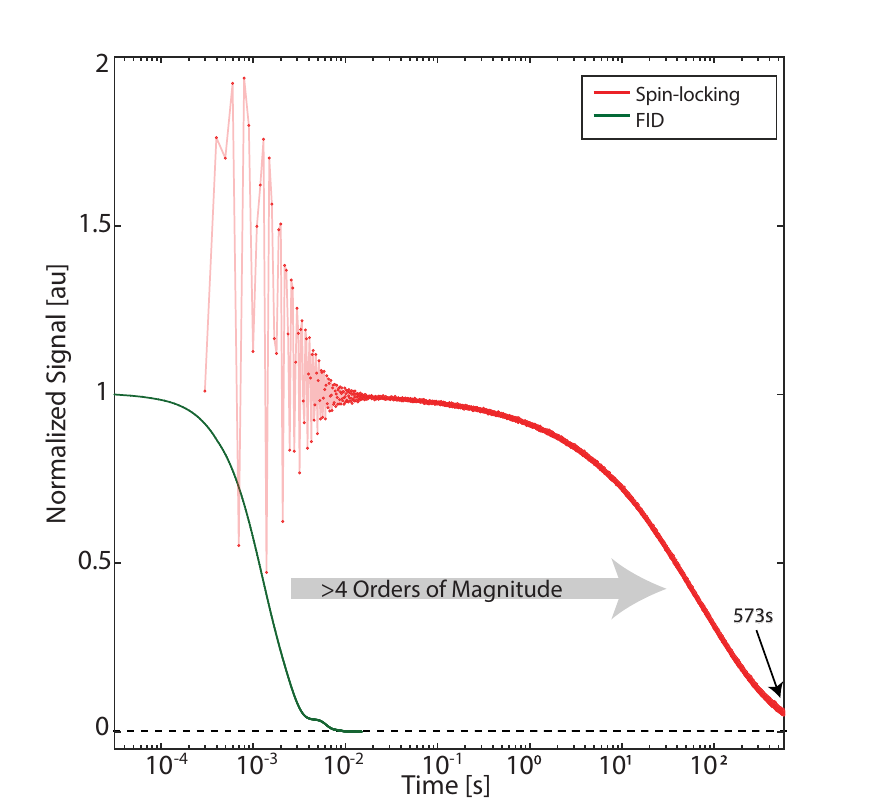}
   \caption{\T{Free induction decay and pulsed spin-lock data. Measurements} shown for free induction decay (green line and points) and for pulsed spin-locking Floquet control (red line and points). Free induction decay lifetime $T_2^*=1.5$ms is extended with high-frequency pulsing by over 5 orders of magnitude, $T_2^{\prime}=90.9$s. Adapated from Ref.~\cite{Beatrez21}. }
\label{fig:lifetime-data}
\end{figure}

 \vspace{-1mm}
The flip angles $(\xt,\xg)$ in the experiments are calibrated using a $\Cs$ nuclear Rabi oscillation.  The first tipping pulse (to be calibrated) is followed by a $\xt{\app}\pi/2$ spin-lock train. We plot the integrated signal in Fig.~\ref{fig:rabi-oscillation}. The SNR is very good because of the long resulting rotating frame lifetimes $T_2^{\prime}$.  Furthermore,  in order to make the experiments less susceptible to precise calibration of the first $\pi/2$ pulse (that tips the spins onto the $\xy$ plane), in Fig.~1B [main text] we apply the first phase $\xg$-kick to the spins after a period of 1s, so that the spins have already prethermalized along $\xhat$ before the $\xg$-kicks are applied. This ensures that the initial state of the spins in the experiment is exactly $\rho_0 \propto \epsilon\mI_x$, with the experimentally measured value $\epsilon=0.68\%$. Fig.~\ref{fig:lifetime-data} shows experimental data for the system's natural free induction decay time and maximum possible lifetime extension under Floquet control. Further relevant system and drive parameters are collected in Tab.~\ref{tab:parameters}.

\vspace{-1mm}
\renewcommand{\arraystretch}{1.5}
\begin{table}[]
    \centering
    \begin{tabular}{|c|c|}
        \hline
        Parameter & Value \\
        \hline\hline
        Inter-spin spacing & $1\mathrm{nm}$ \\
        \hline
        Initial state $\xhat$-polarization $\epsilon$ & $0.68\%$ \\ 
        \hline
        Median interaction strength $\langle J \rangle $ & $0.66\mathrm{kHz}$ \\ 
        \hline
        Drive strength $\Omega$ & $7.6\mathrm{kHz}$ \\
        \hline
        Total spin lifetime $T_2'$ & $90.9\mathrm{s}$ \\
        \hline
        Normalized interaction strength $\langle J\rangle/\gamma_n^2$ & $5.8\times  10^{-12}\mathrm{HzT^2}$ \\ 
        \hline
        Normalized drive strength $\Omega/\gamma_n$ & $7.1\times 10^-4\mathrm{T}$ \\
        \hline
        Normalized spin lifetime $\gamma_n T_2'$ & $9.7\times 10^{-8}\mathrm{T^{-1}}$ \\
        \hline
        Sampling rate & $(0.105\mathrm{ms})^{-1}$ \\
        \hline
        Number of measured points per kick & 299 \\
        \hline
        Total observation time & $14\mathrm{s}$ \\
        \hline
        Signal to noise ratio per point & $>10^2$ \\
        \hline
    \end{tabular}
    \caption{Summary of experimentally relevant system and drive parameters.}
    \label{tab:parameters}
\end{table}

 \vspace{-3mm}
\section{\label{sec:num_details}Details of the numerical simulations}
 \vspace{-1mm}

\begin{figure}[t]
	\centering
	\includegraphics[width=0.3\textwidth]{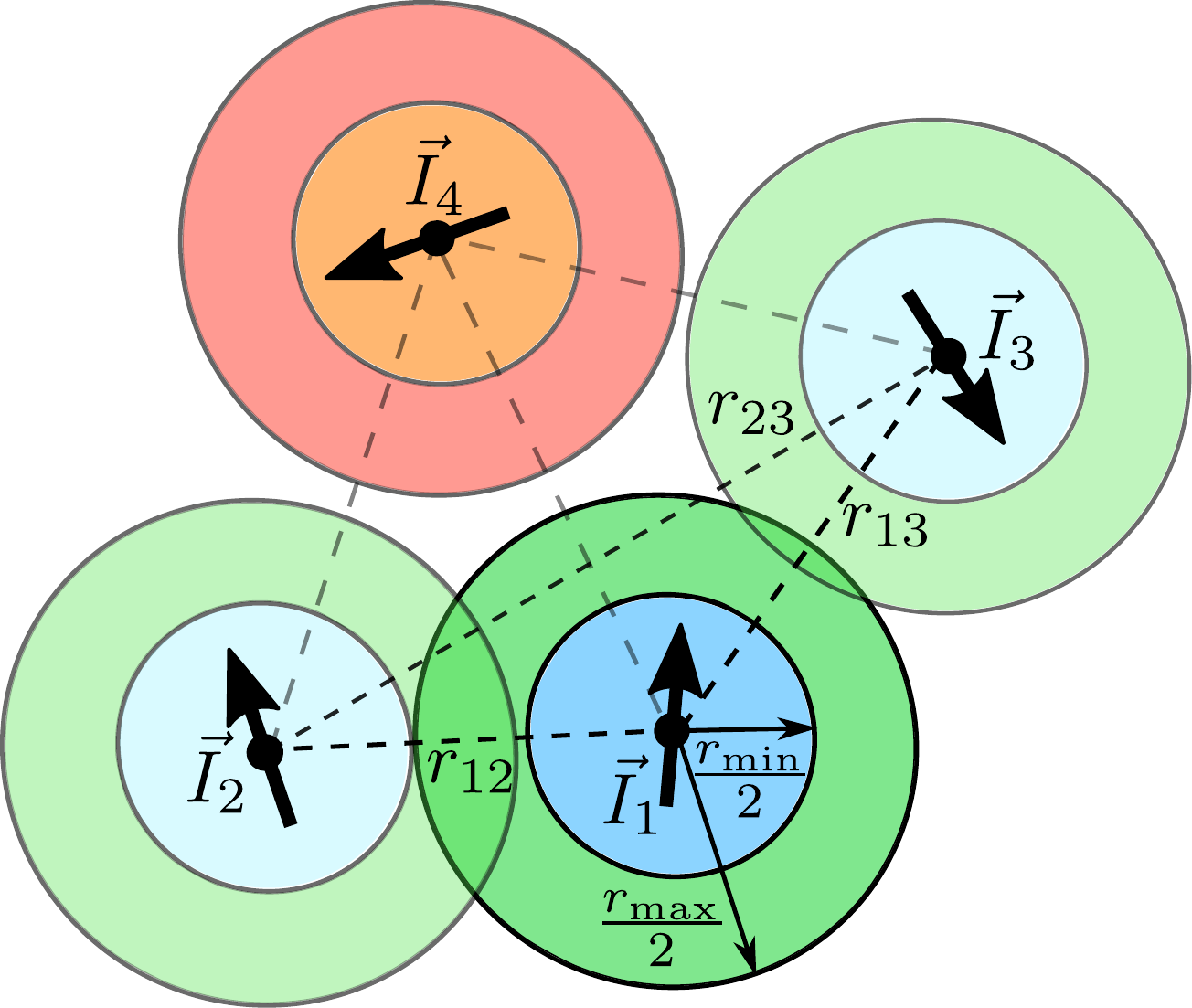}
	\caption{\T{Schematic for the generation of a pseudo-random graph} used in numerical simulations: Starting off with a spin $\vec{I}_1$ in some random position, we propose a new position for the next spin $\vec{I}_2$ [see text]. Since $r_{\mathrm{min}}< r_{12} < r_{\mathrm{max}}$ we accept $\vec{I}_2$ and add it to the graph. Next, we propose $\vec{I}_3$, which also satisfies all requirements since all inter-spin vectors $r_{13},r_{23} > r_{\mathrm{min}}$, while $r_{13}<r_{\mathrm{max}}$. Hence, we also accept $\vec{I}_3$. The next proposal, $\vec{I}_4$, only has inter-spin vectors which are all larger than $r_{\mathrm{max}}$ and, thus, constitutes an invalid choice; hence, it is discarded. If we aim to simulate $L=4$ spins (or more) we would have to continue proposing new spins until a valid fourth spin position is found and added to the graph.}
	\label{fig:random_graph}
\end{figure}

\subsection{Random graph design}

We perform exact numerical simulations of the system with $L$ $\Cs$ nuclear spins, placed on a pseudo-random graph. In creating the lattice graph, we respect two conditions: 
(i) we ensure a certain average spin density at the system size $L$; 
(ii) we avoid isolated spin positions (`outliers'), that would not contribute to the many-body dynamics as their coupling constant becomes negligible in comparison to their ``bulk" spins counterparts due to the spatial decay of the dipolar interactions. 

To iteratively generate the spin positions on the graph, we apply the following rule: first, we randomly generate a new spin position, and we check if it has a minimum distance $r_{\mathrm{min}}$ to all other spins. In addition, we also require that at least one of the mutual interspin vectors satisfies $|\vec{r}_{jk}|< r_{\mathrm{max}}$ [cf.~Fig.~\ref{fig:random_graph}]. If these conditions are met, we accept the new spin position and add it to the list of spins on the graph. Otherwise we discard the proposed spin position and start over with a new proposal. In the main text we use $r_{\mathrm{min}}=0.7$, $r_{\mathrm{max}}=0.8$, both in units of $\sqrt[3]{\mu_0\hbar \gamma_n^2}$.

\subsection{\label{subsec:time_evo}Time evolution and initial state}

In the numerical simulations using this pseudo-random graph of spins, unless stated explicitly otherwise, we initialize the system in the $\xhat$-polarized pure product state $\vert \psi_0\rangle \!=\!\bigotimes_{j=1}^L  \frac{1}{\sqrt{2}}\left(\vert \! \uparrow_j \rangle \!+\! \vert \downarrow_j \rangle\right)$, and perform numerically exact time evolution according to the protocol given in Eq.~\eqref{eq:two_freq}. While most of our results are based on this initial state, we also show that the effects we observe are qualitatively initial-state independent [cf.~Sec.~\ref{sec:thermalization} and Fig.~\ref{fig:init_state}]. 

Since the interaction is long-range and the positions of the nuclei are not spatially ordered, the characteristic energy scale of the model is not immediately obvious, and needs to be extracted from the dynamics. To this end, we define a characteristic energy $J=1/\tau_d$, where $\tau_d$ is the timescale on which single-particle observables decay by a factor of $1/\mathrm e$, as they approach their equilibrium value when evolved under $U_{\mathcal{H}}$. We note in passing that $J$ may vary depending on the initial state, i.e., different values might be obtained for $J$ when probed with other initial states than $\vert \psi_0\rangle$. Therefore, the value of $J$ only serves as a rough estimate of relevant energy scales.

To further increase the ergodicity of the drive and diminish finite-size effects in the dynamics, we add a small uniformly distributed random ``noise'' $\delta\tau \in [-0.05\tau, 0.05\tau ]$ to the duration $\tau$ for which we apply $\mathcal{H}$, i.e. $\tau \rightarrow \tilde{\tau} = \tau +\delta\tau$ so that $\tilde{\tau}$ is slightly different in each cycle of $(U_xU_\mathcal{H})$~\cite{fleckenstein2021prethermalization,fleckenstein2021thermalization}.
After each application of $U_x$, $U_z$, respectively, we compute the expectation value of single particle observables 
\begin{equation*}
  \langle x\rangle =\fr{2}{L} \langle \mathcal{I}_x\rangle,\quad 
  \langle y\rangle =\fr{2}{L} \langle \mathcal{I}_y\rangle,\quad \mathrm{and}\quad 
  \langle z\rangle =\fr{2}{L} \langle \mathcal{I}_z\rangle. 
\end{equation*}

\section{\label{sec:theory_vs_exp}Thermalizing dynamics: simulation versus experiment}

\begin{figure}[t!]
	\centering
	\includegraphics[width=0.49\textwidth]{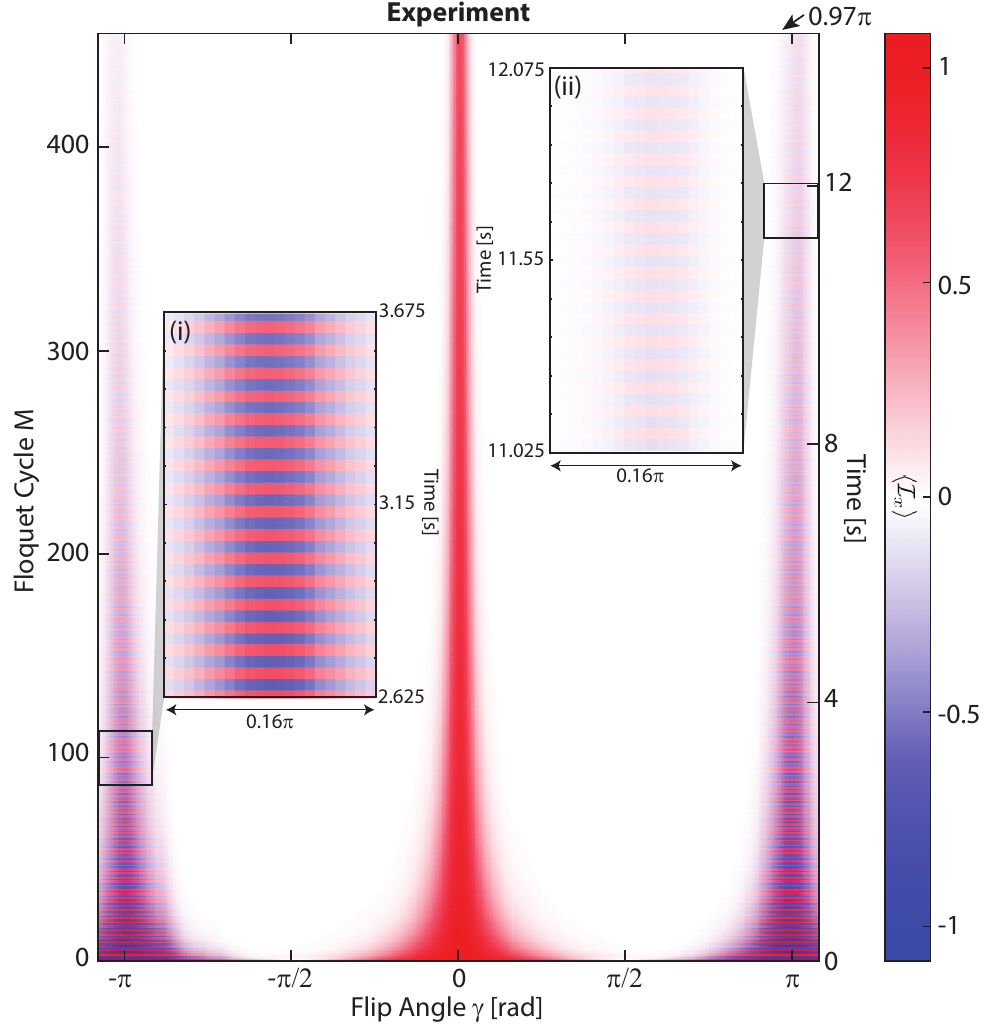}
	\caption{\textbf{Experimental data} showing a linear scale depiction of Fig.~3A of the main text. 285 traces similar to Fig.~2B of main paper are plotted stacked for different values of kick angle $\xg$ in $[-\pi,\pi]$. Colors represent signal $\expec{\mathcal{I}_x}$ (see colorbar). Time $N\qt$ and Floquet cycle number $M$ run vertically and are plotted here on a linear scale [cf.~left and right $\xhat$-axes]. Insets show PDTC behavior (evidenced by oscillations between positive and negative signals every 300 fast pulses) in 0.16$\pi$- radian-wide windows near $-\pi$ (i) and $\pi$ (ii). The PDTC response is strong at early times (i) and persists to long times (ii) despite signal decay caused by heating. See ref.~\cite{PDTC_video25,PDTC_video55,PDTC_video155} for movies of data for first 25, 55 and 155 Floquet cycles, respectively. }
	\label{fig:linear_heatmap}
\end{figure}

\begin{figure*}[t!]
	\centering
	\includegraphics[width=1\textwidth]{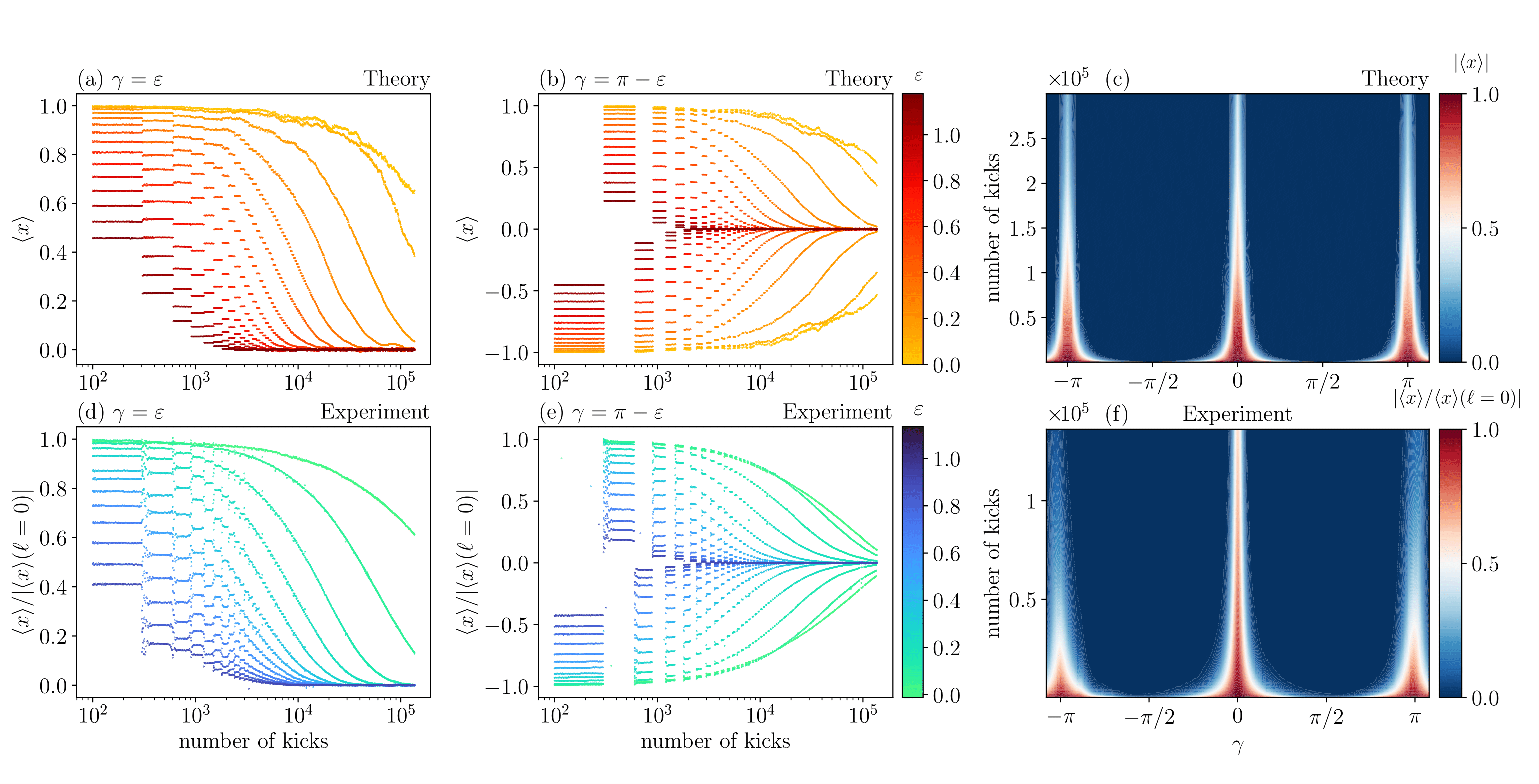}
	\caption{\textbf{Comparison of numerical simulation and experiment --}\textbf{(a)}-\textbf{(b)} Time evolution of the $\xhat$-polarized initial state under the drive generated by Eq.~\eqref{eq:two_freq} at $\gamma=\varepsilon$ (a) and $\gamma=\pi+\varepsilon$ (b) for different values of $\varepsilon$: we plot the expectation value of the $\xhat$-magnetization, $\langle x \rangle =\fr{2}{L} \langle \mathcal{I}_x\rangle$, vs. time given in number of kicks. 
    \textbf{(c)} $\vert \langle x \rangle \vert$ as a function of the $\zhat$-kick angle $\gamma$ and number of kicks. To generate the random graph of spins we used $r_{\mathrm{min}}=0.7$, $r_{\mathrm{max}}=0.8$ (in units of $\sqrt[3]{\mu_0\hbar \gamma_n^2}$) with $L=14$. The single particle energies $c_j$ are normally distributed random numbers with mean $\langle c_j\rangle= \overline{b_{jk}}$ and standard deviation $\sigma_c= 10\times \overline{b_{jk}}$, where $\overline{b_{jk}}$ is the median of all coupling constants of the graph.
	\textbf{(d)}-\textbf{(f)} experimental data corresponding to (a) and (b) and (c) respectively. Parameters are as in Fig.~3 of the main text. In all panels, we plot logarithmically many points in the number of kicks. $N=300$ and  $\tau J=0.07$ for all figures. 
	}
	\label{fig:time_evolution}
\end{figure*}

\begin{figure}[t]
    \centering
    \includegraphics[width=0.48\textwidth]{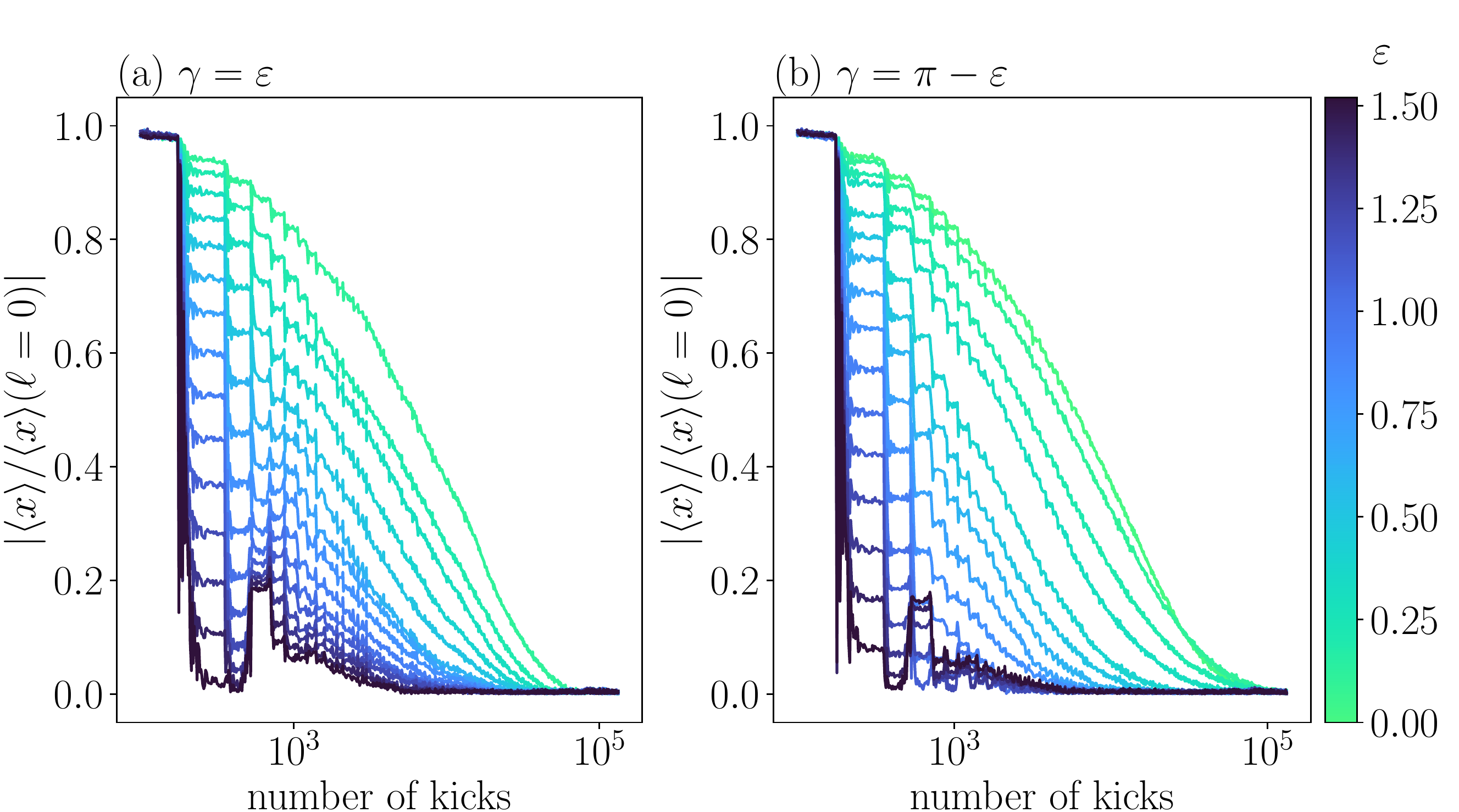}
    \caption{\T{Equivalence of $\zhat$ and $\yhat$ kicks. Experimental} results analogous to the dataset shown in Fig.~\ref{fig:time_evolution}, except here the low-frequency driving consists of composite pulses along $\zhat$-axis rather than simple $\yhat$-pulses (see Fig.~\ref{fig:two_freq_drive} for driving protocol). NMR signal $\expec{\mathcal{I}_x}$ is measured between fast pulses and composite $\zhat$-rotation is applied between every $N=176$ fast pulses. Single-shot experiment yields one line, and experiment is repeated for different values of $\xg$ in $[0,\pi]$. Just as with the experiment with $\yhat$-pulses, the dynamics show long lifetimes for small flip angle deviations $\varepsilon$,  $\gamma=\varepsilon$ (a) and $\gamma=\pi+\varepsilon$ (b). }
    \label{fig:z_drive}
\end{figure}

We now elucidate in greater detail the thermalization dynamics of the spins in a comparison between theory and experiment. First, we begin by displaying Fig.~\ref{fig:linear_heatmap} that shows the data corresponding to the experiments in Fig.~3B of the main text, but with the time-axis plotted in a linear (as opposed to a logarithmic) scale. We plot the survival probability of the spins $\expec{\mI_x}$ under the concatenated action of the slow and fast drives. The stable points around $\xg{=}0$ and $\pm\pi$ are clearly visible, along with the large region in the center where the spins decay rapidly. The insets (reproduced from the main text), show the PDTC behavior in more detail for two ${\app}1$s-long windows centered at $t{=}3.15$s and $t{=}11.55$s. 

Figure~\ref{fig:time_evolution} displays the dynamics of the system, obtained numerically ((a)-(c)) and in the experiment ((d)-(f)).  In the numerical simulations, the initial state $\vert \psi_0\rangle $ is evolved in time up to $5\times 10^4$ kicks for $N=300$ [throughout the supplementary material, we also show results for a few more values of $N$]. Note that here we treat $\xhat$-kicks and $\zhat$-kicks on equal footing: each application of $U_x$, $U_z$, respectively, adds $+1$ to the total number of kicks.
We show the simulated time evolution curves for $\gamma=0+\varepsilon$ (a) and $\gamma=\pi- \varepsilon$ (b), where different curves correspond to different values of $\varepsilon$ (see colorbar): with an increasing value of $\varepsilon$ the heating time gets gradually reduced until no stable magnetization is visible anymore when $\varepsilon\to\pi/2$. 

In the experiment, the initial state after the $\pi/2$-pulse about the $y$-axis [see Fig.~1, main text] is the mixed state $\rho_0\propto \epsilon\mathcal{I}_x$. The time evolution of the $\xhat$-magnetization shown Fig.~\ref{fig:time_evolution} (d) and (e) is measured in the experiment. 
Despite the different initial states, the curves are thoroughly comparable on a qualitative basis, with the notable difference that the experimental curves yield much cleaner results in the small $\varepsilon$ regime. This can be traced back to large differences in the system size in experiment ($L\sim \mathcal{O}(10^4)$) and theory ($L\sim \mathcal{O}(10^1)$). 

In Fig.~\ref{fig:time_evolution} (c)  (Fig.~\ref{fig:time_evolution} (f)) we display theoretical (experimental) results of the absolute value of the $\xhat$-magnetization, $|\langle x\rangle|$, for a wide range of $\gamma$-values. 
In both panels, we can distinguish between three overall regimes: around $\gamma=0$ ($\gamma=2\pi$) heating is suppressed as $U_z$ almost wraps up to the identity so that the slow-frequency $\zhat$-kicks become ineffective, and the effective Hamiltonian (discussed in Sec.~\ref{sec:eff_H}) is engineered solely from the high-frequency $\xhat$-drive. Around $\gamma=\pi$, we observe a similar behaviour; there, the formation of the PDTC [cf.~Sec.~\ref{sec:FDTC_theo}] leads to slow thermalization dynamics. 
In between these regimes the $\zhat$-kicks lead to rapid thermalization to infinite temperature resulting in a decay of the signal within very few cycles (cf.~Fig~\ref{fig:linear_heatmap}).  

We note finally that quantitatively identical results are obtained using a $\zhat$-kick to the spins instead of a $\yhat$-kick. In the experiment, however, this requires each kick to be constructed out of a composite rotations along the $\xhat$ and $\yhat$ axes. Data corresponding to such $\zhat$-kicks are shown in Fig.~\ref{fig:z_drive} around the two stable points, $\xg=\xe$ and $\xg=\pi-\xe$.

Overall, the excellent qualitative agreement between theory and experiment is remarkable considering the three orders of magnitude difference in the system size between the two. From a theory perspective, this justifies the results obtained using exact simulation of the dynamics in the modelled system, and indicates that these numerical techniques can be used to make reliable predictions about experimental systems. Nonetheless, the experiment covers system sizes that are infeasible to reach on any classical computer in the foreseeable future. Therefore, the experiment remains indispensable for probing statistical mechanics concepts related to collective phenomena, such as thermalization dynamics or symmetry breaking, which are only well-defined in the thermodynamic limit.

 \vspace{-3mm}
\section{\label{sec:eff_H} Effective Hamiltonian}
 \vspace{-1mm}

Intuitively, the physics of the two-frequency drive is easy to comprehend when the repetition rates of the two different drives are well separated from one another. In this case, the repetition number of the $\xhat$-drive is $N\gg 1$ [cf.~Fig.~\ref{fig:two_freq_drive}].  In this regime, we can consider the $\xhat$-drive to be fast, and the $\zhat$-drive -- slow, w.r.t.~the energy scale $J$ of the Hamiltonian. 

For the subsequent analysis, it will prove useful to rewrite Eq.~\eqref{eq:two_freq} in the so-called toggling frame:
\begin{equation}
 \left( U_x U_\mathcal{H}\right)^N  U_z =
\left(  U_x U_\mathcal{H} U_x^{-1} \right)  \left(  U_x^2 U_\mathcal{H} U_x^{-2} \right) \dots \times U_x^{N} U_z  .
\end{equation}
Denoting the toggling frame Hamiltonians by $\mathcal{H}_{n} =U_x^n \mathcal{H} U_x^{-n}$ leads to the compact expression 
\begin{equation}
\label{eq:train_of_unitaries}
U_F= \left( U_x U_\mathcal{H}\right)^N  U_z   =
 \left(\prod_{n=1}^N \exp(-i\tau \mathcal{H}_{n})\right)  U_x^{N} U_z.
\end{equation}

The toggling-frame expansion is based on the Magnus expansion~\cite{Magnus54,Haeberlen76,blanes2009} in $N^{-1}$, and provides a tool to compute an approximation for the generator of the $N$-cycle unitary $U^N = \left( U_x U_\mathcal{H}\right)^N$. 

Apriori, the toggling-frame expansion aims to directly approximate the generator of $U^N$ after $N$ cycles, which is different from the $\xhat$-drive Floquet unitary $U=U_x U_\mathcal{H}$ associated with the repetition of the $\xhat$-drive only. With this, we draw the readers' attention to the subtle distinction between the Magnus expansion (which does not require a periodic drive), and the Floquet-Magnus expansion (which is the specialization of the former to Floquet systems). The toggling-frame expansion is based on the Magnus expansion.

Upon adding the second drive, the proper stroboscopic Floquet period is increased. Because of this, rather coincidentally, the toggling frame expansion w.r.t.~$N$ will actually approximate the proper Floquet unitary $U_F =  \left( U_x U_\mathcal{H}\right)^N U_z$, up to the overall single-particle initial kick $U_x^{N} U_z$, cf.~Eq.~\eqref{eq:train_of_unitaries} and the subsequent analysis in Sec.~\ref{subsec:x_drive}. Then, one can consider applying the toggling frame expansion once more, this time w.r.t.~the $\zhat$-drive repetition rate $M$, cf.~Eq.~\eqref{eq:psi_t}. This directly provides us an approximate series expansion in powers of $M^{-1}$ for the generator of $[U_F]^M$, [Sec.~\ref{subsec:z_drive}].

 \vspace{-3mm}
\subsection{\label{subsec:x_drive}Analysis of the fast $\xhat$-drive}
 \vspace{-1mm}

{\bf Toggling-Frame Expansion} -- Let us assume the $\xhat$-kicks to be high-frequency compared to the energy scales of the system, i.e., $\tau J \ll 1$. Then, applying the Baker-Campbell-Hausdorff (BCH) formula to leading-order, the toggling frame unitary can be written as
\begin{eqnarray}
	\prod_{n=1}^N \exp(-i\tau \mathcal{H}_{n}) = \exp\left[-iN \tau \bar{\mathcal{H}}^{(0)}+\mathcal{O}\left((\tau J)^2\right)\right],
\end{eqnarray}
where the effective Hamiltonian $\bar{\mathcal{H}}^{(0)}$ to order $\mathcal{O}\left((\tau J)^2\right)$, is given by \cite{ajoy2020}
\begin{eqnarray}
	\label{eq:toggling_H}
	\bar{\mathcal{H}}^{(0)}=\frac{1}{N}\sum_{n=1}^{N} \mathcal{H}_{n} = \bar{\mathcal{H}}^{(0)}_\mathrm{dd} + \bar{\mathcal{H}}^{(0)}_z, 
\end{eqnarray}
where
\begin{eqnarray}
	\label{eq:Hdd_Hz}
	\bar{\mathcal{H}}^{(0)}_\mathrm{dd}  &=& \sum_{j<k}b_{jk} \bigg( \frac{3}{2}\bigg[ \mathcal{H}_\mathrm{ff}+\mathcal{G}_\mathrm{c}(N,\vartheta)\mathcal{H}_\mathrm{dq} - \tilde{\mathcal{H}}_\mathrm{ff} \mathcal{G}_\mathrm{s}(N,\vartheta) \bigg]- \vec{I}_j\vec{I}_k\bigg), \nonumber \\
	\bar{\mathcal{H}}^{(0)}_z &=& \sum_j c_j \left( \mathcal{G}_\mathrm{c}(N,\vartheta/2)  I_{jz} - \mathcal{G}_\mathrm{s}(N,\vartheta/2)  I_{jy} \right),
\end{eqnarray}
with
\begin{eqnarray}
	\mathcal{H}_\mathrm{ff/dq} &=& I_{jz}I_{kz}\pm I_{jy}I_{ky }, \nonumber \\
	\tilde{\mathcal{H}}_\mathrm{ff} &=& I_{jz}I_{ky}+ I_{jy}I_{kz },
\end{eqnarray}
and 
\begin{eqnarray}
\label{eq:Gs}
	\mathcal{G}_\mathrm{c}(N,\vartheta)  &=& \frac{1}{N} \frac{\sin(N\vartheta)}{\sin(\vartheta)}\cos\left( (N+1) \vartheta \right), \\
	\mathcal{G}_\mathrm{s}(N,\vartheta)  &=& \frac{1}{N} \frac{\sin(N\vartheta)}{\sin(\vartheta)}\sin\left( (N+1) \vartheta \right). \nonumber
\end{eqnarray}
Notice that 
$\mathcal{G}_\mathrm{c/s}(N,\vartheta)$ decays as $1/N$ for large $N$. Thus, as long as $\vartheta \not\approx l\pi$ ($l\in\mathbb{N}$), we find $\mathcal{G}_\mathrm{c/s}(N,\vartheta) \rightarrow 0$ in the $N\gg 1$ regime, which simplifies to,
\begin{eqnarray}
	\label{eq:H0}
	\bar{\mathcal{H}}^{(0)}\simeq \overline{\mathcal{H}} = \sum_{j<k}b_{jk}\bigg(\frac{3}{2}\mathcal{H}_\mathrm{ff}-\vec{I}_j\cdot\vec{I}_k\bigg).
\end{eqnarray}

Importantly, we find $\left[ \mathcal{I}_x, \overline{\mathcal{H}}\right]=0$, which implies that the drive preserves any polarization in $\xhat$ direction to leading order in $\tau J$. One can readily convince oneself that the smaller higher-order terms break this emergent conservation law.
However, at least to order $\mathcal{O}((J\tau)^2)$, these terms are additionally suppressed by a factor of $1/N$ \cite{ajoy2020}. Thus, we expect the leading order effective Hamiltonian to capture well all the interesting physics that plays out in the prethermal state.

Note that the above analysis breaks down as $\vartheta \rightarrow l\pi$. Indeed, in this limit we lose the quasi-conservation of the $\xhat$-magnetization already in the leading-order Hamiltonian (cf.~Fig.~\ref{fig:flipangle_graph}). This implies that initially polarized states will decay rapidly to their equilibrium value close to zero magnetization. In the present experiment, the state of the system would then appear indistinguishable from an infinite-temperature state, since we can only measure single-particle operators, whose expectation values match the infinite-temperature value in the quench dynamics governed by $\overline{\mathcal{H}}$. Fortunately, this behavior can be prevented by setting  $\vartheta$ sufficiently far away from the $0,\pi,2\pi$ by employing a two-frequency drive. In this work, we therefore work at $\vartheta = \pi/2$.

{\bf Replica Expansion} -- An alternative approach to the dynamics generated by the fast $\xhat$-drive is given by the Floquet Hamiltonian $\mathcal{H}_{F,x}$. Unlike the effective Hamiltonian, $\mathcal{H}_{F,x}$ only governs the dynamics over a single period of $\xhat$ driving, $U_xU_{\mathcal{H}}$. Formally, $\mathcal{H}_{F,x}$ is defined by, 
\begin{equation}
\label{eq:H_F_def}
U_xU_{\mathcal{H}}=\mathrm e^{-i(\vartheta + \tau) \mathcal{H}_{F,x}}=\mathrm e^{-i\vartheta \mathcal{I}_x}\mathrm e^{-i\tau \mathcal{H}}.
\end{equation}
Typically, $\mathcal{H}_{F,x}$ can be obtained via a BCH expansion. However, having $\tau J \ll \vartheta $ complicates the standard procedure. In fact, here, the BCH expansion produces infinitely many terms $\propto \tau \vartheta$. A closed form expression requires a resummation of these terms. A neat way how this can be achieved was recently put forward in Ref. \cite{vajna2018replica} by using a replica resummation. This trick utilizes a replica expression of the logarithm
\begin{eqnarray}
\label{eq:replica_1}
(\vartheta+\tau)\mathcal{H}_{F,x} = i\log (U_xU_{\mathcal{H}}) = \lim_{\rho \rightarrow 0} \frac{i}{\rho}\left((U_xU_{\mathcal{H}})^{\rho}-1\right).
\end{eqnarray}
In the next step, the right hand side of Eq.~\eqref{eq:replica_1} can be expanded in powers of $\tau$. Eventually, to linear order in $\tau$ this yields
\begin{eqnarray}
(\vartheta+\tau)\mathcal{H}_{F,x} &=& \vartheta \mathcal{I}_x +\tau \bigg[ \sum_{j<k}b_{jk}\bigg(\frac{1}{2}(1\!-\!3\vartheta \cot(\vartheta))I_{jy}I_{ky} \nonumber\\  &+&\frac{1}{2}(1\!+\!3\vartheta\cot(\vartheta))I_{jz}I_{kz}\!-\!I_{jx}I_{kx}\!-\!\frac{3}{2}\vartheta(I_{jz}I_{ky}\!+\!I_{jy}I_{kz})\bigg) \nonumber\\
&-&\sum_j c_j \left(\frac{\vartheta}{2}I_{jy}\!+\!\frac{\vartheta}{2}\cot(\vartheta/2)I_{jz}\right)
\bigg]+\mathcal{O}(\tau^2).
\end{eqnarray}
At $\vartheta=\pi/2$, the above expression simplifies to
\begin{eqnarray}
\label{eq:H_F}
\left(\frac{\pi}{2}+\tau\right)\mathcal{H}_{F,x} &=& \frac{\pi}{2} \mathcal{I}_x + \tau \overline{\mathcal{H}}  -\tau \bigg(\frac{3}{4}\pi  \sum_{j<k} b_{jk} \tilde{\mathcal{H}}_\mathrm{ff}+\frac{\pi}{4}\sum_j c_j (I_{jy} - I_{jz})\bigg) \nonumber \\
\end{eqnarray}
Equation~\eqref{eq:H_F} shines a more microscopic light on the dynamics generated by the fast drive. With this approach, we even obtain the expansion up to first order in $\tau$. Notice that these corrections are all linear in $\tau$, however they can be highly non-linear in $\vartheta$ as they emerge from resummations of commutators. The appearance of such terms can be traced back to the particular structure of the drive where $J\tau \ll 1$ while $\vartheta \sim \mathcal{O}(1)$. Moreover, Eq.~\eqref{eq:H_F} allows us to study the limit $\tau \rightarrow 0$. There, the system effectively becomes a single particle problem where no notion of (pre-)thermalization exists. Likewise, PDTCs become ill-defined since they require many-body interactions [cf.~Sec.~\ref{sec:FDTC_theo}]. This renders the $\tau \rightarrow 0$ regime unfavorable for us as it decreases the regime of rigidity  [see Fig.~\ref{fig:lab-frame-DTC} (D)]. 
On the other hand, the large-$\tau$ regime is equally bad, since the onset of higher-order terms spoils the quasi-conservation of $\xhat$-magnetization which protects the PDTC order, cf.~Sec.~\ref{sec:FDTC}.
Hence, eventually we are left with a window of suitable $\tau$ values so that the many-body nature of the effective Hamiltonian is guaranteed while simultaneously the $\xhat$-magnetization remains quasi-conserved.

 \vspace{-3mm}
\subsection{\label{subsec:z_drive}Analysis of the slow $\zhat$-drive.} 
 \vspace{-1mm}

To sum up, in the $N\gg 1$ regime, we can use the toggling-frame expansion to approximate Eq.~\eqref{eq:train_of_unitaries} to order $\mathcal{O}((J\tau)^2)$ as
\begin{eqnarray}
	\label{eq:U0}
	U_F \approx \mathrm e^{-iT \overline{\mathcal{H}}}  U_x^{N} U_z = U_0  U_x^{N} U_z,\quad
	U_0=\mathrm e^{-iT \overline{\mathcal{H}}}
\end{eqnarray}
with $T=N\tau$. Describing the effects of the slow $\zhat$-drive requires a careful analysis of the induced modification to the effective Hamiltonian, which is a prerequisite to understand the thermalizing dynamics of the system in the experiment. Therefore, let us now investigate the repeated application of the slow $\zhat$-kicks. 

The unitary which describes the time evolution up to $M$ Floquet cycles is given by
\begin{eqnarray}
	\label{eq:slow_period}
	[U_F]^M = \left(U_0  U_x^{N}  U_z  \right)^M.
\end{eqnarray}
For large values $N\gg 1$, we have $\left[U_0,U_x\right]=0$, where the unitary $U_0$ is defined in Eq.~\eqref{eq:U0}. Moreover, $U_x^{N}$ denotes a rotation by $N\vartheta =N\pi/2$. Thus, we have $U_x^{N}\equiv U_x^{\left(N\;\mathrm{mod}\;8\right)}$. Hence, in the limit $N\gg 1$, the action of the high-frequency $\xhat$-drive can be summarized into a single unitary
\begin{eqnarray}
\label{eq:fast_drive_eff_H}
U_0 U_x^{N}\!&\equiv&\! U_{0x} 
= \exp \bigg[-iT \overline{\mathcal{H}}-\frac{i\pi}{2}\left(N\;\mathrm{mod}\;8\right) \mathcal{I}_x \!\bigg].
\end{eqnarray}

To derive the leading-order effective Hamiltonian of the combined $\xhat$- and $\zhat$-drive, we apply the toggling frame expansion once again, this time w.r.t.~$M$.
To do so, we multiply Eq.~\eqref{eq:slow_period} from the left by $\id=U_z^M U_z^{-M}$, and re-group the terms:
\begin{eqnarray}
\label{eq:large_N_unitary}
\left(  U_0 U_x^{N} U_z \right)^M &=&   U_z ^M\left(  U_z^{-M} U_{0x}  U_z ^M \right) \times \nonumber \\
&& \times \left(  U_z^{-(M-1)} U_{0x} U_z ^{M-1} \right)  
\times \dots \times \left(  U_z^{-1} U_{0x} U_z  \right) \nonumber \\
& = & U_z^{M} \left( \prod_{m=M}^1 U_z^{-m}  U_{0x} U_z ^{m}  \right), 
\end{eqnarray}
where the $M$-toggling frame Hamiltonians read as
\begin{eqnarray}
\label{eq:large_N_toggling_frame}
\mathcal{H}_m(\gamma) &=&   U_z^{-m} \left(\overline{\mathcal{H}}\!+\!\frac{\pi}{2T}(N~\mathrm{mod}~8)\mathcal{I}_x\right) U_z ^{m} 
\\ &=& \sum_{j<k}b_{jk}\bigg(\frac{3}{2}I_{jz}I_{kz}\nonumber
\!+\!\frac{3}{2}\bigg[\cos^2(m\gamma)I_{jy}I_{ky}\!+\!\sin(m\gamma)^2I_{jx}I_{kx}  \\
&&- \sin(m\gamma)\cos(m\gamma) \left(I_{jx}I_{ky}+I_{jy}I_{kx}\right)\bigg]-\vec{I}_j\vec{I}_k\bigg)\nonumber \\
&&+ \frac{\pi~(N~\mathrm{mod}~8)}{2T} \left(\cos(m\gamma) \mathcal{I}_{jx} - \sin(m\gamma) \mathcal{I}_{jy} \right). \nonumber
\end{eqnarray}

Note that we cannot simply follow similar steps as in the analysis of the high-frequency $\xhat$-drive: since $T J > 1$, the lowest order Magnus expansion of the toggling frame Hamiltonians is a poor approximation to the total effective Hamiltonian that governs the time evolution. However, there is one notable exception: at $\gamma = l  \pi $ all higher orders of the Magnus expansion w.r.t.~$M$ are zero and the lowest order becomes exact (up to corrections from the fast $\xhat$-drive). This can be seen right away from Eq.~\eqref{eq:large_N_toggling_frame} since $\mathcal{H}_m(l \pi)=\overline{\mathcal{H}}+(-1)^{m l }\pi (N~\mathrm{mod}~8)/(2T)\mathcal{I}_x$. Then, the effective unitary describing evolution over $M$ periods becomes (up to corrections from the fast drive),
\begin{eqnarray}
\label{eq:U_FM}
[U_F]^M\big\vert_{\gamma= l\pi} &\approx& U_z^M\exp \big[-iMT \overline{\mathcal{H}}
 + d_N(M,l)\mathcal{I}_x \big], \nonumber\\
 d_N(M,l) &=& \frac{\pi (-1)^l (\!(\!-\! 1)^{l M}\! -\!1)(N~\mathrm{mod}~8)}{2 ((-1)^l-1)}.
\end{eqnarray}
Away from $\gamma= l \pi$, the Magnus expansion cannot be used to obtain generically reliable results.


\section{\label{sec:FDTC} Prethermal Floquet time crystalline order}

For $l$ even and $\gamma=l\pi$, the $\zhat$-kicks are inactive as they give each spin a complete rotation. For these values of $\gamma$, the effective Hamiltonian is time-independent (up to corrections $\mathcal O(1/N)$ emerging from the fast-drive, which we neglect). We leave this case aside for the time being, and will return to study its heating rates in Sec.~\ref{sec:heating}.

 \vspace{-3mm}
\subsection{\label{sec:FDTC_theo} Discrete spatio-temporal symmetry breaking and prethermal time crystalline order}
 \vspace{-1mm}

Much of the intriguing behavior of the system is for odd $l$, where it exhibits prethermal time-crystalline order~\cite{Else2020} [in this case, the effective Hamiltonian depends on the parity of $M$: for odd $M$, we can obtain a contribution from the second term of Eq.~\eqref{eq:U_FM}; for even $M$, the terms in the second line in Eq.~\eqref{eq:U_FM} vanish]. In particular, at kick angles $\gamma \approx \pi$,  we argue that the driven-system satisfies all prerequisites for prethermal discrete time-crystals:

\textit{(i) Spatio-temporal Symmetry Breaking:}
After two Floquet cycles ($M\!=\!2$), the time-evolution operator Eq.~\eqref{eq:U_FM} takes a particularly simple form
\begin{eqnarray}
\label{eq:U_F2}
[U_F]^2\big\vert_{\gamma=\pi}=\mathrm e^{-i2T \overline{\mathcal{H}}}.
\end{eqnarray}
Notice that $[\overline{\mathcal{H}},\hat{P}_z]=0$, where $\hat{P}_z=\mathrm e^{-i \pi \mathcal{I}_z}$ and thus the two-cycle effective Hamiltonian $\overline{\mathcal{H}}$ obeys a discrete $\mathbb{Z}_2$ symmetry (defined by $\hat{P}_z$), arising as a consequence of the Floquet drive. 
This $\mathbb{Z}_2$ symmetry operation is induced by flipping the $\xhat$-direction of all spins.
In fact, it can be generically demonstrated that this emergent symmetry conspires with the discrete time-translation symmetry to yield a spatio-temporal ordering of eigenstates in the Floquet operator itself \cite{Else17}: at $\gamma\!=\!\pi$, $U_F$ can be approximated exponentially-well by
\begin{eqnarray}
U_F \simeq \mathcal{U} \hat{P}_z \mathrm e^{-iT \hat{D}} \mathcal{U}^{\dagger},
\end{eqnarray}
where $\hat{D}$ is an effective Hamiltonian with the property $[\hat{P}_z,\hat{D}]=0$ and $\mathcal{U}$ is a time-independent (though many-body) rotation close to the identity. Then, the effective drive generated by $\hat{P}_z \mathrm e^{-iT \hat{D}}$ obeys spatio-temporal eigenstate order. While it might be difficult to obtain $\mathcal{U}$ and $\hat{D}$ generically, there is an exception for our long-range interacting model at $N\mathrm{mod}~8 \!=\!0$, where $\mathcal{U}=1$ and $\hat{D}\equiv \overline{\mathcal{H}}$. For this case, let us denote the joint eigenstates by $|n,p\rangle$: then $\overline{\mathcal{H}}|n,p\rangle =E_n |n,p\rangle$, and $\hat{P}_z = p|n,p\rangle$ ($p\!\in\!\{\pm1$\}). The Floquet unitary can be eigen-decomposed as 
\begin{eqnarray}
\label{eq:PDTC}
U_F &=&\!\!\! \sum_{n,p\in\{\pm 1\}} \!\!\! \exp\left[-iT \left (E_n + \frac{\pi(1-p)}{2T} \right)\right]\vert n,p \rangle \langle n,p \vert , 
\end{eqnarray}
where each eigenstate $\vert n, p \rangle $ has a partner $\vert n, -p \rangle $ whose quasi-energy is shifted by exactly $\pi/T$.
In particular, this implies that \textit{any} symmetry-broken initial state will necessarily also break the discrete time-translation invariance of the Floquet drive, and instead oscillate with a period of $2T$, exhibiting period-doubling~\cite{Else2020}.

\textit{(ii) Insensitivity to the initial state} appears as consequence of spatio-temporal eigenstate order. In fact, as we will demonstrate below in Sec.~\ref{sec:thermalization}, \textit{any} initial state which is not an eigenstate of the Floquet unitary and possesses a finite $\xhat$-magnetization, even mixed states, is subject to stable long-lived period-doubled oscillations.

\textit{(iii) A parametrically long-lived prethermal time-window}, as analyzed in detail in Sec.~\ref{sec:heating}.

\textit{(iv) Robustness to perturbations:} when $\gamma=\pi+\varepsilon$ is detuned from the sweet-spot ($\xg=\pi$), the system remains symmetry-broken due to the presence of interactions. We have verified this numerically, using a Fourier-transform of the time-evolution curves of observables in units of stroboscopic cycles $M$
\begin{eqnarray}
\langle O \rangle (\omega) = \sum_{j=0}^{M-1} \mathrm e^{- i \omega j T}  \langle O \rangle (jT) ,
\end{eqnarray}
where $\omega = 2\pi k/(MT)$ with $k= 0,\dots, (M-1)$ and $\langle O \rangle (jT)$ is the expectation value of the observable of interest at stroboscopic times. The results are displayed in Fig.~3 B and D of the main text for experimental and theoretical data. Around $\gamma = \pi$ we clearly observe a single rigid peak at $\omega = \pi/T$ of finite extent ($\app \pi/5$), corresponding to a period-doubled oscillation in both panels.
In addition, before the system heats up to infinite temperature, the prethermal time-crystalline order is also robust to small random perturbations in the time duration $\tau$, as we demonstrate numerically in Sec.~\ref{sec:heating}.

It follows therefore that our $\Cs$ nuclear spin system, subject to the two-frequency drive at $\gamma\approx \pi$, features prethermal discrete time crystalline (PDTC) order.

 \vspace{-3mm}
\subsection{\label{sec:thermalization} Thermalizing dynamics in the presence of emergent (quasi-)conservation laws}
 \vspace{-1mm}

Consider again the two-cycle time-evolution operator of Eq.~\eqref{eq:U_F2}. We leave the discussion of unconstrained heating to infinite temperature for Sec.~\ref{sec:heating}. Since the effective Hamiltonian $\overline{\mathcal{H}}$ is ergodic, according to the Eigenstate Thermalization Hypothesis (ETH)~\cite{dalessio2016quantum}, we expect an initial state to thermalize with respect to $\overline{\mathcal{H}}$. After the non-universal initial transient dynamics, the state of the system is described by a Gibbs state, which is determined by the corresponding Lagrange multipliers (the obvious one being the inverse temperature $\beta$). 
Thermal states are time-translation invariant, which implies Lagrange multipliers that are constant in time. 
Similarly, Floquet prethermal states are described by slowly and continuously changing Lagrange multipliers~\cite{fleckenstein2021prethermalization,fleckenstein2021thermalization}.

At first sight, this appears at odds with the period-doubling oscillations of observables, induced by the broken spatio-temporal symmetry near $\gamma=\pi$. To resolve this conundrum, note first that the thermal properties of the state are defined w.r.t.~the period-doubled unitary, Eq.~\eqref{eq:U_F2}: indeed, on the timescale $2M$ (discrete) time-translation invariance is inherent. Yet, this does not \textit{a priori} imply time-invariance of the state on smaller timescales, i.e.~over a single Floquet cycle $M$, where we observe a strong alternating pattern in time. As we explain below, the intriguing mechanism that enables a time-dependence in the (approximate prethermal) Gibbs state can be understood to arise from a fine interplay between two conspiring emergent symmetries of $\overline{\mathcal{H}}$, both induced by the two-frequency drive: (i) the spatio-temporal Ising $\mathbb{Z}_2$ symmetry in combination with (ii) the $\xhat$-magnetization quasi-conservation.

\begin{figure*}
    \centering
    \includegraphics[width=1\textwidth]{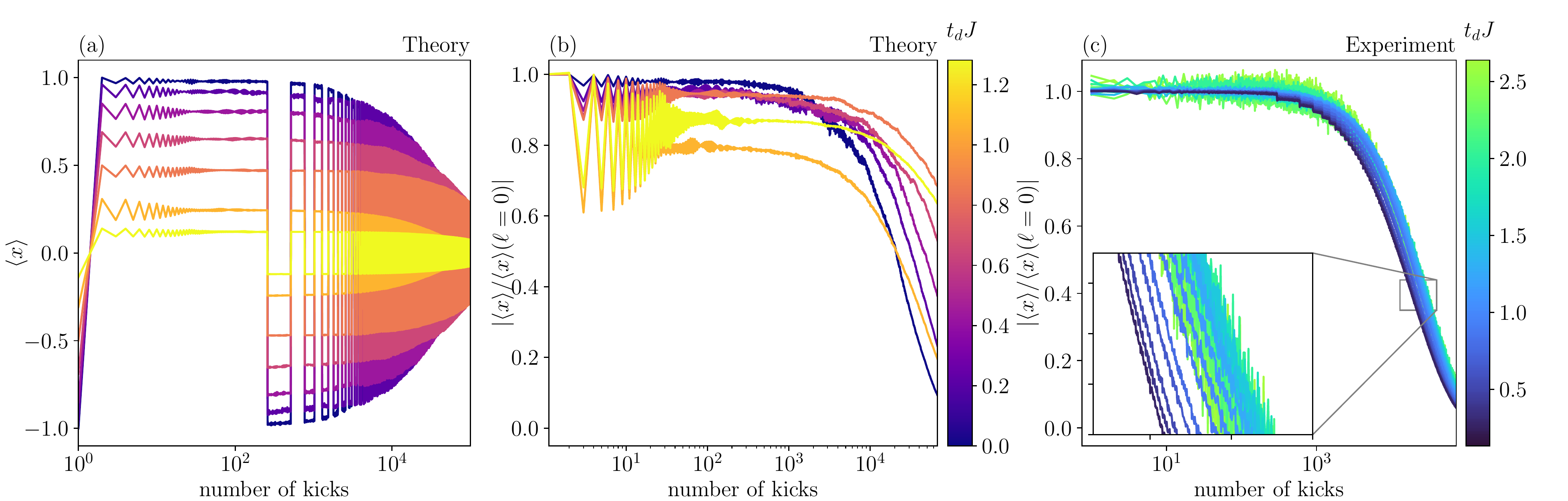}
    \caption{
    \textbf{Numerical Simulation: (a)}-\textbf{(b)} 
    Time evolution of different initial states $\vert \psi \rangle_i^{t_d}$ prepared by evolving the fully polarized state for a time $t_d$ with $\mathcal{H}$: $\vert \psi \rangle_{i}^{t_d} = \mathrm e^{-it_d\mathcal{H}} \bigotimes_{j=1}^L  \frac{1}{\sqrt{2}}\left(\vert \! \uparrow_j \rangle - \vert \downarrow_j \rangle\right)$. 
    In (b) we display the absolute value of the $\xhat$-magnetization using the data from (a), normalized to its initial magnitude. The simulation parameters are $L=16$, $N=255$, $\gamma=\pi+0.02$, $J\tau=0.2$, and the remaining parameters are the same as in Fig.~\ref{fig:time_evolution}.
    \textbf{Experiment: (c)} Corresponding experimental time-evolution curves for different initial states obtained using the same procedure: the initially prepared density matrix (see main text) is evolved with the bare dipolar Hamiltonian $\mathcal{H}$ up to time $t_d$ before the two-frequency drive is started. The inset provides a zoom in of the marked region to visualize individual curves. The parameters of the experiment are the same as in Fig. 3 of the main text, and $\gamma = 0.96\pi$. }
    \label{fig:init_state}
\end{figure*}

If we initialize the system in a $\mathbb{Z}_2$-symmetry-broken state with non-zero magnetization, we expect the system to prethermalize only w.r.t.~the corresponding symmetry-broken sector \cite{Machado2020}. In the grand-canonical formulation of ETH, the subsystem density matrix $\hat{\rho}_{A}$ in the prethermal plateau resembles a thermal state with respect to this symmetry-broken sector of the subsystem (effective) Hamiltonian $\overline{\mathcal{H}}_A$~\cite{fleckenstein2021prethermalization,ikeda2021fermi}. Notice that, in this case, $\hat{\rho}_{A}$ is not time-translation invariant under evolution with $U_F$: instead, $U_F\hat{\rho}_{A}U_F^{\dagger}= \hat{P}_z \hat{\rho}_{A}\hat{P}_z^{\dagger} = \overline{\hat{\rho}_{A}} $, where $\overline{\hat{\rho}_{A}}$ corresponds to the (prethermal) subsystem density matrix in the complementary $\mathbb{Z}_2$-symmetry sector. Indeed, the fact that $\hat{\rho}_A$ is thermal only w.r.t.~one of the two $\mathbb{Z}_2$ symmetry sectors is equivalent to having non-zero magnetization expectations of $\hat{\rho}_A$ and $\overline{\hat{\rho}_{A}}$. 

More precisely, according to ETH, any given initial state $\vert \psi_0 \rangle$ exposed to the drive of Eq.~\eqref{eq:PDTC} is expected to locally resemble a Gibbs state. In the thermodynamic limit, a subsystem $A$ will then be described by 
\begin{eqnarray}
\label{eq:GGE}
\hat{\rho}_A=\frac{1}{\mathcal Z}\exp\left(-\beta \overline{\mathcal{H}}_A - \mu\mathcal{I}_{x,A} \right),
\end{eqnarray}
where $\mathcal Z = \mathrm{Tr}[\exp(-\beta \overline{\mathcal{H}}_A-\mu \mathcal{I}_{x,A} )]$ with the inverse temperature $\beta$. Here $\overline{\mathcal{H}}_A$ and $\mathcal{I}_{x,A}$ are the Hamiltonian and $\xhat$-magnetization restricted to subsystem $A$, respectively. The magnetization potential $\mu$ is determined self-consistently from the initial state: 
\begin{eqnarray}
\label{eq:ETH_beta}
\mathrm{Tr}[\overline{\mathcal{H}}\hat{\rho}(t=0)]&=&\frac{1}{\mathcal Z}\mathrm{Tr}\big[\overline{\mathcal{H}}_A\mathrm e^{-\beta \overline{\mathcal{H}}_A-\mu \mathcal{I}_{x,A} }\big],\\
\label{eq:ETH_mu}
\mathrm{Tr}[\mathcal{I}_{x}\hat{\rho}(t=0)]&=&\frac{1}{\mathcal Z}\mathrm{Tr}\big[\mathcal{I}_{x,A}\mathrm e^{-\beta \overline{\mathcal{H}}_A-\mu \mathcal{I}_{x,A} }\big].
\end{eqnarray}
So long as $\mu\neq 0$, the state described by Eq.~\eqref{eq:GGE} is not invariant under time evolution with $U_F$. Instead, time evolution over a single Floquet cycle maps $\mu \rightarrow - \mu$ as a result of the spatio-temporal $\mathbb{Z}_2$ symmetry.

Now, consider initializing the system at $t=0$ in a state where, $\mathrm{Tr}[\mathcal{I}_{x}\hat{\rho}(t=0)]=0$. For such an initial condition, let us solve Eq.~\eqref{eq:ETH_mu} for $\mu$. Because of the $\xhat$-magnetization symmetry,
$\overline{\mathcal{H}}_A$ and $\mathcal{I}_{x,A}$ possess a common set of eigenvectors, so that $\overline{\mathcal{H}}_A\vert s,m \rangle = E_{s,m}\vert s,m \rangle$, and $\mathcal{I}_{x,A}\vert s,m \rangle = m \vert s,m\rangle$ with $m$ fixing the magnetization sector. Then, Eq.~\eqref{eq:ETH_mu} (assuming vanishing LHS) can be rewritten as
\begin{eqnarray}
\label{eq:ETH_to_solve}
0 = -\frac{\partial }{\partial \mu} \ln\bigg(\sum_{s,m}\exp[-\beta E_{s,n} -\mu m ]\bigg).
\end{eqnarray}
Since $\overline{\mathcal{H}}_A$ has the additional $\mathbb{Z}_2$ Ising symmetry $\hat{P}_z$ [cf.~Sec.~\ref{sec:FDTC}], we know that $E_{s,m}=E_{s,-m}$ and thus Eq.~\eqref{eq:ETH_to_solve} boils down to 
\begin{eqnarray}
\sum_{s,m>0}2m\mathrm e^{-\beta E_{s,m}} \sinh(\mu m) = 0,
\end{eqnarray}
which has a unique solution, namely $\mu\equiv 0$.
To sum up, states that obey $\mathrm{Tr}[\mathcal{I}_{x}\hat{\rho}(t=0)]=0$, are invariant under time-evolution with $U_F$. In particular all eigenstates of $U_F$ satisfy this condition, since they possess a spatio-temporal ordering. 

As an example, consider the following initial state:
\begin{eqnarray}
\vert C_{\pm}\rangle =\frac{1}{\sqrt{2}}\bigg(\bigotimes_{j=1}^L \vert \! \rightarrow_j \rangle \pm \bigotimes_{j=1}^L \vert \! \leftarrow_j \rangle \bigg),
\end{eqnarray}
where $\vert \! \rightarrow_j \rangle$ ($\vert \! \leftarrow_j \rangle$) represents an eigenstate of $I_{jx}$ with eigenvalue $+1/2$ ($-1/2$). One can convince onself that $\vert C_{\pm} \rangle $ obeys the required ordering and is indeed an eigenstate of $U_F$ with $\mathrm{Tr}[\mathcal{I}_{x}\hat{\rho}(t=0)]=0$.
Hence, when starting from the state $\vert C_{\pm} \rangle $ (or likewise any other cat-like linear combination of two $\xhat$-magnetization eigenstates), no subharmonic oscillations can appear: the state is time-translation invariant under evolution with $U_F$. In contrast, any symmetry-broken initial state which obeys $\mu\neq 0$ necessarily breaks the discrete time-translation symmetry.

To validate this conclusion numerically, we initialize the system in various different states by time-evolving the $\xhat$-polarized state for some transient time $t_d$ with $\mathcal{H}$, i.e. $\vert \psi \rangle_{i}^{t_d} = \mathrm e^{-it_d\mathcal{H}} \bigotimes_{j=1}^L  \vert \leftarrow_j \rangle$. In that way we obtain experimentally accessible, yet highly non-trivial, initial states with non-zero $\xhat$-magnetization, provided $t_d J \lesssim 1 $. The results are displayed in Fig.~\ref{fig:init_state} (a) and (b). We plot the absolute value of the signal $|\expec{x}|$ for clarity.  All chosen initial states lead to long-lived subharmonic oscillations of the signal; yet, we find differences in their lifetime: while states with large initial $\xhat$-magnetization tend to increase their lifetime with increasing $t_d$, at small initial $\xhat$-magnetization, the lifetime behavior becomes less systematic. For values very close to zero initial $\xhat$-magnetization we observe no PDTC order (data not shown). 

Experimentally, we pursue the same strategy. We create different initial states by evolving the rotated starting density matrix $\xr_0$ (see main text) with the bare Hamiltonian $\mathcal{H}$ for different times $t_d$ to obtain a set of different and highly non-trivial initial states, which are subsequently subject to two-frequency driving. Each state corresponds to a different run of the experiment. We display the results in Fig.~\ref{fig:init_state} (c). We show only absolute values, yet all tested initial states create persistent (in time) and rigid (in $\varepsilon$) subharmonic oscillations with similar heating times. This is a notable difference to the data obtained using numerical simulations, where different initial states can induce considerably different heating times. We hypothesize that the difference in the heating behavior of various initial states is related to the relatively small system sizes that can be reached in the numerical simulations.

\subsection{Classification of discrete time crystalline order}

As outlined in detail in the previous section, time crystalline order is expected to emerge for all states which initially possess a finite $\xhat$-magnetization expectation value: this is because, within the prethermal plateau, thermalization happens with respect to $\overline{\mathcal{H}}_A$, which preserves the initial $\xhat$-magnetization. Similar to the discussion of Ref.~\cite{luitz2020prethermalization}, this emergent $\xhat$-magnetization conservation additionally constrains thermalization, which implies that no thermal phase transition in $\overline{\mathcal{H}}_A$ can preclude the PDTC order to form. This mechanism of stabilizing prethermal time-crystalline order is different from the conventional approach where the formation of PDTC order is solely dependent on the energy density and the associated inverse temperature $\beta$ of the initial state with no additional protection. In our Floquet system, such a phenomenon can occur when the repetition number $N$ is decreased and $\xhat$-non-conserving terms $\propto 1/N$ start gaining importance, cf.~Eq.~\eqref{eq:Gs}.

Importantly, the extent to which $\overline{\mathcal{H}}$ describes the dynamics of the system is parametrically controlled by the frequency of the fast $\xhat$-drive: increasing the drive frequency
(i) ensures a suppression of $\xhat$-magnetization conservation breaking terms appearing in higher order corrections of the inverse frequency expansion of the two-cycle effective Hamiltonian, and (ii) it parametrically increases the lifetime of the prethermal plateau for which the DTC order can be observed (see Sec.~\ref{sec:heating} below). However, since our system exhibits long-range interactions with a critical exponent, we obtain a power-law suppression of heating rates as a function of frequency (see Sec.~\ref{sec:heating}). The DTC order in our system thus combines features of prethermal time-crystalline order~\cite{Machado2020,Kyprianidis21} and critical time-crystalline order \cite{Choi17,ho2017critical}.



 \vspace{-3mm}
\section{\label{sec:heating}Heating timescales and duration of the prethermal plateau}
 \vspace{-1mm}

The lifetime of a PDTC, like that of any prethermal order, is predominantly determined by the amount of energy absorbed from the nonequilibrium drive. To quantify this energy absorption, we define the heating time empirically, as follows. Given the time evolution curve of some observable $O$, the heating time is the time required to reach $1/\mathrm e$ of the initial value of that observable. 
In principle, one would have to define the heating time with respect to the \textit{prethermal} value of a given observable. However, since we only investigate the quasi-conserved  $\xhat$-magnetization, the initial and prethermal expectation values coincide.

The concatenated two-frequency drive offers two obvious mechanisms for the system to heat up to infinite temperature: the slow and fast drives independently cause energy absorption,  each with a potentially different rate corresponding to its own timescale. Numerically and experimentally, we observe an interplay of both timescales. However, theoretically we can disentangle them and analyze each effect separately. To this end, in what follows, we discuss the influence of heating caused by the slow $\zhat$-drive, and then the fast $\xhat$-drive, before we conclude with the general case for the two-frequency drive.

 \vspace{-3mm}
\subsection{\label{sec:heaating_slow}Heating timescales associated with the slow $\zhat$-drive}
 \vspace{-1mm}

To analyze heating caused by the slow drive, we assume the (approximate) effective Hamiltonian emerging from the fast drive, Eq.~\eqref{eq:fast_drive_eff_H}, to be exact. This allows us to numerically study a periodic drive of the form of Eq.~\eqref{eq:U0}.

We further increase the ergodicity of the relatively small systems accessible in our simulations and  diminish the related finite-size effects by adding a small uniformly distributed noise $\delta$ to the driving period~\cite{fleckenstein2021prethermalization, fleckenstein2021thermalization} 
\begin{eqnarray}
\label{eq:slow_drive_exact}
U_F = \mathrm e^{-i N(\tau +\delta\tau) \overline{\mathcal{H}}}U_z,
\end{eqnarray}
where $\delta$ takes a different value in each Floquet cycle.

\begin{figure}[t!]
    \centering
    \includegraphics[scale=0.49]{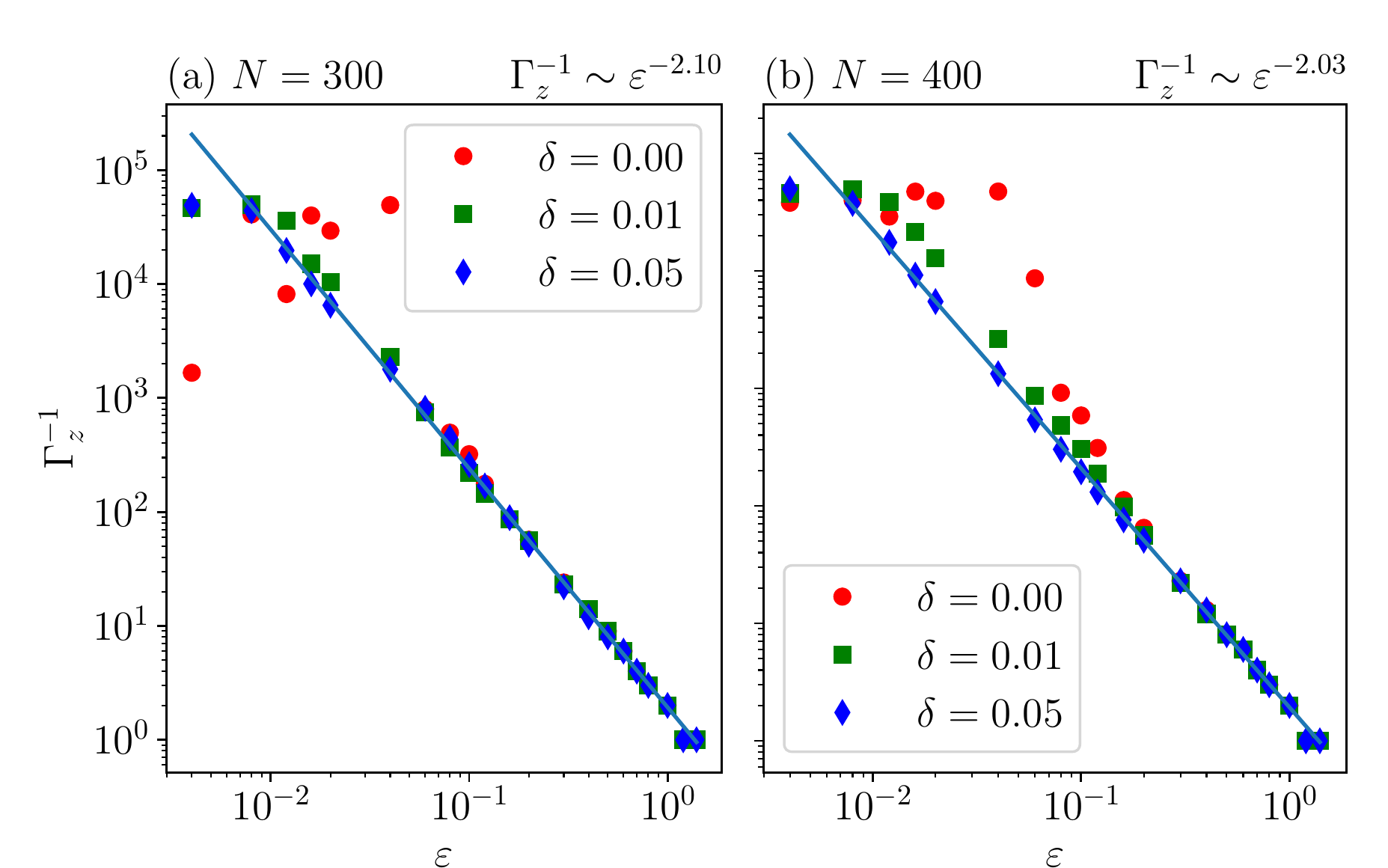}
    \caption{\textbf{Numerical simulation --} $\varepsilon$ dependence of the heating timescales $\Gamma_z^{-1}$ associated with the slow $\zhat$-drive: (a) $N=300$, (b) $N=400$ around  $\gamma = \pi + \varepsilon$. The system size is $L=14$, and $J\tau =0.2$.
    Further simulation parameters are the same as in Fig.~\ref{fig:time_evolution}.
}
    \label{fig:slow_drive_heating_N_dep}
\end{figure}

We then evolve the fully $\xhat$-polarized initial state under Eq.~\eqref{eq:slow_drive_exact} up to $5\times 10^4$ kicks. The results are displayed in Fig.~\ref{fig:slow_drive_heating_N_dep}. We find a clear power-law scaling of the heating times $\Gamma_z^{-1}$ which is close to a Fermi-golden-rule scaling, $\Gamma^{-1}_z\sim \varepsilon^{-2}$. This can be understood as the $\zhat$-drive is effectively operating in the low-frequency regime for $N\gg 1$ [the short-range interacting exponential suppression typically occurs in the high-frequency regime]. 
Note that, at $\varepsilon \!=\! 0$, although the energies of many-body states can differ by multiples of the drive frequency, heating is suppressed by the complete absence of matrix elements between these states. Finite $\varepsilon\!>\!0$ induces finite matrix elements which in turn results in intense energy absorption according to Fermi's Golden rule. 
Note that we observe approximately the same golden rule $\varepsilon$-scaling for different values of $N$ (Fig. \ref{fig:slow_drive_heating_N_dep}a-b). 

\begin{figure}[t!]
    \centering
    \includegraphics[scale=0.49]{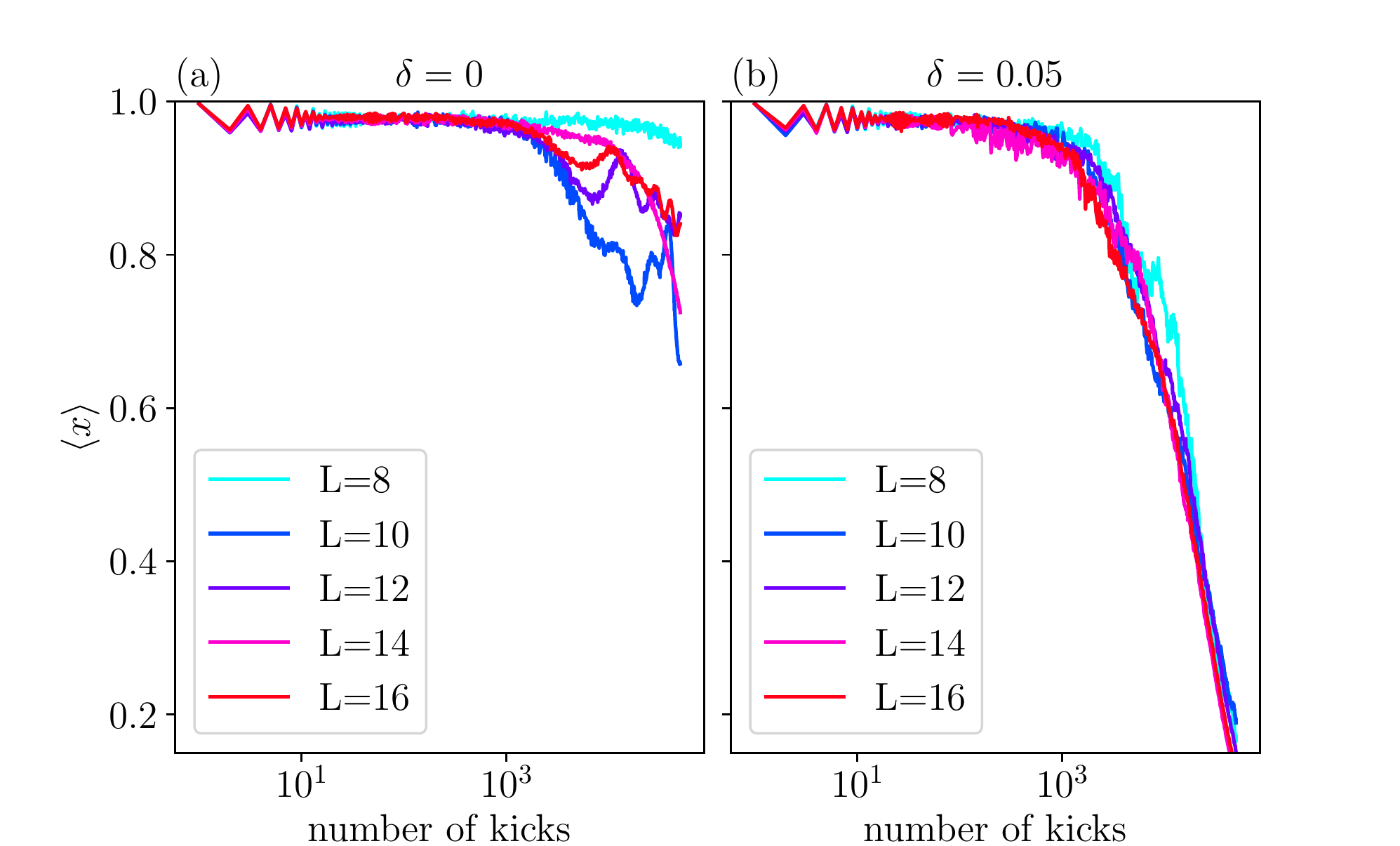}
    \caption{\textbf{Numerical simulation --} Time-evolution for random graphs of different system sizes at $\gamma = - 0.039$ for $\delta=0$ (a) and $\delta=0.05$ (b). $N=300$ and $J\tau=0.2$. Remaining parameters are as in Fig.~\ref{fig:time_evolution}.
}
    \label{fig:finite_size}
\end{figure}

The power-law suppressed heating in our system, causes the finite stable regions around $\gamma = 0,\pi$ to shrink much faster, as compared to the exponential suppression of heating at high-frequencies.
However, in practice the timescales are still parametrically controlled by $\varepsilon$ and the prethermal physics is governed by the effective Hamiltonian derived in Sec.~\ref{sec:eff_H}. In fact, the experimentally observed lifetimes for our PDTC readily exceed state-of-art lifetimes of PDTCs in the high-frequency regime~\cite{Rovny18,Kyprianidis21}.

We note in passing that finite noise $\delta$ in the driving protocol has no influence on the scaling of heating times, but it is capable of removing long-time finite size synchronization effects \cite{fleckenstein2021prethermalization,fleckenstein2021thermalization}. This can be seen in Fig.~\ref{fig:slow_drive_heating_N_dep} where finite $\delta$ causes more points to align on the straight line without changing its slope, whereas points that are already in-line remain unchanged. 
Yet another important aspect of finite $\delta$ is shown in Fig.~\ref{fig:finite_size}: while the curves of the noise-free evolution in (a) -- especially their long-time heating behaviour -- are affected by finite-size effects, the addition of a small random $\delta$ almost collapses all curves so that the heating times become basically insensitive to the system size $L$. By contrast, the experiment offers the advantage of working with a large enough system size which eliminates finite-size effects for practical purposes, and hence the extra $\delta$-noise is neither necessary, nor used.

 \vspace{-3mm}
\subsection{Heating timescales associated with the fast $\xhat$-drive}
 \vspace{-1mm}

The physics of the thermalization dynamics associated with the fast $\xhat$-drive only, is not immediately obvious due to the interplay between three effects:
(i) the power of the dipolar interaction term is critical in three dimensions, which implies a logarithmic divergence of the total energy in the thermodynamic limit.
(ii) In the limit $\tau \rightarrow 0 $, the effective Hamiltonian of the system approaches an integrable model, for which the notion of thermalization itself is not well-defined. Integrability breaking, though, occurs at order $\mathcal{O}(\tau)$ in the effective Hamiltonian, although its not pronounced at the accessible system sizes in the numerical simulations. 
(iii) Even when $J\tau\ll 1 $, we still have $\vartheta \sim \mathcal{O}(1)$ so that the high-frequency regime ($\vartheta\ll1$ and $J\tau\ll 1$) is practically inaccessible.

\begin{figure}
    \centering
    \includegraphics[scale=0.5]{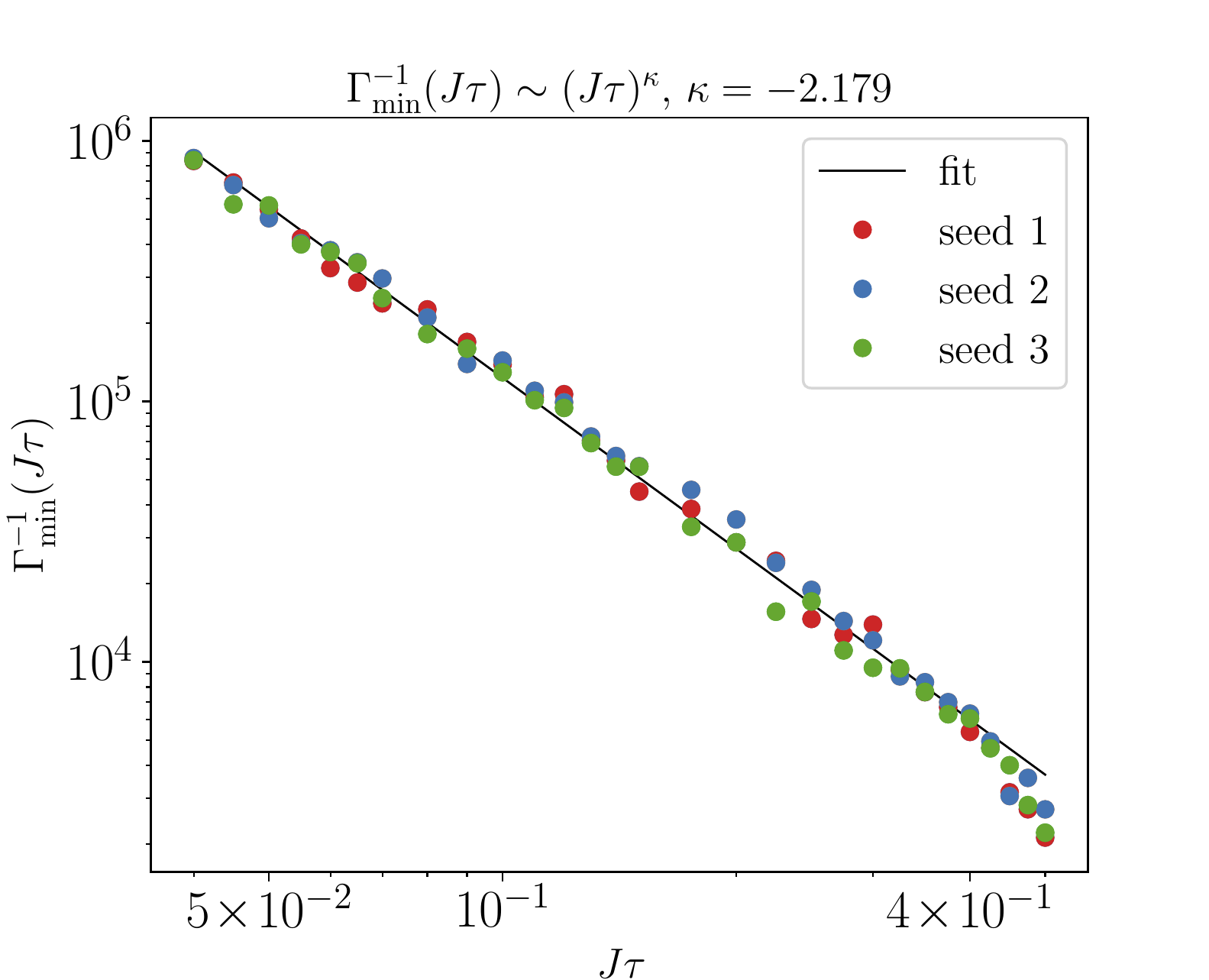}
    \caption{\textbf{Numerical simulation --} Heating times $\Gamma^{-1}_{\mathrm{min}}(J\tau)$ associated with the fast $\xhat$-drive, extracted from corresponding time-evolution curves (data not shown), as a function of $J\tau$. Different colors (red, green and blue) correspond to different random graph realizations. The system size is $L=14$. The remaining simulation parameters are the same as in Fig.~\ref{fig:time_evolution}.
    }
    \label{fig:fast_drive}
\end{figure}

For these reasons, it is difficult to make  predictions for rigorous bounds on the heating rates, given the above properties of the model~\cite{Machado2020}. Nonetheless, it is possible to study the energy absorption numerically. To this end, we investigate the drive generated by the repeated application of $U_x U_\mathcal{H}$ at $\gamma=0$ with fixed $\vartheta=\pi/2$ for different values of $J\tau$. From the time-evolution curves we extract the heating times $\Gamma^{-1}_{\mathrm{min}}(J\tau)$ and display them as a function of $\tau$ in Fig.~\ref{fig:fast_drive} for different random graph realizations. We find that -- similar to the slow $\zhat$-drive -- the fast $\xhat$-drive shows a power-law scaling, $\Gamma^{-1}_\mathrm{min}\propto (J\tau)^\kappa$ with a Golden-rule exponent $\kappa\approx -2$. This result is in agreement with recent experimental measurements reported in Ref.~\cite{Beatrez21}.

 \vspace{-3mm}
\subsection{\label{sec:heating_combined}Heating model for the concatenated two-frequency drive}
 \vspace{-1mm}

Now that we have analyzed the fast and slow drives independently, we can move on with the discussion of the heating rates of the more complex concatenated two-frequency drive. 

\begin{figure}[t]
\centering
\includegraphics[scale=0.5]{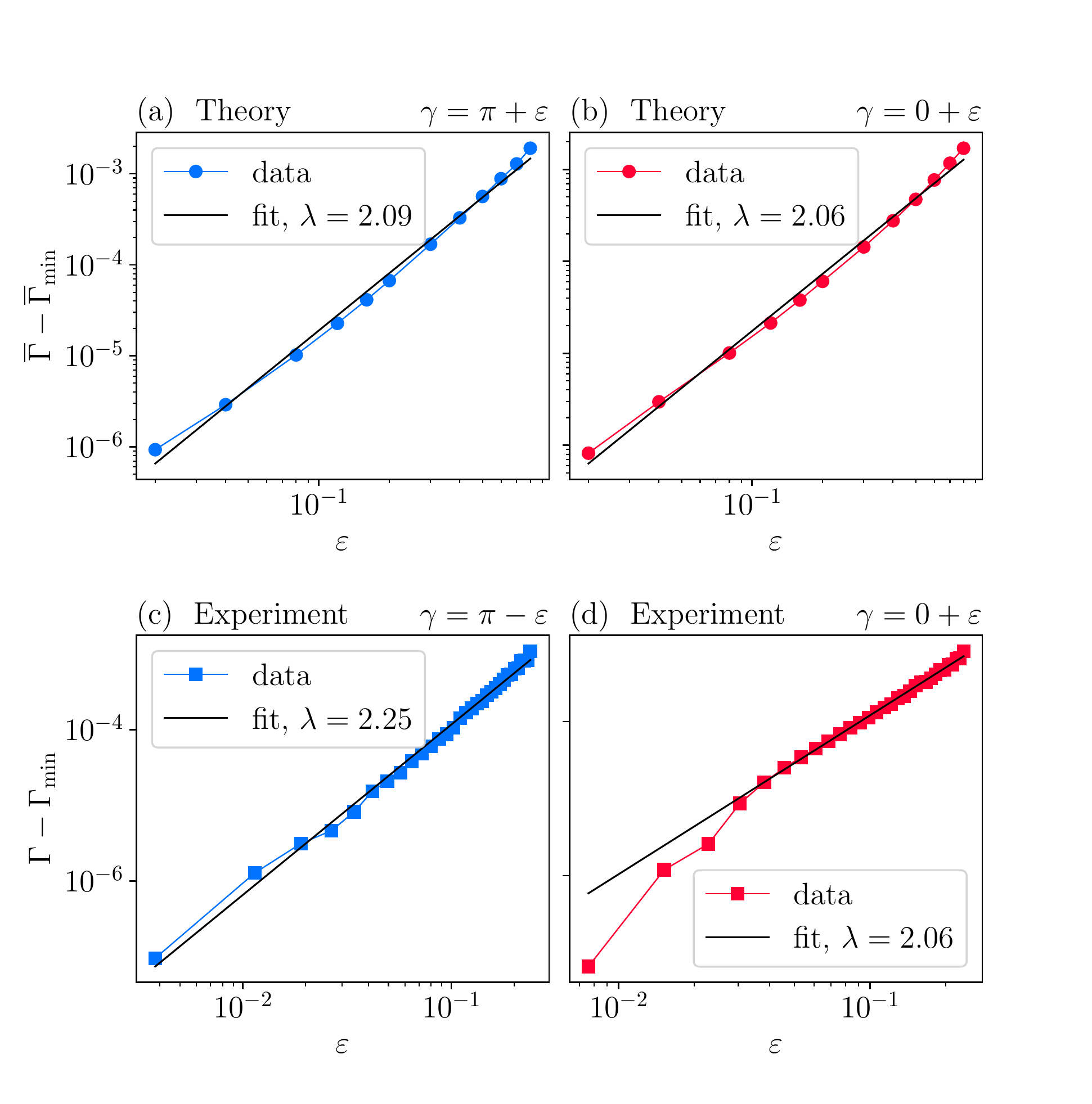}
\caption{
    \textbf{Comparison of scaling behaviour in theory and experiment:} \textbf{(a)}-\textbf{(b)} \textbf{Numerical simulation --} Numerically extracted heating times averaged over 20 graph realizations, $\overline{\Gamma}^{-1}=N_{\mathrm{seed}}^{-1}\sum_{s=1}^{N_{\mathrm{seed}}} \Gamma_s^{-1}$ with $N_{\mathrm{seed}}=20$, as function of $\varepsilon$ at $\gamma=\pi+\varepsilon$ (a) and $\gamma = 0+ \varepsilon$ (b). We plot $\overline{\Gamma}-\overline{\Gamma}_{\mathrm{min}}$ on the $y$ axis to check for corresponding scaling. $\overline{\Gamma}_{\mathrm{min}}^{-1} $ is defined as the heating time at $\varepsilon=0$: $\overline{\Gamma}_{\mathrm{min}}^{-1}\equiv \overline{\Gamma}^{-1}(\varepsilon=0)$. We fit the data to $\overline{\Gamma} = g/N\varepsilon^{\lambda}+\overline{\Gamma}_{\mathrm{min}}$. 
    (fit values of (a): $g=0.56$, $\overline{\Gamma}_{\mathrm{min}}=3.67\times 10 ^{-5}$, fit values of (b): $g=0.52$, $\overline{\Gamma}_{\mathrm{min}}=3.40\times 10 ^{-5}$). The simulation parameters are $L=16$, $N=255$, $J\tau=0.2$.
    Remaining parameters are as in Fig.~\ref{fig:time_evolution}. \textbf{(c)}-\textbf{(d)} \textbf{Experiment --} Corresponding heating rates extracted from experimental data for $\gamma = \pi - \varepsilon$ (c) and $\gamma = \varepsilon$ (d). In (d) the three smallest data points are not included in the fit since these curves did not reach the $1/\mathrm e$-threshold. We fit the data to $\Gamma = g/N\varepsilon^{\lambda}+\Gamma_{\mathrm{min}}$ (fit parameters of (c): $g=6.15$, $\Gamma_{\mathrm{min}}=1.75\times 10 ^{-5}$, fit parameters of (d): $g=4.17$, $\Gamma_{\mathrm{min}}=7.34\times 10 ^{-6}$). The remaining parameters are the same as in Fig. 3 of the main text.
}
\label{fig:lorentzian}
\end{figure}

To understand the combined heating timescales we deploy a simple model: assuming that there are no correlation effects between the different drives (which applies in the time-scale separated regime $N \gg 1$), a simple model that approximates the dynamics of the magnetization decay can be formulated using a discretized-in-time rate equation that captures the influence of the $MN$-th kick on the magnetization:
\begin{equation}
\label{eq:diff_eq_heating}
\langle x \rangle ((MN \! + \! 1)\tau ) - \langle x \rangle (MN\tau) = - \bigg[\frac{\Gamma_z(\varepsilon)}{N}  + \Gamma_{\mathrm{min}}(\!J\tau\!)\bigg] \langle x \rangle (MN\tau).
\end{equation}
Here we denote the heating rates of the slow $\zhat$-drive and the fast $\xhat$-drive by $\Gamma_z(\varepsilon)$ and $\Gamma_{\mathrm{min}}(J\tau)$, respectively.
Recalling that $N$ and $M$ are the repetition numbers of the $\xhat$- and $\zhat$- drives, Eq.~\eqref{eq:diff_eq_heating} has the simple solution 
\begin{equation*}
	\langle x \rangle (MN\tau) = \exp\left[-\left(\frac{\Gamma_z(\varepsilon)}{N} +\Gamma_{\mathrm{min}}(J\tau)\right) MN\tau\right].
\end{equation*} 
Notice that the exact functional form of the true time-dependence of the magnetization may not resemble a simple exponential. However, if we are only interested in heating timescales and taking into account the experimentally observed mono-exponential behavior, Eq.~\eqref{eq:diff_eq_heating} provides a valid approximation, since the time required to reach a value of $1/\mathrm e$ is exactly given by $\Gamma^{-1}= [\Gamma_z(\varepsilon)/N +\Gamma_{\mathrm{min}}(J\tau)]^{-1}$.
At $\varepsilon\!=\!0$ the slow drive is ineffective so that $ \Gamma^{-1}\vert_{\varepsilon=0}= \Gamma_{\mathrm{min}}^{-1}(J\tau)$, and we recover the heating rate of the fast drive. For a fixed fast-drive frequency, we can thus identify the minimum heating rate as $ \Gamma_{\mathrm{min}}(J\tau)$. Together with the conclusions from Sec.~\ref{sec:heaating_slow} we, therefore, expect a scaling approximately given by
\begin{eqnarray}
\label{eq:scaling_combined}
\Gamma^{-1} = \frac{1}{\frac{g}{N}\varepsilon^{2}+\Gamma_{\mathrm{min}}(J\tau)},
\end{eqnarray}
with some numerical constant $g$. Rearranging this expression as ${\Gamma} = g/N\varepsilon^{\lambda}+{\Gamma}_{\mathrm{min}}$ we fit a straight line on a log-log plot to both numerical and experimental data, with fitting parameters $g$, $\lambda$, and $\Gamma_{\mathrm{min}}$.

In Fig.~\ref{fig:lorentzian} we present fits of the heating times extracted from both numerical simulations (Fig.~\ref{fig:lorentzian} (a)-(b)) and experimental data (Fig.~\ref{fig:lorentzian} (c)-(d)). The left (right) columns in the figure show data taken in the vicinity of $\gamma=\pi$ ($\gamma=0$).
The axes are chosen such that a straight line indicates scaling according to Eq.~\eqref{eq:scaling_combined}. For all datasets we see a straight line over at least two decades, which renders the fit results trustworthy, and in agreement with our simplified theoretical model. That said, in the simulated data, we observe a slight upward bending in the small $\varepsilon$-regime. This is to be expected as Eq.~\eqref{eq:diff_eq_heating} constitutes only a minimal approximate model, which completely disregards the interplay between the two constituent drives that are likely to contribute an additional $\varepsilon$ dependence. However, these interplay effects are expected to fade away as time-scale separation becomes more pronounced in the limit $N\gg 1$. This behavior is corroborated by the experimental data in Fig.~\ref{fig:lorentzian} (c) and (d)  obtained with $N=300$, where almost no bending is visible anymore. 
A few experimental data points for small $\varepsilon$ in (d) were left out of consideration when computing the fitting line, since the corresponding time-traces did not reach the $1/\mathrm e$ threshold by the time the experiment ended. 
Overall, simulation and experimental data yield very similar heating rates, described by a power-law with an exponent close to $2$, consistent with our theoretical analysis.

\subsection{\label{sec:freq_dep}Frequency dependence of heating timescales}

\begin{figure}
    \centering
    \includegraphics[scale=0.48]{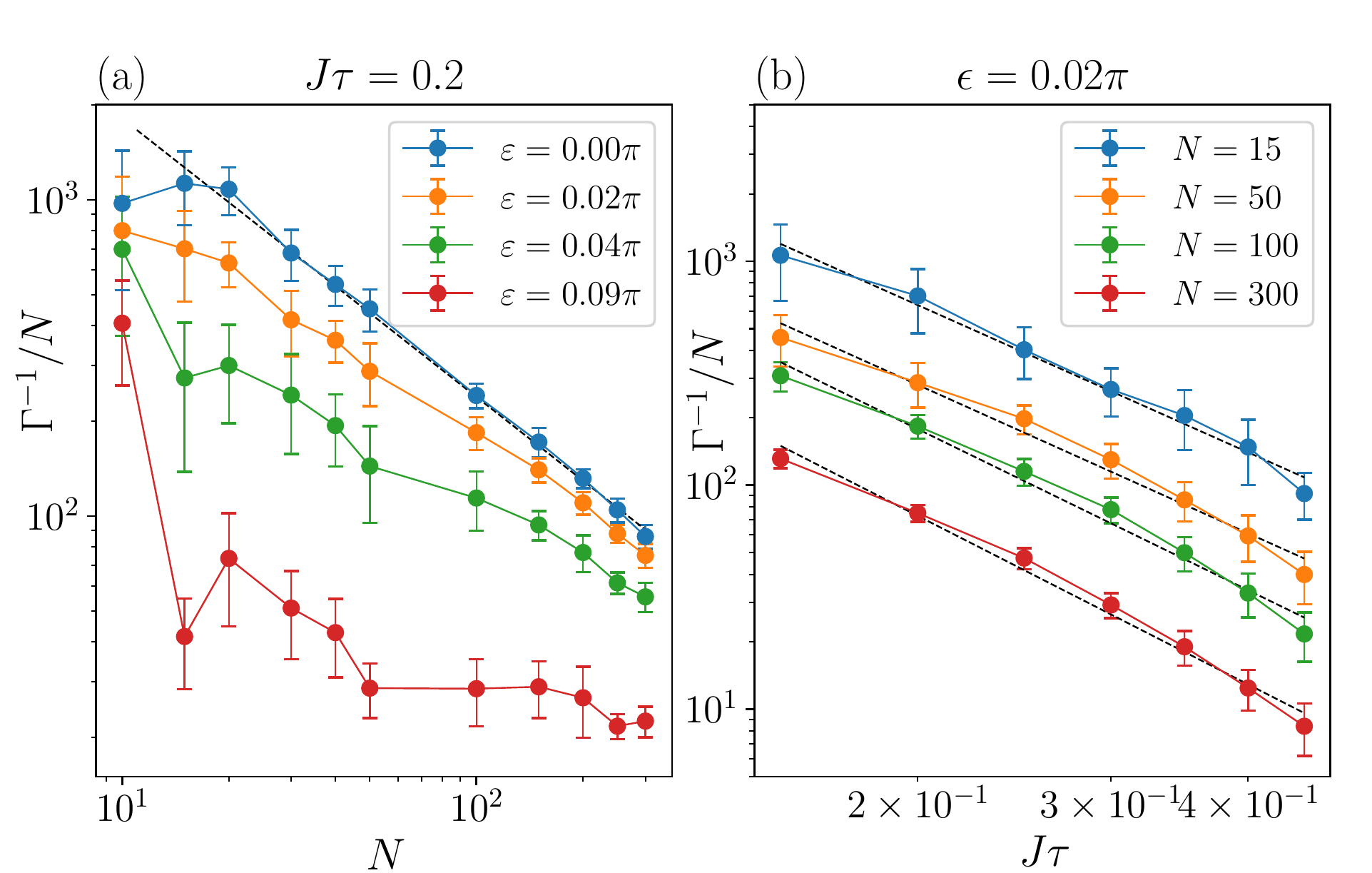}
    \caption{\textbf{Numerical simulation --} (a) $N$-dependence of the heating time $\Gamma^{-1}$ in units of Floquet cycles for 4 different values of $\varepsilon$. The dashed lines correspond to a least-square fit. We find a scaling of $\Gamma^{-1}/N \sim N^{-0.88}$ (b) $J\tau$-dependence of the heating time $\Gamma^{-1}$ in units of Floquet cycles for 4 different $N$. In both panels all points are averaged over 10 random graph realizations. Dashed lines correspond to least-square fits: we find a scaling of We find a scaling of $\Gamma^{-1}/N \sim (J\tau)^{-\alpha}$ with $\alpha=2.18$ ($N=15$), $\alpha=2.20$ ($N=50$), $\alpha=2.39$ ($N=100$) and $\alpha=2.50$ ($N=300$). The system size is $L=16$. Remaining parameters are as in Fig.~\ref{fig:time_evolution}.}
    \label{fig:freq_dep}
\end{figure}

In Floquet systems, heating is predominantly controlled by the frequency of the employed drive, where increased frequencies lead to a suppression of heating rates and a corresponding increase of heating timescales. In our system the frequency of the Floquet drive is given by $\omega_F=2\pi/(N\tau)$. An increase of driving frequency can thus be accomplished but tuning either $\tau$ or $N$. Since, finite $\varepsilon$ has vanishing influence on the overall frequency of the Floquet drive, let us first investigate the limit $\varepsilon\rightarrow 0$. Then, the model of Eq.~\eqref{eq:scaling_combined} predicts heating timescales (in units of Floquet cycles) to scale as $\Gamma^{-1}/N \sim 1/(N(J\tau)^2)$ which we confirm numerically (Fig.~\ref{fig:freq_dep}).
With increasing $\varepsilon$, the increase of $\Gamma^{-1}/N$ as a function of $N$ is slowed down until a break-even point is reached where $\varepsilon$ becomes the dominant source of heating (Fig.~\ref{fig:freq_dep} (a)). Notice that as $N\approx 10$ timescale separation is gradually lost while interplay effects arising from interference/resonance between the two drives start becoming relevant; this implies that Eq.~\eqref{eq:scaling_combined} does not capture the heating dynamics anymore in the non-timescale-separated regime. In fact, in this regime, the drive induces new effective Hamiltonians whose properties depend sensitively on the specific value of $N$. 

\section{\label{sec:two-freqs}Two-frequency Floquet engineering}

\begin{figure}[t!]
  \centering
  {\includegraphics[width=0.49\textwidth]{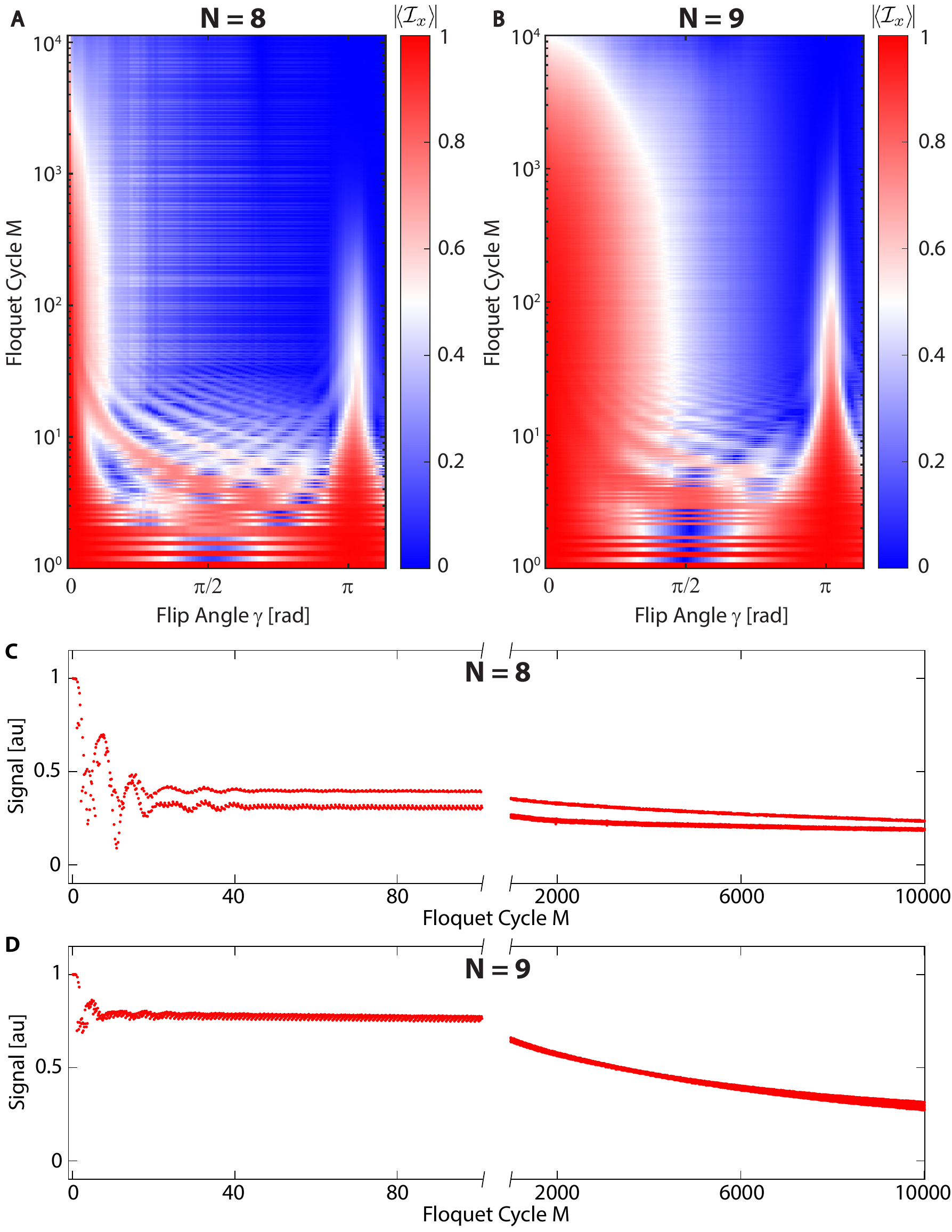}}
  \caption{\T{Experimental characterisation of the heating dynamics at small N.} 
  	Complex interplay between the fast and slow periodic drives is seen when the condition $N=T/\tau \gg 1$ is not met. (A) Movie showing time-series data from 103 experiments with different values of flip angle $\gamma$ in $[0,1.1\pi]$ and $N = 8$ $\xhat$-pulses between $\yhat$-pulses. 
	Colors represent absolute value of signal $\vert \langle \mathcal{I}_x\rangle \vert$ (see colorbar). Floquet cycle number $M$ runs vertically on a logarithmic scale. Data are taken past $10^4$ Floquet cycles. (B) Movie similar to (A) but with  $N=9$  $\xhat$-pulses between the $\yhat$-pulses. In both movies (A) and (B), $\xhat$-pulse flip angle was calibrated to $\vartheta = \pi/2$. (C-D) Line cut of (A), (B), respectively, at $\gamma=0.244\pi$. In (C), different plateaus correspond to micromotion within a Floquet cycle. The data shown in (A) and (B) is identical to the data shown in the main text \zfr{fig5}.}

	\zfl{fig5_supp}
\end{figure}

\begin{figure}[t!]
    \centering
    \includegraphics[scale=0.52]{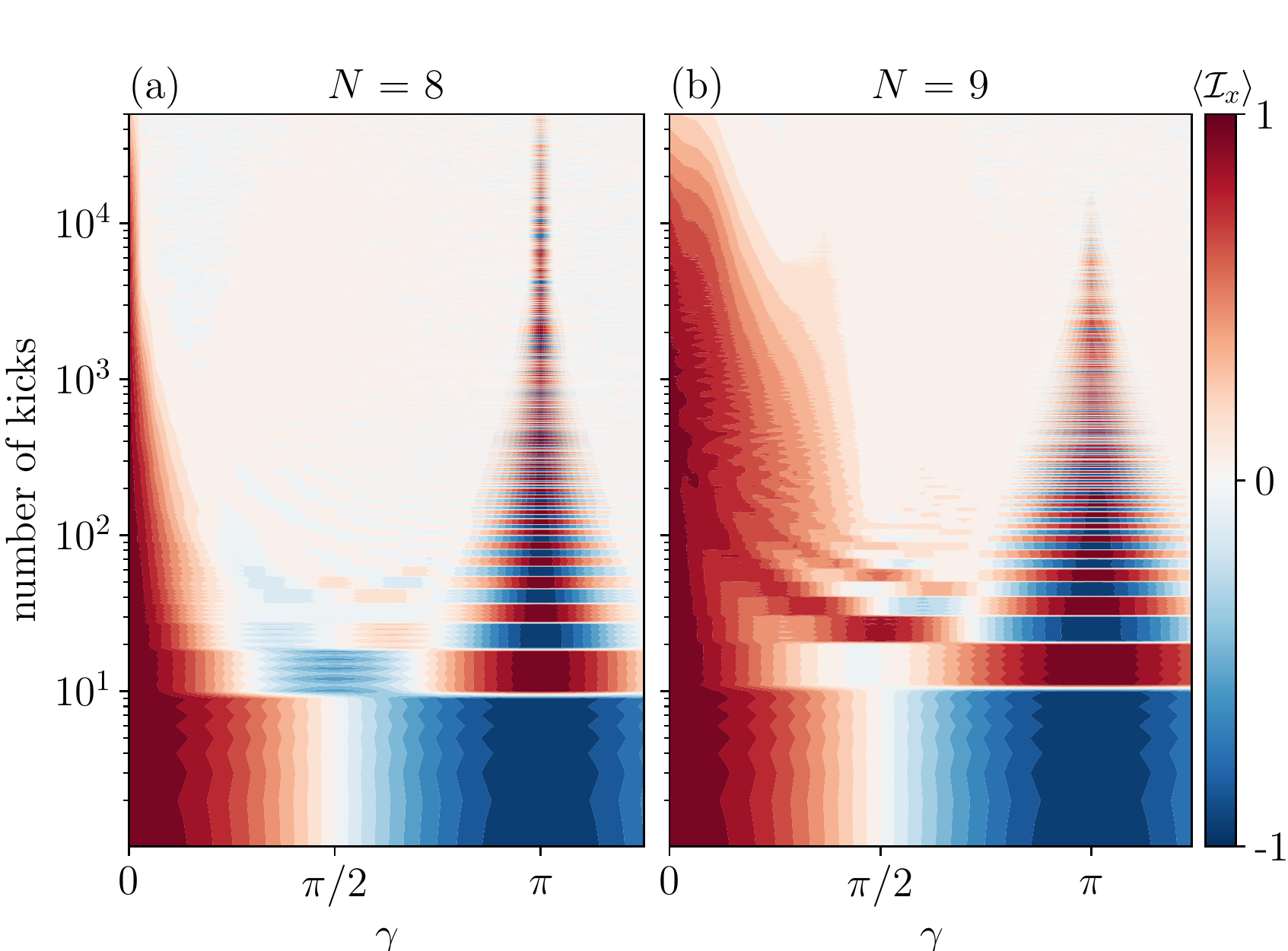}
    \caption{\textbf{Numerical simulation --} Magnetization $\langle \mathcal{I}_x \rangle$ as a function of the number of kicks and  $\gamma$ for $N=8$ (a) and $N=9$ (b). The system size is $L=16$. Remaining parameters are as in Fig.~\ref{fig:time_evolution}.}
    \label{fig:small_N}
\end{figure}

\begin{figure}[t!]
    \centering
    \includegraphics[scale=0.52]{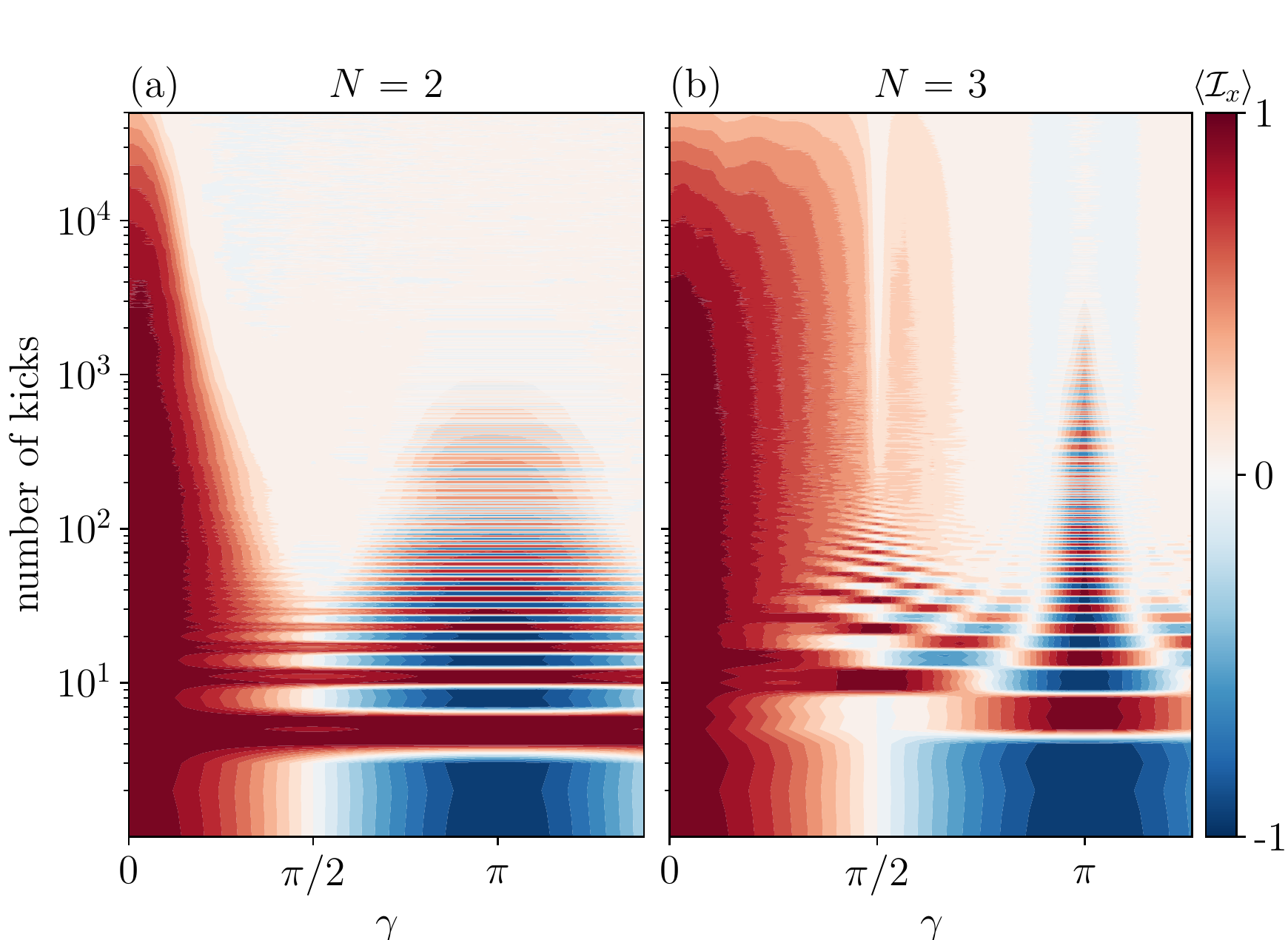}
    \caption{\textbf{Numerical simulation --} Same as Fig.~\ref{fig:small_N} for $N=2$ (a) and $N=3$ (b). The system size is $L=16$. Remaining parameters are as in Fig.~\ref{fig:time_evolution}.
    }
    \label{fig:small_N_2_3}
\end{figure}

When timescale separation between the fast and slow drive is lost, both drives mutually affect each other, and new effective Hamiltonians arise. To understand the relevant mechanisms let us investigate once more the effective Hamiltonian that captures the daynamics generated by the fast $\xhat$ drive (Eq.~\eqref{eq:toggling_H}). 

For large $N$, $\xhat$ magnetization conservation is approximately present in the effective Hamiltonian, as terms that violate it are suppressed as $1/N$. In the small $N$ regime, this is generically not true. As a consequence $\left[U_0, U_x\right]\neq 0$ (cf.~Eqs.~\eqref{eq:U0} and \eqref{eq:slow_period}) and the dynamics over $M$ Floquet cycles is governed by 
\begin{eqnarray}
\label{eq:small_N_unitary}
U_F^M \simeq \left(\mathrm e^{-iT\bar{\mathcal{H}}^{(0)}}U_x^NU_z\right)^M =  \left(\mathrm  e^{-iT\bar{\mathcal{H}}^{(0)}} U_{\vec{n}}\right)^M ,
\end{eqnarray}
where $\bar{\mathcal{H}}^{(0)}$ (given in Eq.~\eqref{eq:toggling_H}) depends on $N$.
During each driving period we now effectively rotate the system around a new axis $\vec{n}$ which is composed of an $\xhat$- and a $\zhat$-rotation: $U_{\vec{n}}= U_x^{N} U_z \equiv \mathrm e^{-i\alpha \sum_j \vec{n}\cdot\vec{I}_j}$, where 
\begin{eqnarray}
\alpha &=& \frac{c}{\sin(c)} \sqrt{1+\sin(\vartheta/2N)^2\cos(\gamma/2)^2+\sin(\gamma/2)^2},\nonumber \\
\vec{n}\! &=&\!
\left[ \!\sin(\vartheta\!/2N)\!\cos(\gamma\!/2) ,-\!\sin(\!\vartheta\!/2N)\!\sin(\!\gamma\!/2),\sin(\!\gamma\!/2)\!\cos(\!\vartheta\!/2N)\!\right]^T \nonumber \\
 & \times & 1/\sqrt{1+\sin(\vartheta/2N)^2\cos(\gamma/2)^2+\sin(\gamma/2)^2}
\end{eqnarray}
with $c= \mathrm{arccos}(\cos(\vartheta/2N)\cos(\gamma/2))$.
Note how a change in $\gamma$ now modifies both the value of $\alpha$ and the direction $\vec{n}$ of the rotation. Thus, the (single- and two-cycle) effective Hamiltonian around $\gamma =\pi $ can be significantly different from that obtained close to $\gamma=0$. Remember that especially the two-cycle effective Hamiltonian for $\gamma=0$ and $\gamma=\pi$ emerged to be identical in the time-scale separated regime leading to very similar heating characteristics around these two points of interest. To illustrate the difference and the sensitivity of the effective Hamiltonian as a function of $\gamma$ and $N$ in the non-time-scale separated regime, let us exemplary  discuss the two cases $N=8$ and $N=9$. 

For $N=8$, the total accumulated kick angle $\vartheta/2N = 2\pi$ so that $\alpha = \gamma $ and $\vec{n}=(0,0,1)^T$. Moreover, $N=8$ yields $\bar{\mathcal{H}}^{(0)}=\overline{\mathcal{H}}$.
Thus, the toggling frame expansion of Eq.~\eqref{eq:small_N_unitary} for the two-cycle effective Hamiltonian is identical (to leading order) at $\gamma=0,\pi$ eventually inducing
similar heating behaviour around these two points (see Fig.~\ref{fig:small_N} (a)).

In contrast, for $N=9$ we obtain $\alpha \rightarrow \pi/2$ and $\vec{n}\rightarrow(1,0,0)^T$ as $\gamma \rightarrow 0$, whereas for $\gamma \rightarrow \pi $ we find $\alpha \rightarrow \pi$ with $\vec{n}\rightarrow 1/\sqrt{2}(0,-1,1)^T$. A single such rotation induces the following transformation:
\begin{eqnarray}
I_{jz} \rightarrow I_{jy},~~~ 
I_{jy} \rightarrow I_{jz},~~~
I_{jx} \rightarrow -I_{jx}.
\end{eqnarray}
The toggling-frame expansion of Eq.~\eqref{eq:small_N_unitary} now yields the toggling frame Hamiltonians 
\begin{eqnarray}
\mathcal{H}_{m}  &=& \sum_{j<k}b_{jk} \bigg( \frac{3}{2}\bigg[ \mathcal{H}_{\mathrm{ff}}+(-1)^m\mathcal{G}_\mathrm{c}(N,\pi/2)\mathcal{H}_{\mathrm{dq}} \nonumber \\
&&- \tilde{\mathcal{H}}_{\mathrm{ff}} \mathcal{G}_\mathrm{s}(N,\pi/2) \bigg]- \vec{I}_j\vec{I}_k\bigg) \\
&&+ \sum_j c_j \left( \mathcal{F}_{m,+}(N,\pi/4)  I_{jz} + \mathcal{F}_{m,-}(N,\pi/4)  I_{jy} \right),\nonumber
\end{eqnarray}
where 
\begin{eqnarray}
\mathcal{F}_{m,\pm}(N,\pi/4) & = &\frac{(-1)^m}{2}\left( \mathcal{G}_{\mathrm{c}}(N,\pi/4) - \mathcal{G}_\mathrm{s}(N,\pi/4)  \right)  \nonumber \\
&&\pm \frac{1}{2} \left( \mathcal{G}_{\mathrm{c}}(N,\pi/4) +  \mathcal{G}_{\mathrm{s}}(N,\pi/4)  \right) .
\end{eqnarray}
The leading order two-cycle effective Hamiltonian is readily obtained from $\mathcal{H}_1+\mathcal{H}_2$ and yields
\begin{equation}
\label{eq:small_N_2}
 \mathcal{H}_1+\mathcal{H}_2 = \overline{\mathcal{H}}+
\frac{1}{2N}\sum_j \! c_j \left( I_{jz}\! -  I_{jy} \right). 
\end{equation}


Although $\overline{\mathcal{H}}$ preserves $\xhat$-magnetization, the single particle terms $\frac{1}{2N}\sum_j \! c_j \left( I_{jz}\! -  I_{jy} \right)$ do not. Thus, around $\gamma=\pi$, we expect to find a decreased lifetime of the corresponding DTC order for $N=9$ as compared to $N=8$. We confirm this result numerically (cf. Fig.~\ref{fig:small_N}). Contrasting Fig.~\ref{fig:small_N} (a) with (b) allows to visualize the differences in the heating dynamics induced by a minimal change in $N$ and the corresponding change of the effective Hamiltonian not only around $\gamma=\pi$ but in particular also around $\gamma=0$. In comparing Fig.~\ref{fig:small_N} to the corresponding experimental results in \zfr{fig5_supp} and \zfr{fig5} of the main text, we find a qualitative agreement also for small values of $N$. The reduced lifetime of the DTC order in the experimental data for $N=8$ as compared to the simulations might be attributed to drive imperfections: due to the small region of rigidity, already small imperfections might induce a decay of the DTC order.

Note that for our choice of $\vartheta=\pi/2$ we can engineer $8$ different families of effective Hamiltonians characterised by the $8$ different rotations $U_x^NU_z$. Other choices of $\vartheta$ can yield even more different families. Interestingly, the different effective Hamiltonians show quite different behaviour with respect to their prethermal properties as well as the formation of DTC order, which not only concerns the lifetime but also the rigidity away from $\gamma = \pi$. Most prominently, this can be observed for $N=2$ (cf. Fig.~\ref{fig:small_N_2_3} (a)) where the region of rigidity is significantly increased as compared to (for instance) $N=3$ (Fig.~\ref{fig:small_N_2_3} (b)). 

In conclusion, our experimental and numerical studies show that control over $N$ and $\gamma$ is sufficient to case-specifically design effective Hamiltonians, heating dynamics and non-equilibrium order.

\begin{figure}[t!]
    \centering
    \includegraphics[scale=0.6]{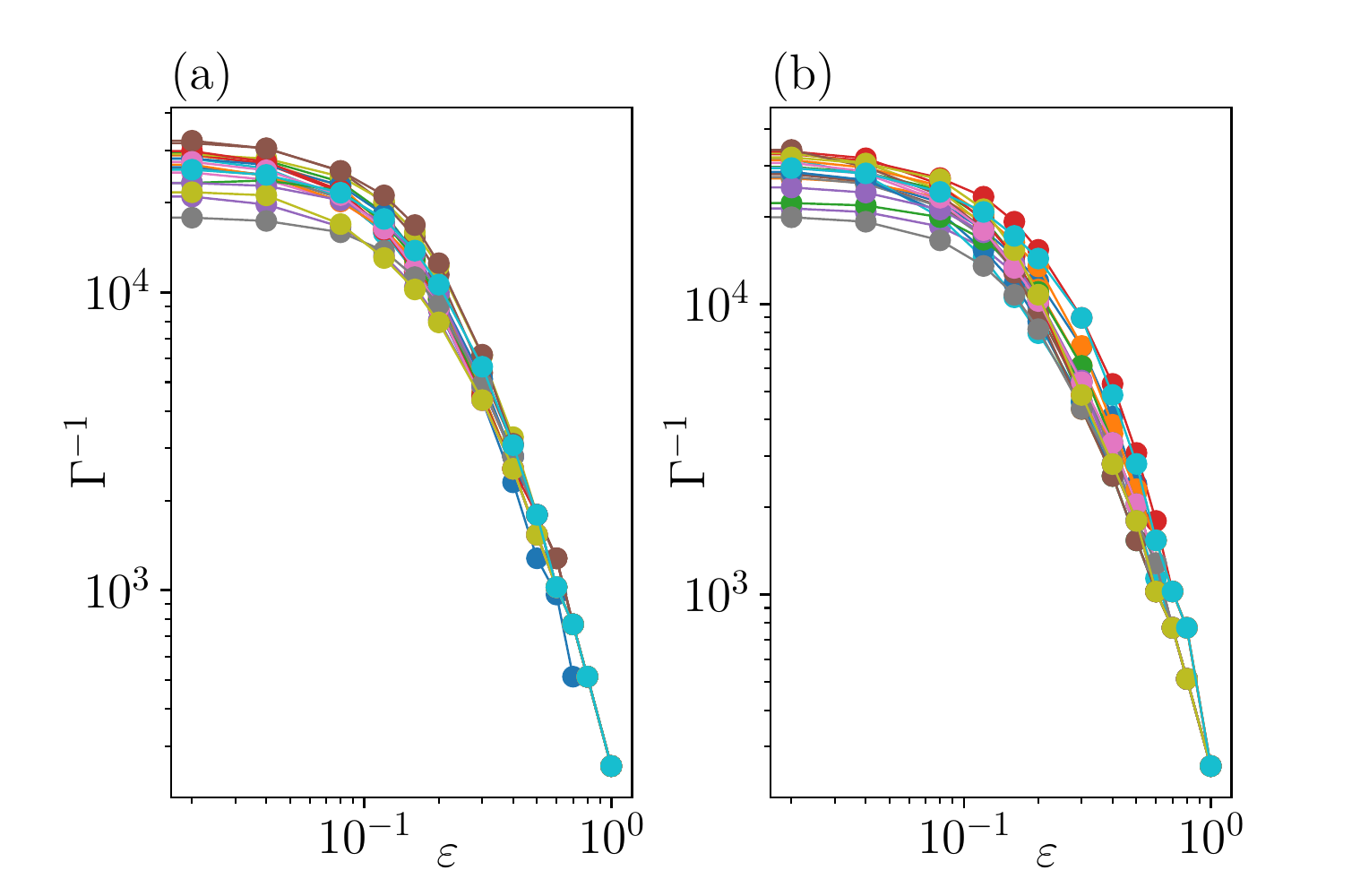}
     \caption{\textbf{Numerical simulation --} Random graph dependence of the heating time $\Gamma^{-1}$ as a function of $\varepsilon$ for $\gamma=\pi+\varepsilon$ in (a) and $\gamma = 0+\varepsilon$ in (b) for $N=255$. We show $20$ different realizations. Different colors represent different random graphs, each with $L=16$, generated using $r_{\mathrm{min}}=0.7$ and $r_{\mathrm{min}}=0.8$. For each graph we independently determine $J$ and set $\tau J=0.2$. Solid lines serve as guide to the eye.}
    \label{fig:seed_dep}
\end{figure}

\section{ Random graph independence}
\label{sec:random}

To ensure that our results are not affected by the choice of the random graph realization, we show
in Fig.~\ref{fig:seed_dep} the heating times for $20$ different random graph realizations around the two special points of interest $\gamma\approx 0$ and $\gamma \approx \pi$. In both regimes, the heating rates show no significant dependence on the specific random graph realization. Instead the long-range interaction terms appear to have a self-averaging effect already at moderate system sizes. This is additionally enhanced by the ergodicity of the dynamics, promoted by adding small random noise to $\tau$.

\end{document}

%% file: Commands3.tex
\newcommand{\id}{\mathds{1}}

\newcommand{\xd}{\delta}
\newcommand{\xe}{\epsilon}
\newcommand{\vxe}{\varepsilon}
\newcommand{\tm}{{\text -}}

\newcommand{\tacq}{t_{\R{acq}}}
\newcommand{\xg}{\gamma}
\newcommand{\xt}{\vartheta}

\newcommand{\xl}{\lambda}
\newcommand{\xr}{\rho}
\newcommand{\xo}{\omega}
\newcommand{\xph}{\phi}

\newcommand{\app}{\approx}

\newcommand{\tpol}{t_{\R{pol}}}

\newcommand{\Cs}{{}^{13}\R{C}}

\newcommand{\mI}[0]{\mathcal{I}}

\newcommand{\mHdd}[0]{\mH_{\R{dd}}}

\newcommand{\mHeff}[0]{\mH_{\R{eff}}}

\newcommand{\xy}[0]{\xhat\tm\yhat}

\newcommand{\xG}{\Gamma}

\newcommand{\fr}[2]{\frac{#1}{#2}}
\newcommand{\ov}[1]{\overline{#1}}

\newcommand{\mH}[0]{\mathcal{H}}

\newcommand{\rt}{\rightarrow}

\newcommand{\beq}{\begin{equation}}
\newcommand{\eeq}{\end{equation}}
                  
\newcommand{\benum}{\begin{enumerate}}
\newcommand{\eenum}{\end{enumerate}}
                    
\newcommand{\bit}{\begin{itemize}}
\newcommand{\eit}{\end{itemize}}
\newcommand{\xhat}{\hat{\T{x}}}
\newcommand{\yhat}{\hat{\T{y}}}
\newcommand{\zhat}{\hat{\T{z}}}

\newcommand{\bea}{\begin{eqnarray}}
\newcommand{\eea}{\end{eqnarray}}

\newcommand{\bev}{\begin{verbatim}}
\newcommand{\eev}{\end{verbatim}}

\newcommand{\zt}{\times}

\newcommand{\qt}{\tau}

\newcommand{\T}[1]{\textbf{#1}}
\newcommand{\I}[1]{\textit{#1}}
\newcommand{\R}[1]{\textrm{#1}}

\newcommand{\zfl}[1]{\protect\label{fig:#1}}
\newcommand{\zfr}[1]{\figurename\,\ref{fig:#1}}

\newcommand{\expec}[1]{\left\langle #1\right\rangle}

\newcommand{\ba}{\left\{ \begin{array}{lr}}
\newcommand{\ea}{\end{array}\right.}

\newcommand{\blist}[1]{
 \begin{list}{#1}
 \begin{align}
	 arrow
 \end{align}
 $\checkmark\star
  { \setlength{\itemsep}{3pt}
     \setlength{\parsep}{2pt}
     \setlength{\topsep}{3pt}
     \setlength{\partopsep}{0pt}
     \setlength{\leftmargin}{1em}
     \setlength{\labelwidth}{1em}
     \setlength{\labelsep}{0.5em} } }
\newcommand{\elist}{
  \end{list}  }

\DeclareMathSymbol{\vartheta}{\mathalpha}{letters}{"12}
\DeclareMathSymbol{\theta}{\mathalpha}{letters}{"23}
\DeclareMathSymbol{\phi}{\mathalpha}{letters}{"27}
\DeclareMathSymbol{\varphi}{\mathalpha}{letters}{"1E}

\newcommand{\bef}
{
\begin{figure}[htbp]
\centering
}

\newcommand{\eef}{\end{figure}}


%% file: main.bbl
\begin{thebibliography}{58}%
\makeatletter
\providecommand \@ifxundefined [1]{%
 \@ifx{#1\undefined}
}%
\providecommand \@ifnum [1]{%
 \ifnum #1\expandafter \@firstoftwo
 \else \expandafter \@secondoftwo
 \fi
}%
\providecommand \@ifx [1]{%
 \ifx #1\expandafter \@firstoftwo
 \else \expandafter \@secondoftwo
 \fi
}%
\providecommand \natexlab [1]{#1}%
\providecommand \enquote  [1]{``#1''}%
\providecommand \bibnamefont  [1]{#1}%
\providecommand \bibfnamefont [1]{#1}%
\providecommand \citenamefont [1]{#1}%
\providecommand \href@noop [0]{\@secondoftwo}%
\providecommand \href [0]{\begingroup \@sanitize@url \@href}%
\providecommand \@href[1]{\@@startlink{#1}\@@href}%
\providecommand \@@href[1]{\endgroup#1\@@endlink}%
\providecommand \@sanitize@url [0]{\catcode `\\12\catcode `\$12\catcode
  `\&12\catcode `\#12\catcode `\^12\catcode `\_12\catcode `\%12\relax}%
\providecommand \@@startlink[1]{}%
\providecommand \@@endlink[0]{}%
\providecommand \url  [0]{\begingroup\@sanitize@url \@url }%
\providecommand \@url [1]{\endgroup\@href {#1}{\urlprefix }}%
\providecommand \urlprefix  [0]{URL }%
\providecommand \Eprint [0]{\href }%
\providecommand \doibase [0]{https://doi.org/}%
\providecommand \selectlanguage [0]{\@gobble}%
\providecommand \bibinfo  [0]{\@secondoftwo}%
\providecommand \bibfield  [0]{\@secondoftwo}%
\providecommand \translation [1]{[#1]}%
\providecommand \BibitemOpen [0]{}%
\providecommand \bibitemStop [0]{}%
\providecommand \bibitemNoStop [0]{.\EOS\space}%
\providecommand \EOS [0]{\spacefactor3000\relax}%
\providecommand \BibitemShut  [1]{\csname bibitem#1\endcsname}%
\let\auto@bib@innerbib\@empty
\bibitem [{\citenamefont {Sacha}\ and\ \citenamefont
  {Zakrzewski}(2017)}]{sacha2017time}%
  \BibitemOpen
  \bibfield  {author} {\bibinfo {author} {\bibfnamefont {K.}~\bibnamefont
  {Sacha}}\ and\ \bibinfo {author} {\bibfnamefont {J.}~\bibnamefont
  {Zakrzewski}},\ }\bibfield  {title} {\bibinfo {title} {Time crystals: a
  review},\ }\href
  {https://iopscience.iop.org/article/10.1088/1361-6633/aa8b38} {\bibfield
  {journal} {\bibinfo  {journal} {Reports on Progress in Physics}\ }\textbf
  {\bibinfo {volume} {81}},\ \bibinfo {pages} {016401} (\bibinfo {year}
  {2017})}\BibitemShut {NoStop}%
\bibitem [{\citenamefont {Khemani}\ \emph {et~al.}(2019)\citenamefont
  {Khemani}, \citenamefont {Moessner},\ and\ \citenamefont
  {Sondhi}}]{khemani2019brief}%
  \BibitemOpen
  \bibfield  {author} {\bibinfo {author} {\bibfnamefont {V.}~\bibnamefont
  {Khemani}}, \bibinfo {author} {\bibfnamefont {R.}~\bibnamefont {Moessner}},\
  and\ \bibinfo {author} {\bibfnamefont {S.}~\bibnamefont {Sondhi}},\
  }\bibfield  {title} {\bibinfo {title} {A brief history of time crystals},\
  }\href {https://arxiv.org/abs/1910.10745} {\bibfield  {journal} {\bibinfo
  {journal} {arXiv preprint arXiv:1910.10745}\ } (\bibinfo {year}
  {2019})}\BibitemShut {NoStop}%
\bibitem [{\citenamefont {Else}\ \emph
  {et~al.}(2020{\natexlab{a}})\citenamefont {Else}, \citenamefont {Monroe},
  \citenamefont {Nayak},\ and\ \citenamefont {Yao}}]{Else2020}%
  \BibitemOpen
  \bibfield  {author} {\bibinfo {author} {\bibfnamefont {D.~V.}\ \bibnamefont
  {Else}}, \bibinfo {author} {\bibfnamefont {C.}~\bibnamefont {Monroe}},
  \bibinfo {author} {\bibfnamefont {C.}~\bibnamefont {Nayak}},\ and\ \bibinfo
  {author} {\bibfnamefont {N.~Y.}\ \bibnamefont {Yao}},\ }\bibfield  {title}
  {\bibinfo {title} {Discrete time crystals},\ }\href
  {https://doi.org/https://doi.org/10.1146/annurev-conmatphys-031119-050658}
  {\bibfield  {journal} {\bibinfo  {journal} {Annual Review of Condensed Matter
  Physics}\ }\textbf {\bibinfo {volume} {11}},\ \bibinfo {pages} {467}
  (\bibinfo {year} {2020}{\natexlab{a}})}\BibitemShut {NoStop}%
\bibitem [{\citenamefont {Sacha}(2015)}]{Sacha2015}%
  \BibitemOpen
  \bibfield  {author} {\bibinfo {author} {\bibfnamefont {K.}~\bibnamefont
  {Sacha}},\ }\bibfield  {title} {\bibinfo {title} {Modeling spontaneous
  breaking of time-translation symmetry},\ }\href
  {https://doi.org/10.1103/PhysRevA.91.033617} {\bibfield  {journal} {\bibinfo
  {journal} {Phys. Rev. A}\ }\textbf {\bibinfo {volume} {91}},\ \bibinfo
  {pages} {033617} (\bibinfo {year} {2015})}\BibitemShut {NoStop}%
\bibitem [{\citenamefont {Khemani}\ \emph {et~al.}(2016)\citenamefont
  {Khemani}, \citenamefont {Lazarides}, \citenamefont {Moessner},\ and\
  \citenamefont {Sondhi}}]{khemani2016phase}%
  \BibitemOpen
  \bibfield  {author} {\bibinfo {author} {\bibfnamefont {V.}~\bibnamefont
  {Khemani}}, \bibinfo {author} {\bibfnamefont {A.}~\bibnamefont {Lazarides}},
  \bibinfo {author} {\bibfnamefont {R.}~\bibnamefont {Moessner}},\ and\
  \bibinfo {author} {\bibfnamefont {S.~L.}\ \bibnamefont {Sondhi}},\ }\bibfield
   {title} {\bibinfo {title} {Phase structure of driven quantum systems},\
  }\href {https://doi.org/10.1103/PhysRevLett.116.250401} {\bibfield  {journal}
  {\bibinfo  {journal} {Phys. Rev. Lett.}\ }\textbf {\bibinfo {volume} {116}},\
  \bibinfo {pages} {250401} (\bibinfo {year} {2016})}\BibitemShut {NoStop}%
\bibitem [{\citenamefont {Else}\ \emph {et~al.}(2016)\citenamefont {Else},
  \citenamefont {Bauer},\ and\ \citenamefont {Nayak}}]{else2016floquet}%
  \BibitemOpen
  \bibfield  {author} {\bibinfo {author} {\bibfnamefont {D.~V.}\ \bibnamefont
  {Else}}, \bibinfo {author} {\bibfnamefont {B.}~\bibnamefont {Bauer}},\ and\
  \bibinfo {author} {\bibfnamefont {C.}~\bibnamefont {Nayak}},\ }\bibfield
  {title} {\bibinfo {title} {Floquet time crystals},\ }\href
  {https://doi.org/10.1103/PhysRevLett.117.090402} {\bibfield  {journal}
  {\bibinfo  {journal} {Phys. Rev. Lett.}\ }\textbf {\bibinfo {volume} {117}},\
  \bibinfo {pages} {090402} (\bibinfo {year} {2016})}\BibitemShut {NoStop}%
\bibitem [{\citenamefont {von Keyserlingk}\ \emph {et~al.}(2016)\citenamefont
  {von Keyserlingk}, \citenamefont {Khemani},\ and\ \citenamefont
  {Sondhi}}]{keyserlingk2016absolute}%
  \BibitemOpen
  \bibfield  {author} {\bibinfo {author} {\bibfnamefont {C.~W.}\ \bibnamefont
  {von Keyserlingk}}, \bibinfo {author} {\bibfnamefont {V.}~\bibnamefont
  {Khemani}},\ and\ \bibinfo {author} {\bibfnamefont {S.~L.}\ \bibnamefont
  {Sondhi}},\ }\bibfield  {title} {\bibinfo {title} {Absolute stability and
  spatiotemporal long-range order in floquet systems},\ }\href
  {https://doi.org/10.1103/PhysRevB.94.085112} {\bibfield  {journal} {\bibinfo
  {journal} {Phys. Rev. B}\ }\textbf {\bibinfo {volume} {94}},\ \bibinfo
  {pages} {085112} (\bibinfo {year} {2016})}\BibitemShut {NoStop}%
\bibitem [{\citenamefont {Yao}\ \emph {et~al.}(2017)\citenamefont {Yao},
  \citenamefont {Potter}, \citenamefont {Potirniche},\ and\ \citenamefont
  {Vishwanath}}]{yao2017discrete}%
  \BibitemOpen
  \bibfield  {author} {\bibinfo {author} {\bibfnamefont {N.~Y.}\ \bibnamefont
  {Yao}}, \bibinfo {author} {\bibfnamefont {A.~C.}\ \bibnamefont {Potter}},
  \bibinfo {author} {\bibfnamefont {I.-D.}\ \bibnamefont {Potirniche}},\ and\
  \bibinfo {author} {\bibfnamefont {A.}~\bibnamefont {Vishwanath}},\ }\bibfield
   {title} {\bibinfo {title} {Discrete time crystals: Rigidity, criticality,
  and realizations},\ }\href {https://doi.org/10.1103/PhysRevLett.118.030401}
  {\bibfield  {journal} {\bibinfo  {journal} {Phys. Rev. Lett.}\ }\textbf
  {\bibinfo {volume} {118}},\ \bibinfo {pages} {030401} (\bibinfo {year}
  {2017})}\BibitemShut {NoStop}%
\bibitem [{\citenamefont {Ho}\ \emph {et~al.}(2017)\citenamefont {Ho},
  \citenamefont {Choi}, \citenamefont {Lukin},\ and\ \citenamefont
  {Abanin}}]{ho2017critical}%
  \BibitemOpen
  \bibfield  {author} {\bibinfo {author} {\bibfnamefont {W.~W.}\ \bibnamefont
  {Ho}}, \bibinfo {author} {\bibfnamefont {S.}~\bibnamefont {Choi}}, \bibinfo
  {author} {\bibfnamefont {M.~D.}\ \bibnamefont {Lukin}},\ and\ \bibinfo
  {author} {\bibfnamefont {D.~A.}\ \bibnamefont {Abanin}},\ }\bibfield  {title}
  {\bibinfo {title} {Critical time crystals in dipolar systems},\ }\href
  {https://doi.org/10.1103/PhysRevLett.119.010602} {\bibfield  {journal}
  {\bibinfo  {journal} {Phys. Rev. Lett.}\ }\textbf {\bibinfo {volume} {119}},\
  \bibinfo {pages} {010602} (\bibinfo {year} {2017})}\BibitemShut {NoStop}%
\bibitem [{\citenamefont {Liao}\ \emph {et~al.}(2019)\citenamefont {Liao},
  \citenamefont {Smits}, \citenamefont {van~der Straten},\ and\ \citenamefont
  {Stoof}}]{Liao2019}%
  \BibitemOpen
  \bibfield  {author} {\bibinfo {author} {\bibfnamefont {L.}~\bibnamefont
  {Liao}}, \bibinfo {author} {\bibfnamefont {J.}~\bibnamefont {Smits}},
  \bibinfo {author} {\bibfnamefont {P.}~\bibnamefont {van~der Straten}},\ and\
  \bibinfo {author} {\bibfnamefont {H.~T.~C.}\ \bibnamefont {Stoof}},\
  }\bibfield  {title} {\bibinfo {title} {Dynamics of a space-time crystal in an
  atomic bose-einstein condensate},\ }\href
  {https://doi.org/10.1103/PhysRevA.99.013625} {\bibfield  {journal} {\bibinfo
  {journal} {Phys. Rev. A}\ }\textbf {\bibinfo {volume} {99}},\ \bibinfo
  {pages} {013625} (\bibinfo {year} {2019})}\BibitemShut {NoStop}%
\bibitem [{\citenamefont {Choi}\ \emph {et~al.}(2017)\citenamefont {Choi},
  \citenamefont {Choi}, \citenamefont {Landig}, \citenamefont {Kucsko},
  \citenamefont {Zhou}, \citenamefont {Isoya}, \citenamefont {Jelezko},
  \citenamefont {Onoda}, \citenamefont {Sumiya}, \citenamefont {Khemani} \emph
  {et~al.}}]{Choi17}%
  \BibitemOpen
  \bibfield  {author} {\bibinfo {author} {\bibfnamefont {S.}~\bibnamefont
  {Choi}}, \bibinfo {author} {\bibfnamefont {J.}~\bibnamefont {Choi}}, \bibinfo
  {author} {\bibfnamefont {R.}~\bibnamefont {Landig}}, \bibinfo {author}
  {\bibfnamefont {G.}~\bibnamefont {Kucsko}}, \bibinfo {author} {\bibfnamefont
  {H.}~\bibnamefont {Zhou}}, \bibinfo {author} {\bibfnamefont {J.}~\bibnamefont
  {Isoya}}, \bibinfo {author} {\bibfnamefont {F.}~\bibnamefont {Jelezko}},
  \bibinfo {author} {\bibfnamefont {S.}~\bibnamefont {Onoda}}, \bibinfo
  {author} {\bibfnamefont {H.}~\bibnamefont {Sumiya}}, \bibinfo {author}
  {\bibfnamefont {V.}~\bibnamefont {Khemani}}, \emph {et~al.},\ }\bibfield
  {title} {\bibinfo {title} {Observation of discrete time-crystalline order in
  a disordered dipolar many-body system},\ }\href
  {https://doi.org/https://doi.org/10.1038/nature21426} {\bibfield  {journal}
  {\bibinfo  {journal} {Nature}\ }\textbf {\bibinfo {volume} {543}},\ \bibinfo
  {pages} {221} (\bibinfo {year} {2017})}\BibitemShut {NoStop}%
\bibitem [{\citenamefont {Zhang}\ \emph {et~al.}(2017)\citenamefont {Zhang},
  \citenamefont {Hess}, \citenamefont {Kyprianidis}, \citenamefont {Becker},
  \citenamefont {Lee}, \citenamefont {Smith}, \citenamefont {Pagano},
  \citenamefont {Potirniche}, \citenamefont {Potter}, \citenamefont
  {Vishwanath} \emph {et~al.}}]{Zhang17}%
  \BibitemOpen
  \bibfield  {author} {\bibinfo {author} {\bibfnamefont {J.}~\bibnamefont
  {Zhang}}, \bibinfo {author} {\bibfnamefont {P.}~\bibnamefont {Hess}},
  \bibinfo {author} {\bibfnamefont {A.}~\bibnamefont {Kyprianidis}}, \bibinfo
  {author} {\bibfnamefont {P.}~\bibnamefont {Becker}}, \bibinfo {author}
  {\bibfnamefont {A.}~\bibnamefont {Lee}}, \bibinfo {author} {\bibfnamefont
  {J.}~\bibnamefont {Smith}}, \bibinfo {author} {\bibfnamefont
  {G.}~\bibnamefont {Pagano}}, \bibinfo {author} {\bibfnamefont {I.-D.}\
  \bibnamefont {Potirniche}}, \bibinfo {author} {\bibfnamefont {A.~C.}\
  \bibnamefont {Potter}}, \bibinfo {author} {\bibfnamefont {A.}~\bibnamefont
  {Vishwanath}}, \emph {et~al.},\ }\bibfield  {title} {\bibinfo {title}
  {Observation of a discrete time crystal},\ }\href
  {https://doi.org/https://doi.org/10.1038/nature21413} {\bibfield  {journal}
  {\bibinfo  {journal} {Nature}\ }\textbf {\bibinfo {volume} {543}},\ \bibinfo
  {pages} {217} (\bibinfo {year} {2017})}\BibitemShut {NoStop}%
\bibitem [{\citenamefont {Pal}\ \emph {et~al.}(2018)\citenamefont {Pal},
  \citenamefont {Nishad}, \citenamefont {Mahesh},\ and\ \citenamefont
  {Sreejith}}]{pal2018temporal}%
  \BibitemOpen
  \bibfield  {author} {\bibinfo {author} {\bibfnamefont {S.}~\bibnamefont
  {Pal}}, \bibinfo {author} {\bibfnamefont {N.}~\bibnamefont {Nishad}},
  \bibinfo {author} {\bibfnamefont {T.~S.}\ \bibnamefont {Mahesh}},\ and\
  \bibinfo {author} {\bibfnamefont {G.~J.}\ \bibnamefont {Sreejith}},\
  }\bibfield  {title} {\bibinfo {title} {Temporal order in periodically driven
  spins in star-shaped clusters},\ }\href
  {https://doi.org/10.1103/PhysRevLett.120.180602} {\bibfield  {journal}
  {\bibinfo  {journal} {Phys. Rev. Lett.}\ }\textbf {\bibinfo {volume} {120}},\
  \bibinfo {pages} {180602} (\bibinfo {year} {2018})}\BibitemShut {NoStop}%
\bibitem [{\citenamefont {Smits}\ \emph {et~al.}(2018)\citenamefont {Smits},
  \citenamefont {Liao}, \citenamefont {Stoof},\ and\ \citenamefont {van~der
  Straten}}]{Smits2018}%
  \BibitemOpen
  \bibfield  {author} {\bibinfo {author} {\bibfnamefont {J.}~\bibnamefont
  {Smits}}, \bibinfo {author} {\bibfnamefont {L.}~\bibnamefont {Liao}},
  \bibinfo {author} {\bibfnamefont {H.~T.~C.}\ \bibnamefont {Stoof}},\ and\
  \bibinfo {author} {\bibfnamefont {P.}~\bibnamefont {van~der Straten}},\
  }\bibfield  {title} {\bibinfo {title} {Observation of a space-time crystal in
  a superfluid quantum gas},\ }\href
  {https://doi.org/10.1103/PhysRevLett.121.185301} {\bibfield  {journal}
  {\bibinfo  {journal} {Phys. Rev. Lett.}\ }\textbf {\bibinfo {volume} {121}},\
  \bibinfo {pages} {185301} (\bibinfo {year} {2018})}\BibitemShut {NoStop}%
\bibitem [{\citenamefont {Rovny}\ \emph
  {et~al.}(2018{\natexlab{a}})\citenamefont {Rovny}, \citenamefont {Blum},\
  and\ \citenamefont {Barrett}}]{Rovny2018}%
  \BibitemOpen
  \bibfield  {author} {\bibinfo {author} {\bibfnamefont {J.}~\bibnamefont
  {Rovny}}, \bibinfo {author} {\bibfnamefont {R.~L.}\ \bibnamefont {Blum}},\
  and\ \bibinfo {author} {\bibfnamefont {S.~E.}\ \bibnamefont {Barrett}},\
  }\bibfield  {title} {\bibinfo {title} {$^{31}\mathrm{P}$ nmr study of
  discrete time-crystalline signatures in an ordered crystal of ammonium
  dihydrogen phosphate},\ }\href {https://doi.org/10.1103/PhysRevB.97.184301}
  {\bibfield  {journal} {\bibinfo  {journal} {Phys. Rev. B}\ }\textbf {\bibinfo
  {volume} {97}},\ \bibinfo {pages} {184301} (\bibinfo {year}
  {2018}{\natexlab{a}})}\BibitemShut {NoStop}%
\bibitem [{\citenamefont {Randall}\ \emph {et~al.}()\citenamefont {Randall},
  \citenamefont {Bradley}, \citenamefont {van~der Gronden}, \citenamefont
  {Galicia}, \citenamefont {Abobeih}, \citenamefont {Markham}, \citenamefont
  {Twitchen}, \citenamefont {Machado}, \citenamefont {Yao},\ and\ \citenamefont
  {Taminiau}}]{Randall21}%
  \BibitemOpen
  \bibfield  {author} {\bibinfo {author} {\bibfnamefont {J.}~\bibnamefont
  {Randall}}, \bibinfo {author} {\bibfnamefont {C.}~\bibnamefont {Bradley}},
  \bibinfo {author} {\bibfnamefont {F.}~\bibnamefont {van~der Gronden}},
  \bibinfo {author} {\bibfnamefont {A.}~\bibnamefont {Galicia}}, \bibinfo
  {author} {\bibfnamefont {M.}~\bibnamefont {Abobeih}}, \bibinfo {author}
  {\bibfnamefont {M.}~\bibnamefont {Markham}}, \bibinfo {author} {\bibfnamefont
  {D.}~\bibnamefont {Twitchen}}, \bibinfo {author} {\bibfnamefont
  {F.}~\bibnamefont {Machado}}, \bibinfo {author} {\bibfnamefont
  {N.}~\bibnamefont {Yao}},\ and\ \bibinfo {author} {\bibfnamefont
  {T.}~\bibnamefont {Taminiau}},\ }\bibfield  {title} {\bibinfo {title}
  {Many-body-localized discrete time crystal with a programmable spin-based
  quantum simulator},\ }\href {https://doi.org/10.1126/science.abk0603}
  {\bibfield  {journal} {\bibinfo  {journal} {Science}\ }\textbf {\bibinfo
  {volume} {374}},\ \bibinfo {pages} {1474}}\BibitemShut {NoStop}%
\bibitem [{\citenamefont {Mi}\ \emph {et~al.}(2021)\citenamefont {Mi},
  \citenamefont {Ippoliti}, \citenamefont {Quintana}, \citenamefont {Greene},
  \citenamefont {Chen}, \citenamefont {Gross}, \citenamefont {Arute},
  \citenamefont {Arya}, \citenamefont {Atalaya}, \citenamefont {Babbush} \emph
  {et~al.}}]{Mi21}%
  \BibitemOpen
  \bibfield  {author} {\bibinfo {author} {\bibfnamefont {X.}~\bibnamefont
  {Mi}}, \bibinfo {author} {\bibfnamefont {M.}~\bibnamefont {Ippoliti}},
  \bibinfo {author} {\bibfnamefont {C.}~\bibnamefont {Quintana}}, \bibinfo
  {author} {\bibfnamefont {A.}~\bibnamefont {Greene}}, \bibinfo {author}
  {\bibfnamefont {Z.}~\bibnamefont {Chen}}, \bibinfo {author} {\bibfnamefont
  {J.}~\bibnamefont {Gross}}, \bibinfo {author} {\bibfnamefont
  {F.}~\bibnamefont {Arute}}, \bibinfo {author} {\bibfnamefont
  {K.}~\bibnamefont {Arya}}, \bibinfo {author} {\bibfnamefont {J.}~\bibnamefont
  {Atalaya}}, \bibinfo {author} {\bibfnamefont {R.}~\bibnamefont {Babbush}},
  \emph {et~al.},\ }\bibfield  {title} {\bibinfo {title} {Time-crystalline
  eigenstate order on a quantum processor},\ }\href
  {https://doi.org/https://doi.org/10.1038/s41586-021-04257-w} {\bibfield
  {journal} {\bibinfo  {journal} {Nature}\ ,\ \bibinfo {pages} {1}} (\bibinfo
  {year} {2021})}\BibitemShut {NoStop}%
\bibitem [{\citenamefont {Singh}\ \emph {et~al.}(2019)\citenamefont {Singh},
  \citenamefont {Fujiwara}, \citenamefont {Geiger}, \citenamefont {Simmons},
  \citenamefont {Lipatov}, \citenamefont {Cao}, \citenamefont {Dotti},
  \citenamefont {Rajagopal}, \citenamefont {Senaratne}, \citenamefont
  {Shimasaki}, \citenamefont {Heyl}, \citenamefont {Eckardt},\ and\
  \citenamefont {Weld}}]{singh2019quantifying}%
  \BibitemOpen
  \bibfield  {author} {\bibinfo {author} {\bibfnamefont {K.}~\bibnamefont
  {Singh}}, \bibinfo {author} {\bibfnamefont {C.~J.}\ \bibnamefont {Fujiwara}},
  \bibinfo {author} {\bibfnamefont {Z.~A.}\ \bibnamefont {Geiger}}, \bibinfo
  {author} {\bibfnamefont {E.~Q.}\ \bibnamefont {Simmons}}, \bibinfo {author}
  {\bibfnamefont {M.}~\bibnamefont {Lipatov}}, \bibinfo {author} {\bibfnamefont
  {A.}~\bibnamefont {Cao}}, \bibinfo {author} {\bibfnamefont {P.}~\bibnamefont
  {Dotti}}, \bibinfo {author} {\bibfnamefont {S.~V.}\ \bibnamefont
  {Rajagopal}}, \bibinfo {author} {\bibfnamefont {R.}~\bibnamefont
  {Senaratne}}, \bibinfo {author} {\bibfnamefont {T.}~\bibnamefont
  {Shimasaki}}, \bibinfo {author} {\bibfnamefont {M.}~\bibnamefont {Heyl}},
  \bibinfo {author} {\bibfnamefont {A.}~\bibnamefont {Eckardt}},\ and\ \bibinfo
  {author} {\bibfnamefont {D.~M.}\ \bibnamefont {Weld}},\ }\bibfield  {title}
  {\bibinfo {title} {Quantifying and controlling prethermal nonergodicity in
  interacting floquet matter},\ }\href
  {https://doi.org/10.1103/PhysRevX.9.041021} {\bibfield  {journal} {\bibinfo
  {journal} {Phys. Rev. X}\ }\textbf {\bibinfo {volume} {9}},\ \bibinfo {pages}
  {041021} (\bibinfo {year} {2019})}\BibitemShut {NoStop}%
\bibitem [{\citenamefont {Rubio-Abadal}\ \emph {et~al.}(2020)\citenamefont
  {Rubio-Abadal}, \citenamefont {Ippoliti}, \citenamefont {Hollerith},
  \citenamefont {Wei}, \citenamefont {Rui}, \citenamefont {Sondhi},
  \citenamefont {Khemani}, \citenamefont {Gross},\ and\ \citenamefont
  {Bloch}}]{abadal2020floquet}%
  \BibitemOpen
  \bibfield  {author} {\bibinfo {author} {\bibfnamefont {A.}~\bibnamefont
  {Rubio-Abadal}}, \bibinfo {author} {\bibfnamefont {M.}~\bibnamefont
  {Ippoliti}}, \bibinfo {author} {\bibfnamefont {S.}~\bibnamefont {Hollerith}},
  \bibinfo {author} {\bibfnamefont {D.}~\bibnamefont {Wei}}, \bibinfo {author}
  {\bibfnamefont {J.}~\bibnamefont {Rui}}, \bibinfo {author} {\bibfnamefont
  {S.~L.}\ \bibnamefont {Sondhi}}, \bibinfo {author} {\bibfnamefont
  {V.}~\bibnamefont {Khemani}}, \bibinfo {author} {\bibfnamefont
  {C.}~\bibnamefont {Gross}},\ and\ \bibinfo {author} {\bibfnamefont
  {I.}~\bibnamefont {Bloch}},\ }\bibfield  {title} {\bibinfo {title} {Floquet
  prethermalization in a bose-hubbard system},\ }\href
  {https://doi.org/10.1103/PhysRevX.10.021044} {\bibfield  {journal} {\bibinfo
  {journal} {Phys. Rev. X}\ }\textbf {\bibinfo {volume} {10}},\ \bibinfo
  {pages} {021044} (\bibinfo {year} {2020})}\BibitemShut {NoStop}%
\bibitem [{\citenamefont {Peng}\ \emph {et~al.}(2021)\citenamefont {Peng},
  \citenamefont {Yin}, \citenamefont {Huang}, \citenamefont {Ramanathan},\ and\
  \citenamefont {Cappellaro}}]{peng2021floquet}%
  \BibitemOpen
  \bibfield  {author} {\bibinfo {author} {\bibfnamefont {P.}~\bibnamefont
  {Peng}}, \bibinfo {author} {\bibfnamefont {C.}~\bibnamefont {Yin}}, \bibinfo
  {author} {\bibfnamefont {X.}~\bibnamefont {Huang}}, \bibinfo {author}
  {\bibfnamefont {C.}~\bibnamefont {Ramanathan}},\ and\ \bibinfo {author}
  {\bibfnamefont {P.}~\bibnamefont {Cappellaro}},\ }\bibfield  {title}
  {\bibinfo {title} {Floquet prethermalization in dipolar spin chains},\ }\href
  {https://www.nature.com/articles/s41567-020-01120-z} {\bibfield  {journal}
  {\bibinfo  {journal} {Nature Physics}\ }\textbf {\bibinfo {volume} {17}},\
  \bibinfo {pages} {444} (\bibinfo {year} {2021})}\BibitemShut {NoStop}%
\bibitem [{\citenamefont {Abanin}\ \emph {et~al.}(2015)\citenamefont {Abanin},
  \citenamefont {De~Roeck},\ and\ \citenamefont {Huveneers}}]{abanin_15}%
  \BibitemOpen
  \bibfield  {author} {\bibinfo {author} {\bibfnamefont {D.~A.}\ \bibnamefont
  {Abanin}}, \bibinfo {author} {\bibfnamefont {W.}~\bibnamefont {De~Roeck}},\
  and\ \bibinfo {author} {\bibfnamefont {F.}~\bibnamefont {Huveneers}},\ }\href
  {http://link.aps.org/doi/10.1103/PhysRevLett.115.256803} {\bibfield
  {journal} {\bibinfo  {journal} {Phys. Rev. Lett.}\ }\textbf {\bibinfo
  {volume} {115}},\ \bibinfo {pages} {256803} (\bibinfo {year}
  {2015})}\BibitemShut {NoStop}%
\bibitem [{\citenamefont {Mori}\ \emph {et~al.}(2016)\citenamefont {Mori},
  \citenamefont {Kuwahara},\ and\ \citenamefont {Saito}}]{mori_15}%
  \BibitemOpen
  \bibfield  {author} {\bibinfo {author} {\bibfnamefont {T.}~\bibnamefont
  {Mori}}, \bibinfo {author} {\bibfnamefont {T.}~\bibnamefont {Kuwahara}},\
  and\ \bibinfo {author} {\bibfnamefont {K.}~\bibnamefont {Saito}},\ }\href
  {http://journals.aps.org/prl/abstract/10.1103/PhysRevLett.116.120401}
  {\bibfield  {journal} {\bibinfo  {journal} {Phys. Rev. Lett.}\ }\textbf
  {\bibinfo {volume} {116}},\ \bibinfo {pages} {120401} (\bibinfo {year}
  {2016})}\BibitemShut {NoStop}%
\bibitem [{\citenamefont {Machado}\ \emph {et~al.}(2020)\citenamefont
  {Machado}, \citenamefont {Else}, \citenamefont {Kahanamoku-Meyer},
  \citenamefont {Nayak},\ and\ \citenamefont {Yao}}]{Machado2020}%
  \BibitemOpen
  \bibfield  {author} {\bibinfo {author} {\bibfnamefont {F.}~\bibnamefont
  {Machado}}, \bibinfo {author} {\bibfnamefont {D.~V.}\ \bibnamefont {Else}},
  \bibinfo {author} {\bibfnamefont {G.~D.}\ \bibnamefont {Kahanamoku-Meyer}},
  \bibinfo {author} {\bibfnamefont {C.}~\bibnamefont {Nayak}},\ and\ \bibinfo
  {author} {\bibfnamefont {N.~Y.}\ \bibnamefont {Yao}},\ }\bibfield  {title}
  {\bibinfo {title} {Long-range prethermal phases of nonequilibrium matter},\
  }\href {https://doi.org/10.1103/PhysRevX.10.011043} {\bibfield  {journal}
  {\bibinfo  {journal} {Phys. Rev. X}\ }\textbf {\bibinfo {volume} {10}},\
  \bibinfo {pages} {011043} (\bibinfo {year} {2020})}\BibitemShut {NoStop}%
\bibitem [{\citenamefont {Else}\ \emph {et~al.}(2017)\citenamefont {Else},
  \citenamefont {Bauer},\ and\ \citenamefont {Nayak}}]{Else17}%
  \BibitemOpen
  \bibfield  {author} {\bibinfo {author} {\bibfnamefont {D.~V.}\ \bibnamefont
  {Else}}, \bibinfo {author} {\bibfnamefont {B.}~\bibnamefont {Bauer}},\ and\
  \bibinfo {author} {\bibfnamefont {C.}~\bibnamefont {Nayak}},\ }\bibfield
  {title} {\bibinfo {title} {Prethermal phases of matter protected by
  time-translation symmetry},\ }\href
  {https://doi.org/10.1103/PhysRevX.7.011026} {\bibfield  {journal} {\bibinfo
  {journal} {Phys. Rev. X}\ }\textbf {\bibinfo {volume} {7}},\ \bibinfo {pages}
  {011026} (\bibinfo {year} {2017})}\BibitemShut {NoStop}%
\bibitem [{\citenamefont {Pizzi}\ \emph
  {et~al.}(2021{\natexlab{a}})\citenamefont {Pizzi}, \citenamefont
  {Nunnenkamp},\ and\ \citenamefont {Knolle}}]{Pizzi21}%
  \BibitemOpen
  \bibfield  {author} {\bibinfo {author} {\bibfnamefont {A.}~\bibnamefont
  {Pizzi}}, \bibinfo {author} {\bibfnamefont {A.}~\bibnamefont {Nunnenkamp}},\
  and\ \bibinfo {author} {\bibfnamefont {J.}~\bibnamefont {Knolle}},\
  }\bibfield  {title} {\bibinfo {title} {Classical prethermal phases of
  matter},\ }\href {https://doi.org/10.1103/PhysRevLett.127.140602} {\bibfield
  {journal} {\bibinfo  {journal} {Phys. Rev. Lett.}\ }\textbf {\bibinfo
  {volume} {127}},\ \bibinfo {pages} {140602} (\bibinfo {year}
  {2021}{\natexlab{a}})}\BibitemShut {NoStop}%
\bibitem [{\citenamefont {Ye}\ \emph {et~al.}(2021)\citenamefont {Ye},
  \citenamefont {Machado},\ and\ \citenamefont {Yao}}]{ye2021floquet}%
  \BibitemOpen
  \bibfield  {author} {\bibinfo {author} {\bibfnamefont {B.}~\bibnamefont
  {Ye}}, \bibinfo {author} {\bibfnamefont {F.}~\bibnamefont {Machado}},\ and\
  \bibinfo {author} {\bibfnamefont {N.~Y.}\ \bibnamefont {Yao}},\ }\bibfield
  {title} {\bibinfo {title} {Floquet phases of matter via classical
  prethermalization},\ }\href {https://doi.org/10.1103/PhysRevLett.127.140603}
  {\bibfield  {journal} {\bibinfo  {journal} {Phys. Rev. Lett.}\ }\textbf
  {\bibinfo {volume} {127}},\ \bibinfo {pages} {140603} (\bibinfo {year}
  {2021})}\BibitemShut {NoStop}%
\bibitem [{\citenamefont {Yao}\ \emph {et~al.}(2020)\citenamefont {Yao},
  \citenamefont {Nayak}, \citenamefont {Balents},\ and\ \citenamefont
  {Zaletel}}]{yao2020classical}%
  \BibitemOpen
  \bibfield  {author} {\bibinfo {author} {\bibfnamefont {N.~Y.}\ \bibnamefont
  {Yao}}, \bibinfo {author} {\bibfnamefont {C.}~\bibnamefont {Nayak}}, \bibinfo
  {author} {\bibfnamefont {L.}~\bibnamefont {Balents}},\ and\ \bibinfo {author}
  {\bibfnamefont {M.~P.}\ \bibnamefont {Zaletel}},\ }\bibfield  {title}
  {\bibinfo {title} {Classical discrete time crystals},\ }\href
  {https://www.nature.com/articles/s41567-019-0782-3} {\bibfield  {journal}
  {\bibinfo  {journal} {Nature Physics}\ }\textbf {\bibinfo {volume} {16}},\
  \bibinfo {pages} {438} (\bibinfo {year} {2020})}\BibitemShut {NoStop}%
\bibitem [{\citenamefont {Rovny}\ \emph
  {et~al.}(2018{\natexlab{b}})\citenamefont {Rovny}, \citenamefont {Blum},\
  and\ \citenamefont {Barrett}}]{Rovny18}%
  \BibitemOpen
  \bibfield  {author} {\bibinfo {author} {\bibfnamefont {J.}~\bibnamefont
  {Rovny}}, \bibinfo {author} {\bibfnamefont {R.~L.}\ \bibnamefont {Blum}},\
  and\ \bibinfo {author} {\bibfnamefont {S.~E.}\ \bibnamefont {Barrett}},\
  }\bibfield  {title} {\bibinfo {title} {Observation of discrete-time-crystal
  signatures in an ordered dipolar many-body system},\ }\href
  {https://doi.org/10.1103/PhysRevLett.120.180603} {\bibfield  {journal}
  {\bibinfo  {journal} {Phys. Rev. Lett.}\ }\textbf {\bibinfo {volume} {120}},\
  \bibinfo {pages} {180603} (\bibinfo {year} {2018}{\natexlab{b}})}\BibitemShut
  {NoStop}%
\bibitem [{\citenamefont {Kyprianidis}\ \emph {et~al.}(2021)\citenamefont
  {Kyprianidis}, \citenamefont {Machado}, \citenamefont {Morong}, \citenamefont
  {Becker}, \citenamefont {Collins}, \citenamefont {Else}, \citenamefont
  {Feng}, \citenamefont {Hess}, \citenamefont {Nayak}, \citenamefont {Pagano},
  \citenamefont {Yao},\ and\ \citenamefont {Monroe}}]{Kyprianidis21}%
  \BibitemOpen
  \bibfield  {author} {\bibinfo {author} {\bibfnamefont {A.}~\bibnamefont
  {Kyprianidis}}, \bibinfo {author} {\bibfnamefont {F.}~\bibnamefont
  {Machado}}, \bibinfo {author} {\bibfnamefont {W.}~\bibnamefont {Morong}},
  \bibinfo {author} {\bibfnamefont {P.}~\bibnamefont {Becker}}, \bibinfo
  {author} {\bibfnamefont {K.~S.}\ \bibnamefont {Collins}}, \bibinfo {author}
  {\bibfnamefont {D.~V.}\ \bibnamefont {Else}}, \bibinfo {author}
  {\bibfnamefont {L.}~\bibnamefont {Feng}}, \bibinfo {author} {\bibfnamefont
  {P.~W.}\ \bibnamefont {Hess}}, \bibinfo {author} {\bibfnamefont
  {C.}~\bibnamefont {Nayak}}, \bibinfo {author} {\bibfnamefont
  {G.}~\bibnamefont {Pagano}}, \bibinfo {author} {\bibfnamefont {N.~Y.}\
  \bibnamefont {Yao}},\ and\ \bibinfo {author} {\bibfnamefont {C.}~\bibnamefont
  {Monroe}},\ }\bibfield  {title} {\bibinfo {title} {Observation of a
  prethermal discrete time crystal},\ }\href
  {https://doi.org/https://doi.org/10.1126/science.abg8102} {\bibfield
  {journal} {\bibinfo  {journal} {Science}\ }\textbf {\bibinfo {volume}
  {372}},\ \bibinfo {pages} {1192} (\bibinfo {year} {2021})}\BibitemShut
  {NoStop}%
\bibitem [{Note1()}]{Note1}%
  \BibitemOpen
  \bibinfo {note} {The term frequency refers to the inverse periods of the two
  superimposed drives, rather than their Fourier decompositions.}\BibitemShut
  {Stop}%
\bibitem [{SM()}]{SM}%
  \BibitemOpen
  \bibinfo {note} {See Supplemental Material.}\BibitemShut {Stop}%
\bibitem [{PDT(2021{\natexlab{a}})}]{PDTC_zoomed}%
  \BibitemOpen
  \href@noop {} {}\bibinfo {howpublished} {Movie of full dataset from Fig. 2A
  of main text: \url{https://youtu.be/61ZqLgbCuyo}} (\bibinfo {year}
  {2021}{\natexlab{a}})\BibitemShut {NoStop}%
\bibitem [{PDT(2021{\natexlab{b}})}]{PDTC_video55}%
  \BibitemOpen
  \href@noop {} {}\bibinfo {howpublished} {Movie of full dataset from Fig. 2B
  and Fig. 3A of main text (first 55 Floquet cycles):
  \url{https://youtu.be/m5iASnBZ9oo}} (\bibinfo {year}
  {2021}{\natexlab{b}})\BibitemShut {NoStop}%
\bibitem [{\citenamefont {Ajoy}\ \emph
  {et~al.}(2018{\natexlab{a}})\citenamefont {Ajoy}, \citenamefont {Liu},
  \citenamefont {Nazaryan}, \citenamefont {Lv}, \citenamefont {Zangara},
  \citenamefont {Safvati}, \citenamefont {Wang}, \citenamefont {Arnold},
  \citenamefont {Li}, \citenamefont {Lin} \emph {et~al.}}]{Ajoy17}%
  \BibitemOpen
  \bibfield  {author} {\bibinfo {author} {\bibfnamefont {A.}~\bibnamefont
  {Ajoy}}, \bibinfo {author} {\bibfnamefont {K.}~\bibnamefont {Liu}}, \bibinfo
  {author} {\bibfnamefont {R.}~\bibnamefont {Nazaryan}}, \bibinfo {author}
  {\bibfnamefont {X.}~\bibnamefont {Lv}}, \bibinfo {author} {\bibfnamefont
  {P.~R.}\ \bibnamefont {Zangara}}, \bibinfo {author} {\bibfnamefont
  {B.}~\bibnamefont {Safvati}}, \bibinfo {author} {\bibfnamefont
  {G.}~\bibnamefont {Wang}}, \bibinfo {author} {\bibfnamefont {D.}~\bibnamefont
  {Arnold}}, \bibinfo {author} {\bibfnamefont {G.}~\bibnamefont {Li}}, \bibinfo
  {author} {\bibfnamefont {A.}~\bibnamefont {Lin}}, \emph {et~al.},\ }\bibfield
   {title} {\bibinfo {title} {Orientation-independent room temperature optical
  13c hyperpolarization in powdered diamond},\ }\href
  {http://advances.sciencemag.org/content/4/5/eaar5492} {\bibfield  {journal}
  {\bibinfo  {journal} {Sci. Adv.}\ }\textbf {\bibinfo {volume} {4}},\ \bibinfo
  {pages} {eaar5492} (\bibinfo {year} {2018}{\natexlab{a}})}\BibitemShut
  {NoStop}%
\bibitem [{\citenamefont {Ajoy}\ \emph
  {et~al.}(2018{\natexlab{b}})\citenamefont {Ajoy}, \citenamefont {Nazaryan},
  \citenamefont {Liu}, \citenamefont {Lv}, \citenamefont {Safvati},
  \citenamefont {Wang}, \citenamefont {Druga}, \citenamefont {Reimer},
  \citenamefont {Suter}, \citenamefont {Ramanathan} \emph {et~al.}}]{Ajoy18}%
  \BibitemOpen
  \bibfield  {author} {\bibinfo {author} {\bibfnamefont {A.}~\bibnamefont
  {Ajoy}}, \bibinfo {author} {\bibfnamefont {R.}~\bibnamefont {Nazaryan}},
  \bibinfo {author} {\bibfnamefont {K.}~\bibnamefont {Liu}}, \bibinfo {author}
  {\bibfnamefont {X.}~\bibnamefont {Lv}}, \bibinfo {author} {\bibfnamefont
  {B.}~\bibnamefont {Safvati}}, \bibinfo {author} {\bibfnamefont
  {G.}~\bibnamefont {Wang}}, \bibinfo {author} {\bibfnamefont {E.}~\bibnamefont
  {Druga}}, \bibinfo {author} {\bibfnamefont {J.}~\bibnamefont {Reimer}},
  \bibinfo {author} {\bibfnamefont {D.}~\bibnamefont {Suter}}, \bibinfo
  {author} {\bibfnamefont {C.}~\bibnamefont {Ramanathan}}, \emph {et~al.},\
  }\bibfield  {title} {\bibinfo {title} {Enhanced dynamic nuclear polarization
  via swept microwave frequency combs},\ }\href
  {https://doi.org/https://doi.org/10.1073/pnas.1807125115} {\bibfield
  {journal} {\bibinfo  {journal} {Proceedings of the National Academy of
  Sciences}\ }\textbf {\bibinfo {volume} {115}},\ \bibinfo {pages} {10576}
  (\bibinfo {year} {2018}{\natexlab{b}})}\BibitemShut {NoStop}%
\bibitem [{\citenamefont {Duer}(2004)}]{Duer04}%
  \BibitemOpen
  \bibfield  {author} {\bibinfo {author} {\bibfnamefont {M.}~\bibnamefont
  {Duer}},\ }\href@noop {} {\emph {\bibinfo {title} {Introduction to
  Solid-State NMR Spectroscopy}}}\ (\bibinfo  {publisher} {John Wiley $\&$
  Sons},\ \bibinfo {year} {2004})\BibitemShut {NoStop}%
\bibitem [{\citenamefont {Beatrez}\ \emph {et~al.}(2021)\citenamefont
  {Beatrez}, \citenamefont {Janes}, \citenamefont {Akkiraju}, \citenamefont
  {Pillai}, \citenamefont {Oddo}, \citenamefont {Reshetikhin}, \citenamefont
  {Druga}, \citenamefont {McAllister}, \citenamefont {Elo}, \citenamefont
  {Gilbert}, \citenamefont {Suter},\ and\ \citenamefont {Ajoy}}]{Beatrez21}%
  \BibitemOpen
  \bibfield  {author} {\bibinfo {author} {\bibfnamefont {W.}~\bibnamefont
  {Beatrez}}, \bibinfo {author} {\bibfnamefont {O.}~\bibnamefont {Janes}},
  \bibinfo {author} {\bibfnamefont {A.}~\bibnamefont {Akkiraju}}, \bibinfo
  {author} {\bibfnamefont {A.}~\bibnamefont {Pillai}}, \bibinfo {author}
  {\bibfnamefont {A.}~\bibnamefont {Oddo}}, \bibinfo {author} {\bibfnamefont
  {P.}~\bibnamefont {Reshetikhin}}, \bibinfo {author} {\bibfnamefont
  {E.}~\bibnamefont {Druga}}, \bibinfo {author} {\bibfnamefont
  {M.}~\bibnamefont {McAllister}}, \bibinfo {author} {\bibfnamefont
  {M.}~\bibnamefont {Elo}}, \bibinfo {author} {\bibfnamefont {B.}~\bibnamefont
  {Gilbert}}, \bibinfo {author} {\bibfnamefont {D.}~\bibnamefont {Suter}},\
  and\ \bibinfo {author} {\bibfnamefont {A.}~\bibnamefont {Ajoy}},\ }\bibfield
  {title} {\bibinfo {title} {Floquet prethermalization with lifetime exceeding
  90 s in a bulk hyperpolarized solid},\ }\href
  {https://doi.org/10.1103/PhysRevLett.127.170603} {\bibfield  {journal}
  {\bibinfo  {journal} {Phys. Rev. Lett.}\ }\textbf {\bibinfo {volume} {127}},\
  \bibinfo {pages} {170603} (\bibinfo {year} {2021})}\BibitemShut {NoStop}%
\bibitem [{\citenamefont {Reynhardt}(2003)}]{Reynhardt03a}%
  \BibitemOpen
  \bibfield  {author} {\bibinfo {author} {\bibfnamefont {E.}~\bibnamefont
  {Reynhardt}},\ }\bibfield  {title} {\bibinfo {title} {Spin lattice relaxation
  of spin-ï¿½ nuclei in solids containing diluted paramagnetic impurity
  centers. i. zeeman polarization of nuclear spin system},\ }\href
  {http://dx.doi.org/10.1002/cmr.a.10077} {\bibfield  {journal} {\bibinfo
  {journal} {Concepts in Magnetic Resonance Part A}\ }\textbf {\bibinfo
  {volume} {19A}},\ \bibinfo {pages} {20} (\bibinfo {year} {2003})}\BibitemShut
  {NoStop}%
\bibitem [{\citenamefont {Ajoy}\ \emph {et~al.}(2019)\citenamefont {Ajoy},
  \citenamefont {Safvati}, \citenamefont {Nazaryan}, \citenamefont {Oon},
  \citenamefont {Han}, \citenamefont {Raghavan}, \citenamefont {Nirodi},
  \citenamefont {Aguilar}, \citenamefont {Liu}, \citenamefont {Cai},
  \citenamefont {Lv}, \citenamefont {Druga}, \citenamefont {Ramanathan},
  \citenamefont {Reimer}, \citenamefont {Meriles}, \citenamefont {Suter},\ and\
  \citenamefont {Pines}}]{Ajoy19relax}%
  \BibitemOpen
  \bibfield  {author} {\bibinfo {author} {\bibfnamefont {A.}~\bibnamefont
  {Ajoy}}, \bibinfo {author} {\bibfnamefont {B.}~\bibnamefont {Safvati}},
  \bibinfo {author} {\bibfnamefont {R.}~\bibnamefont {Nazaryan}}, \bibinfo
  {author} {\bibfnamefont {J.~T.}\ \bibnamefont {Oon}}, \bibinfo {author}
  {\bibfnamefont {B.}~\bibnamefont {Han}}, \bibinfo {author} {\bibfnamefont
  {P.}~\bibnamefont {Raghavan}}, \bibinfo {author} {\bibfnamefont
  {R.}~\bibnamefont {Nirodi}}, \bibinfo {author} {\bibfnamefont
  {A.}~\bibnamefont {Aguilar}}, \bibinfo {author} {\bibfnamefont
  {K.}~\bibnamefont {Liu}}, \bibinfo {author} {\bibfnamefont {X.}~\bibnamefont
  {Cai}}, \bibinfo {author} {\bibfnamefont {X.}~\bibnamefont {Lv}}, \bibinfo
  {author} {\bibfnamefont {E.}~\bibnamefont {Druga}}, \bibinfo {author}
  {\bibfnamefont {C.}~\bibnamefont {Ramanathan}}, \bibinfo {author}
  {\bibfnamefont {J.~A.}\ \bibnamefont {Reimer}}, \bibinfo {author}
  {\bibfnamefont {C.~A.}\ \bibnamefont {Meriles}}, \bibinfo {author}
  {\bibfnamefont {D.}~\bibnamefont {Suter}},\ and\ \bibinfo {author}
  {\bibfnamefont {A.}~\bibnamefont {Pines}},\ }\bibfield  {title} {\bibinfo
  {title} {Hyperpolarized relaxometry based nuclear t1 noise spectroscopy in
  diamond},\ }\href
  {https://doi.org/https://doi.org/10.1038/s41467-019-13042-3} {\bibfield
  {journal} {\bibinfo  {journal} {Nature Communications}\ }\textbf {\bibinfo
  {volume} {10}},\ \bibinfo {pages} {5160} (\bibinfo {year}
  {2019})}\BibitemShut {NoStop}%
\bibitem [{\citenamefont {Luitz}\ \emph {et~al.}(2020)\citenamefont {Luitz},
  \citenamefont {Moessner}, \citenamefont {Sondhi},\ and\ \citenamefont
  {Khemani}}]{luitz2020prethermalization}%
  \BibitemOpen
  \bibfield  {author} {\bibinfo {author} {\bibfnamefont {D.~J.}\ \bibnamefont
  {Luitz}}, \bibinfo {author} {\bibfnamefont {R.}~\bibnamefont {Moessner}},
  \bibinfo {author} {\bibfnamefont {S.~L.}\ \bibnamefont {Sondhi}},\ and\
  \bibinfo {author} {\bibfnamefont {V.}~\bibnamefont {Khemani}},\ }\bibfield
  {title} {\bibinfo {title} {Prethermalization without temperature},\ }\href
  {https://doi.org/10.1103/PhysRevX.10.021046} {\bibfield  {journal} {\bibinfo
  {journal} {Phys. Rev. X}\ }\textbf {\bibinfo {volume} {10}},\ \bibinfo
  {pages} {021046} (\bibinfo {year} {2020})}\BibitemShut {NoStop}%
\bibitem [{\citenamefont {Pizzi}\ \emph
  {et~al.}(2021{\natexlab{b}})\citenamefont {Pizzi}, \citenamefont {Knolle},\
  and\ \citenamefont {Nunnenkamp}}]{pizzi2021higher}%
  \BibitemOpen
  \bibfield  {author} {\bibinfo {author} {\bibfnamefont {A.}~\bibnamefont
  {Pizzi}}, \bibinfo {author} {\bibfnamefont {J.}~\bibnamefont {Knolle}},\ and\
  \bibinfo {author} {\bibfnamefont {A.}~\bibnamefont {Nunnenkamp}},\ }\bibfield
   {title} {\bibinfo {title} {Higher-order and fractional discrete time
  crystals in clean long-range interacting systems},\ }\href
  {https://www.nature.com/articles/s41467-021-22583-5} {\bibfield  {journal}
  {\bibinfo  {journal} {Nature communications}\ }\textbf {\bibinfo {volume}
  {12}},\ \bibinfo {pages} {1} (\bibinfo {year}
  {2021}{\natexlab{b}})}\BibitemShut {NoStop}%
\bibitem [{\citenamefont {Else}\ \emph
  {et~al.}(2020{\natexlab{b}})\citenamefont {Else}, \citenamefont {Ho},\ and\
  \citenamefont {Dumitrescu}}]{else2020longlived}%
  \BibitemOpen
  \bibfield  {author} {\bibinfo {author} {\bibfnamefont {D.~V.}\ \bibnamefont
  {Else}}, \bibinfo {author} {\bibfnamefont {W.~W.}\ \bibnamefont {Ho}},\ and\
  \bibinfo {author} {\bibfnamefont {P.~T.}\ \bibnamefont {Dumitrescu}},\
  }\bibfield  {title} {\bibinfo {title} {Long-lived interacting phases of
  matter protected by multiple time-translation symmetries in quasiperiodically
  driven systems},\ }\href {https://doi.org/10.1103/PhysRevX.10.021032}
  {\bibfield  {journal} {\bibinfo  {journal} {Phys. Rev. X}\ }\textbf {\bibinfo
  {volume} {10}},\ \bibinfo {pages} {021032} (\bibinfo {year}
  {2020}{\natexlab{b}})}\BibitemShut {NoStop}%
\bibitem [{\citenamefont {Weinberg}\ and\ \citenamefont
  {Bukov}(2017)}]{weinberg2017quspin}%
  \BibitemOpen
  \bibfield  {author} {\bibinfo {author} {\bibfnamefont {P.}~\bibnamefont
  {Weinberg}}\ and\ \bibinfo {author} {\bibfnamefont {M.}~\bibnamefont
  {Bukov}},\ }\bibfield  {title} {\bibinfo {title} {{QuSpin: a Python Package
  for Dynamics and Exact Diagonalisation of Quantum Many Body Systems part I:
  spin chains}},\ }\href {https://doi.org/10.21468/SciPostPhys.2.1.003}
  {\bibfield  {journal} {\bibinfo  {journal} {SciPost Phys.}\ }\textbf
  {\bibinfo {volume} {2}},\ \bibinfo {pages} {003} (\bibinfo {year}
  {2017})}\BibitemShut {NoStop}%
\bibitem [{\citenamefont {Khodjasteh}\ and\ \citenamefont
  {Lidar}(2005)}]{khodjasteh2005fault}%
  \BibitemOpen
  \bibfield  {author} {\bibinfo {author} {\bibfnamefont {K.}~\bibnamefont
  {Khodjasteh}}\ and\ \bibinfo {author} {\bibfnamefont {D.~A.}\ \bibnamefont
  {Lidar}},\ }\bibfield  {title} {\bibinfo {title} {Fault-tolerant quantum
  dynamical decoupling},\ }\href
  {https://doi.org/10.1103/PhysRevLett.95.180501} {\bibfield  {journal}
  {\bibinfo  {journal} {Phys. Rev. Lett.}\ }\textbf {\bibinfo {volume} {95}},\
  \bibinfo {pages} {180501} (\bibinfo {year} {2005})}\BibitemShut {NoStop}%
\bibitem [{\citenamefont {Witzel}\ and\ \citenamefont
  {Das~Sarma}(2007)}]{witzel2007concatenated}%
  \BibitemOpen
  \bibfield  {author} {\bibinfo {author} {\bibfnamefont {W.~M.}\ \bibnamefont
  {Witzel}}\ and\ \bibinfo {author} {\bibfnamefont {S.}~\bibnamefont
  {Das~Sarma}},\ }\bibfield  {title} {\bibinfo {title} {Concatenated dynamical
  decoupling in a solid-state spin bath},\ }\href
  {https://doi.org/10.1103/PhysRevB.76.241303} {\bibfield  {journal} {\bibinfo
  {journal} {Phys. Rev. B}\ }\textbf {\bibinfo {volume} {76}},\ \bibinfo
  {pages} {241303} (\bibinfo {year} {2007})}\BibitemShut {NoStop}%
\bibitem [{\citenamefont {Cai}\ \emph {et~al.}(2012)\citenamefont {Cai},
  \citenamefont {Naydenov}, \citenamefont {Pfeiffer}, \citenamefont
  {McGuinness}, \citenamefont {Jahnke}, \citenamefont {Jelezko}, \citenamefont
  {Plenio},\ and\ \citenamefont {Retzker}}]{cai2012robust}%
  \BibitemOpen
  \bibfield  {author} {\bibinfo {author} {\bibfnamefont {J.}~\bibnamefont
  {Cai}}, \bibinfo {author} {\bibfnamefont {B.}~\bibnamefont {Naydenov}},
  \bibinfo {author} {\bibfnamefont {R.}~\bibnamefont {Pfeiffer}}, \bibinfo
  {author} {\bibfnamefont {L.~P.}\ \bibnamefont {McGuinness}}, \bibinfo
  {author} {\bibfnamefont {K.~D.}\ \bibnamefont {Jahnke}}, \bibinfo {author}
  {\bibfnamefont {F.}~\bibnamefont {Jelezko}}, \bibinfo {author} {\bibfnamefont
  {M.~B.}\ \bibnamefont {Plenio}},\ and\ \bibinfo {author} {\bibfnamefont
  {A.}~\bibnamefont {Retzker}},\ }\bibfield  {title} {\bibinfo {title} {Robust
  dynamical decoupling with concatenated continuous driving},\ }\href
  {https://iopscience.iop.org/article/10.1088/1367-2630/14/11/113023/meta}
  {\bibfield  {journal} {\bibinfo  {journal} {New Journal of Physics}\ }\textbf
  {\bibinfo {volume} {14}},\ \bibinfo {pages} {113023} (\bibinfo {year}
  {2012})}\BibitemShut {NoStop}%
\bibitem [{\citenamefont {Hayes}\ \emph {et~al.}(2014)\citenamefont {Hayes},
  \citenamefont {Flammia},\ and\ \citenamefont
  {Biercuk}}]{hayes2014programmable}%
  \BibitemOpen
  \bibfield  {author} {\bibinfo {author} {\bibfnamefont {D.}~\bibnamefont
  {Hayes}}, \bibinfo {author} {\bibfnamefont {S.~T.}\ \bibnamefont {Flammia}},\
  and\ \bibinfo {author} {\bibfnamefont {M.~J.}\ \bibnamefont {Biercuk}},\
  }\bibfield  {title} {\bibinfo {title} {Programmable quantum simulation by
  dynamic hamiltonian engineering},\ }\href
  {https://iopscience.iop.org/article/10.1088/1367-2630/16/8/083027/meta}
  {\bibfield  {journal} {\bibinfo  {journal} {New Journal of Physics}\ }\textbf
  {\bibinfo {volume} {16}},\ \bibinfo {pages} {083027} (\bibinfo {year}
  {2014})}\BibitemShut {NoStop}%
\bibitem [{\citenamefont {Vajna}\ \emph {et~al.}(2018)\citenamefont {Vajna},
  \citenamefont {Klobas}, \citenamefont {Prosen},\ and\ \citenamefont
  {Polkovnikov}}]{vajna2018replica}%
  \BibitemOpen
  \bibfield  {author} {\bibinfo {author} {\bibfnamefont {S.}~\bibnamefont
  {Vajna}}, \bibinfo {author} {\bibfnamefont {K.}~\bibnamefont {Klobas}},
  \bibinfo {author} {\bibfnamefont {T.}~\bibnamefont {Prosen}},\ and\ \bibinfo
  {author} {\bibfnamefont {A.}~\bibnamefont {Polkovnikov}},\ }\bibfield
  {title} {\bibinfo {title} {Replica resummation of the
  baker-campbell-hausdorff series},\ }\href
  {https://doi.org/10.1103/PhysRevLett.120.200607} {\bibfield  {journal}
  {\bibinfo  {journal} {Phys. Rev. Lett.}\ }\textbf {\bibinfo {volume} {120}},\
  \bibinfo {pages} {200607} (\bibinfo {year} {2018})}\BibitemShut {NoStop}%
\bibitem [{\citenamefont {Fleckenstein}\ and\ \citenamefont
  {Bukov}(2021{\natexlab{a}})}]{fleckenstein2021prethermalization}%
  \BibitemOpen
  \bibfield  {author} {\bibinfo {author} {\bibfnamefont {C.}~\bibnamefont
  {Fleckenstein}}\ and\ \bibinfo {author} {\bibfnamefont {M.}~\bibnamefont
  {Bukov}},\ }\bibfield  {title} {\bibinfo {title} {Prethermalization and
  thermalization in periodically driven many-body systems away from the
  high-frequency limit},\ }\href {https://doi.org/10.1103/PhysRevB.103.L140302}
  {\bibfield  {journal} {\bibinfo  {journal} {Phys. Rev. B}\ }\textbf {\bibinfo
  {volume} {103}},\ \bibinfo {pages} {L140302} (\bibinfo {year}
  {2021}{\natexlab{a}})}\BibitemShut {NoStop}%
\bibitem [{\citenamefont {Fleckenstein}\ and\ \citenamefont
  {Bukov}(2021{\natexlab{b}})}]{fleckenstein2021thermalization}%
  \BibitemOpen
  \bibfield  {author} {\bibinfo {author} {\bibfnamefont {C.}~\bibnamefont
  {Fleckenstein}}\ and\ \bibinfo {author} {\bibfnamefont {M.}~\bibnamefont
  {Bukov}},\ }\bibfield  {title} {\bibinfo {title} {Thermalization and
  prethermalization in periodically kicked quantum spin chains},\ }\href
  {https://doi.org/10.1103/PhysRevB.103.144307} {\bibfield  {journal} {\bibinfo
   {journal} {Phys. Rev. B}\ }\textbf {\bibinfo {volume} {103}},\ \bibinfo
  {pages} {144307} (\bibinfo {year} {2021}{\natexlab{b}})}\BibitemShut
  {NoStop}%
\bibitem [{PDT(2021{\natexlab{c}})}]{PDTC_video25}%
  \BibitemOpen
  \href@noop {} {}\bibinfo {howpublished} {Movie of full dataset from Fig. 2B
  and Fig. 3A of main text (first 25 Floquet cycles):
  \url{https://youtu.be/_PRlBEPO_54}} (\bibinfo {year}
  {2021}{\natexlab{c}})\BibitemShut {NoStop}%
\bibitem [{PDT(2021{\natexlab{d}})}]{PDTC_video155}%
  \BibitemOpen
  \href@noop {} {}\bibinfo {howpublished} {Movie of full dataset from Fig. 2B
  and Fig. 3A of main text (first 155 Floquet cycles):
  \url{https://youtu.be/YHm4wRz0sfQ}} (\bibinfo {year}
  {2021}{\natexlab{d}})\BibitemShut {NoStop}%
\bibitem [{\citenamefont {Magnus}(1954)}]{Magnus54}%
  \BibitemOpen
  \bibfield  {author} {\bibinfo {author} {\bibfnamefont {W.}~\bibnamefont
  {Magnus}},\ }\bibfield  {title} {\bibinfo {title} {On the exponential
  solution of differential equations for a linear operator},\ }\href@noop {}
  {\bibfield  {journal} {\bibinfo  {journal} {Communications on Pure and
  Applied Mathematics}\ }\textbf {\bibinfo {volume} {7}},\ \bibinfo {pages}
  {649} (\bibinfo {year} {1954})}\BibitemShut {NoStop}%
\bibitem [{\citenamefont {Haeberlen}(1976)}]{Haeberlen76}%
  \BibitemOpen
  \bibfield  {author} {\bibinfo {author} {\bibfnamefont {U.}~\bibnamefont
  {Haeberlen}},\ }\href@noop {} {\emph {\bibinfo {title} {High Resolution NMR
  in Solids: Selective Averaging}}}\ (\bibinfo  {publisher} {Academic Press
  Inc., New York},\ \bibinfo {year} {1976})\BibitemShut {NoStop}%
\bibitem [{\citenamefont {Blanes}\ \emph {et~al.}(2009)\citenamefont {Blanes},
  \citenamefont {Casas}, \citenamefont {Oteo},\ and\ \citenamefont
  {Ros}}]{blanes2009}%
  \BibitemOpen
  \bibfield  {author} {\bibinfo {author} {\bibfnamefont {S.}~\bibnamefont
  {Blanes}}, \bibinfo {author} {\bibfnamefont {F.}~\bibnamefont {Casas}},
  \bibinfo {author} {\bibfnamefont {J.}~\bibnamefont {Oteo}},\ and\ \bibinfo
  {author} {\bibfnamefont {J.}~\bibnamefont {Ros}},\ }\bibfield  {title}
  {\bibinfo {title} {The magnus expansion and some of its applications},\
  }\href {https://doi.org/https://doi.org/10.1016/j.physrep.2008.11.001}
  {\bibfield  {journal} {\bibinfo  {journal} {Physics Reports}\ }\textbf
  {\bibinfo {volume} {470}},\ \bibinfo {pages} {151} (\bibinfo {year}
  {2009})}\BibitemShut {NoStop}%
\bibitem [{\citenamefont {Ajoy}\ \emph {et~al.}(2020)\citenamefont {Ajoy},
  \citenamefont {Nirodi}, \citenamefont {Sarkar}, \citenamefont {Reshetikhin},
  \citenamefont {Druga}, \citenamefont {Akkiraju}, \citenamefont {McAllister},
  \citenamefont {Maineri}, \citenamefont {Le}, \citenamefont {Lin},
  \citenamefont {Souza}, \citenamefont {Meriles}, \citenamefont {Gilbert},
  \citenamefont {Suter}, \citenamefont {Reimer},\ and\ \citenamefont
  {Pines}}]{ajoy2020}%
  \BibitemOpen
  \bibfield  {author} {\bibinfo {author} {\bibfnamefont {A.}~\bibnamefont
  {Ajoy}}, \bibinfo {author} {\bibfnamefont {R.}~\bibnamefont {Nirodi}},
  \bibinfo {author} {\bibfnamefont {A.}~\bibnamefont {Sarkar}}, \bibinfo
  {author} {\bibfnamefont {P.}~\bibnamefont {Reshetikhin}}, \bibinfo {author}
  {\bibfnamefont {E.}~\bibnamefont {Druga}}, \bibinfo {author} {\bibfnamefont
  {A.}~\bibnamefont {Akkiraju}}, \bibinfo {author} {\bibfnamefont
  {M.}~\bibnamefont {McAllister}}, \bibinfo {author} {\bibfnamefont
  {G.}~\bibnamefont {Maineri}}, \bibinfo {author} {\bibfnamefont
  {S.}~\bibnamefont {Le}}, \bibinfo {author} {\bibfnamefont {A.}~\bibnamefont
  {Lin}}, \bibinfo {author} {\bibfnamefont {A.~M.}\ \bibnamefont {Souza}},
  \bibinfo {author} {\bibfnamefont {C.~A.}\ \bibnamefont {Meriles}}, \bibinfo
  {author} {\bibfnamefont {B.}~\bibnamefont {Gilbert}}, \bibinfo {author}
  {\bibfnamefont {D.}~\bibnamefont {Suter}}, \bibinfo {author} {\bibfnamefont
  {J.~A.}\ \bibnamefont {Reimer}},\ and\ \bibinfo {author} {\bibfnamefont
  {A.}~\bibnamefont {Pines}},\ }\bibfield  {title} {\bibinfo {title} {Dynamical
  decoupling in interacting systems: applications to signal-enhanced
  hyperpolarized readout},\ }\href@noop {} {\bibfield  {journal} {\bibinfo
  {journal} {arXiv:2008.08323}\ } (\bibinfo {year} {2020})}\BibitemShut
  {NoStop}%
\bibitem [{\citenamefont {D'Alessio}\ \emph {et~al.}(2016)\citenamefont
  {D'Alessio}, \citenamefont {Kafri}, \citenamefont {Polkovnikov},\ and\
  \citenamefont {Rigol}}]{dalessio2016quantum}%
  \BibitemOpen
  \bibfield  {author} {\bibinfo {author} {\bibfnamefont {L.}~\bibnamefont
  {D'Alessio}}, \bibinfo {author} {\bibfnamefont {Y.}~\bibnamefont {Kafri}},
  \bibinfo {author} {\bibfnamefont {A.}~\bibnamefont {Polkovnikov}},\ and\
  \bibinfo {author} {\bibfnamefont {M.}~\bibnamefont {Rigol}},\ }\bibfield
  {title} {\bibinfo {title} {From quantum chaos and eigenstate thermalization
  to statistical mechanics and thermodynamics},\ }\href
  {https://doi.org/https://doi.org/10.1080/00018732.2016.1198134} {\bibfield
  {journal} {\bibinfo  {journal} {Advances in Physics}\ }\textbf {\bibinfo
  {volume} {65}},\ \bibinfo {pages} {239} (\bibinfo {year} {2016})}\BibitemShut
  {NoStop}%
\bibitem [{\citenamefont {Ikeda}\ and\ \citenamefont
  {Polkovnikov}(2021)}]{ikeda2021fermi}%
  \BibitemOpen
  \bibfield  {author} {\bibinfo {author} {\bibfnamefont {T.~N.}\ \bibnamefont
  {Ikeda}}\ and\ \bibinfo {author} {\bibfnamefont {A.}~\bibnamefont
  {Polkovnikov}},\ }\bibfield  {title} {\bibinfo {title} {Fermi's golden rule
  for heating in strongly driven floquet systems},\ }\href
  {https://doi.org/10.1103/PhysRevB.104.134308} {\bibfield  {journal} {\bibinfo
   {journal} {Phys. Rev. B}\ }\textbf {\bibinfo {volume} {104}},\ \bibinfo
  {pages} {134308} (\bibinfo {year} {2021})}\BibitemShut {NoStop}%
\end{thebibliography}%
